\documentstyle[preprint,harvard,aps,epsf]{revtex}

\newcommand{\id}{{\sf 1 \hspace{-0.3ex} \rule{0.1ex}{1.52ex}
\rule[-.01ex]{0.3ex}{0.1ex}}}
\newcommand{\pr}{{\,\sf P \hspace{-1.45ex} \rule{0.1ex}{1.54ex}
\hspace{1.25ex}}}
\newcommand{\biggreec} [1]{\mbox{\boldmath $#1$\unboldmath}}
\newcommand{\pro}{{\,\sf P \hspace{-1.45ex}
\rule{0.1ex}{1.54ex}\hspace{1.25ex}}_{0}}
\newcommand{\heff}{H_{\mbox{\scriptsize eff}}}
\newcommand{\eqn}[1]{Eq.\,(\ref{#1})}

\citationstyle{dcu}
\begin{document}
\title{The Quantum jump approach to dissipative dynamics in quantum
optics}
\author{M.B. Plenio and P.L. Knight}
\address{Optics Section, Blackett Laboratory, Imperial College 
London SW7 2BZ, UK}
\maketitle
\begin{abstract}
Dissipation, the irreversible loss of energy and coherence, from a
microsystem, is the result of coupling to a much larger macrosystem
(or reservoir) which is so large that one has no chance of keeping 
track of all of its degrees of freedom. The microsystem evolution is 
then described by tracing over the reservoir states, resulting in an 
irreversible decay as excitation leaks out of the initially excited 
microsystems into the outer reservoir environment. Earlier treatments
of this dissipation described an {\bf ensemble} of microsystems using
density matrices, either in Schr{\"o}dinger picture with Master equations, 
or in Heisenberg picture with Langevin equations. The development of
experimental techniques to study single quantum systems (for example 
single trapped ions, or cavity radiation field modes) has stimulated 
the construction of theoretical methods to describe individual 
realizations conditioned on a particular observation record of the 
decay channel, in the environment. These methods, variously described 
as Quantum Jump, Monte Carlo Wavefunction and Quantum Trajectory methods 
are the subject of this review article. We discuss their derivation, apply 
them to a number of current problems in quantum optics and relate them 
to ensemble descriptions.
\end{abstract}
\newpage
\tableofcontents
\newpage
\section{Introduction}
\label{Introduction}
Quantum mechanics is usually introduced as a theory for ensembles, 
but the invention of ion traps , for example, offers the possibility to 
observe and manipulate single particles, where observability of 
quantum jumps, which are not be seen directly in the ensemble, lead
to conceptual problems of how to describe single realizations of these 
systems. Usually Bloch equations or Einstein rate equations are used to 
describe the time evolution of ensembles of atoms or ions driven by light. 
New approaches via conditional time evolution, given say when no photon has 
been emitted, have been developed to describe single experimental 
realizations of quantum systems.   
This leads to a description of the system via wave functions instead 
of density matrices. This conditional "quantum trajectory" approach
is still an ensemble description, but for a sub ensemble where we know
when photons have been emitted. 

The jumps that occur in this description can be considered as due to the
increase of our knowledge about the system which is represented by the 
wave-function (or the density operator) describing the system. 
In the formalism to be presented one usually imagines that
gedanken measurements are performed in a rapid succession , for example, on 
the emitted radiation field. These will either have the result that a 
photon has been found in the environment or that no photon has been found. 
A sudden change in our information about the
radiation field (for example through the detection of a photon emitted by
the system into the environment) leads to a sudden change 
of the wave-function of the system. However, not only the detection of a
photon leads to an increase of information but also the failure to detect 
a photon (i.e. a null result). 
New insights have been obtained into atomic dynamics and into dissipative
processes, and new powerful theoretical approaches developed. Apart from
the new insights into physics these methods also allow the simulation of
complicated problems, e.g., in laser cooling that were completely intractable
using the master equation approach. In general it can be applied to all
master equations that are of Lindblad form, which is in fact the most general 
form of a master equation.

This article reviews the various quantum jump approaches developed over 
the past few years. We focus on the theoretical description of basic 
dynamics and on simple instructive examples
rather than the application to various numerical simulation methods.

Some of the topics covered here can also be found in earlier summaries
\cite{Cook4,Erber2,Molmer3} and more recent summer school lectures 
\cite{Knight1,Molmer2,Zoller2}.
\section{Intermittent Fluorescence}
\label{Howitbegan}
Quantum mechanics is a statistical theory which makes 
probabilistic predictions of the behaviour of ensembles (an 
ideally infinite number of identically prepared quantum systems)
using density operators. 
This description was completely sufficient for the first 
60 years of the existence of quantum mechanics because it was 
generally regarded as completely impossible to observe and 
manipulate single quantum systems. For example, Schr{\"o}dinger,
in 1952 wrote \cite{Schroedinger1}. \\
{\em ... we never experiment with just one electron or atom or 
(small) molecule. In thought-experiments we sometimes assume 
that we do; this invariably entails ridiculous consequences. 
\{...\} In the first place it is fair to state that we are 
not experimenting with single particles, any more than we can 
raise Ichthyosauria in the zoo.}\\[1ex]
This (rather extreme) opinion was challenged by a remarkable idea 
of Dehmelt which he first made public in 1975 \cite{Dehmelt1,Dehmelt2}. 
He considered the problem of high precision spectroscopy, where 
one wants to measure the transition frequency of an optical 
transition as accurately as possible, e.g., by observing the 
resonance fluorescence from that transition as part (say) of an 
optical frequency standard. 
However, the accuracy of such a measurement is fundamentally 
limited by the spectral width of the observed transition. The 
spectral width is due to spontaneous emission from the upper 
level of the transition which leads to a finite lifetime $\tau$ 
of the upper level. Basic Fourier considerations then 
imply a spectral width of the scattered photons of the order 
of $\tau^{-1}$. To obtain a precise value of the transition 
frequency, it would therefore be advantageous to excite a metastable 
transition which scatters only a few photons within the measurement 
time. On the other hand one then has the problem of detecting these few 
photons and this turns out to be practically impossible by  
direct observation. So obviously one has arrived at a major dilemma here. 
Dehmelt's proposal however suggests a solution to these problems, 
provided one would be able to observe and manipulate single 
ions or atoms which became possible with the invention of single ion 
traps \cite{Paul1,Paul2} (for a review see \cite{Horvath1}. We illustrate
Dehmelts idea in its original simplified rate equation picture. It 
runs as follows.

\begin{figure}[hbt]
\epsfxsize12.cm
\centerline{\epsffile{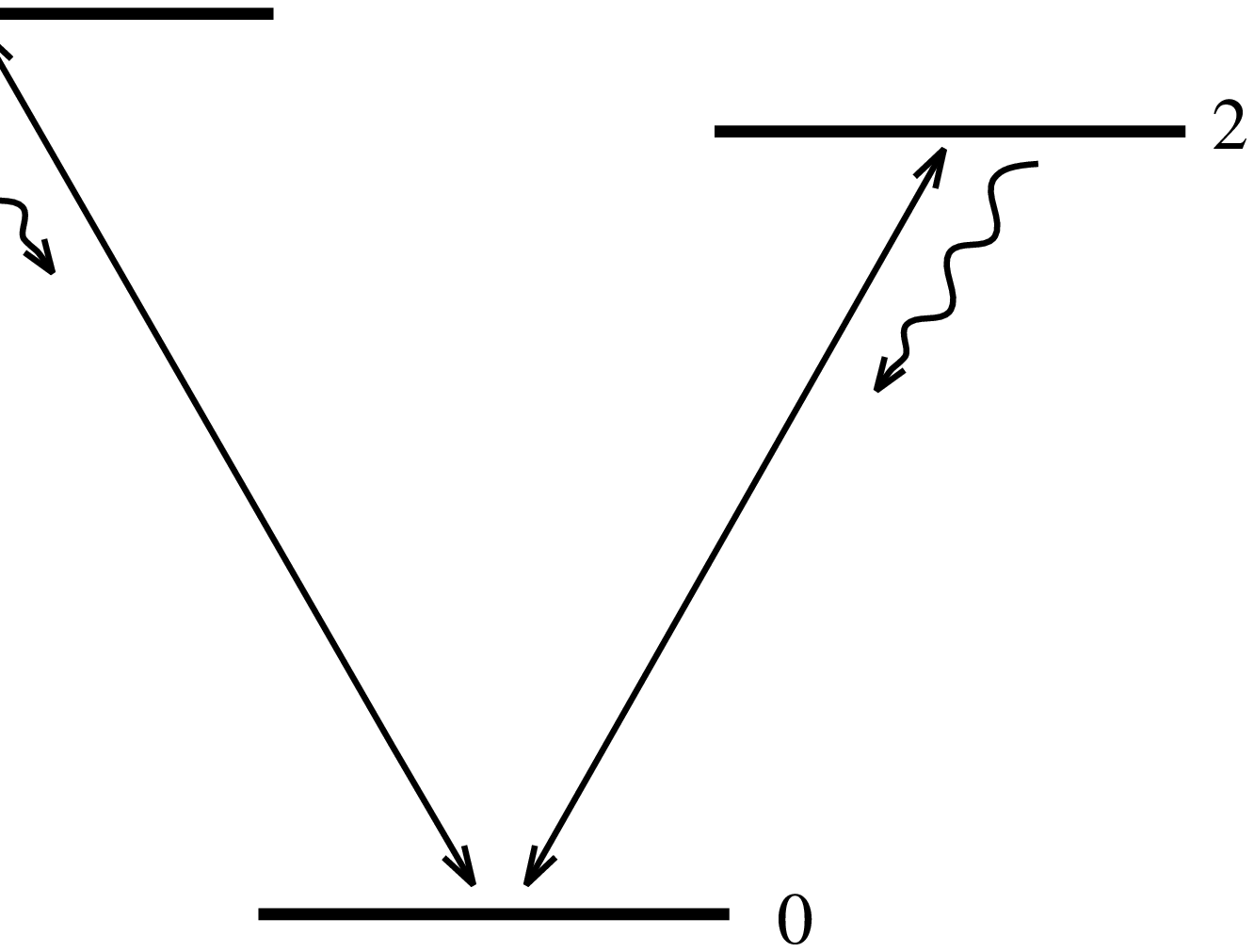}}
\caption{\label{fig2.1} The V system. Two upper levels $1$ and $2$ 
couple to a common ground state $0$. The transition frequencies 
are assumed to be largely different so that each of the two 
lasers driving the system couples to only one of the transitions. 
The $1\leftrightarrow 0$ transition is assumed to be strong while 
the $2\leftrightarrow 0$ transition is weak. }
\end{figure}

\begin{figure}[hbt]
\epsfxsize14.cm
\centerline{{\epsffile{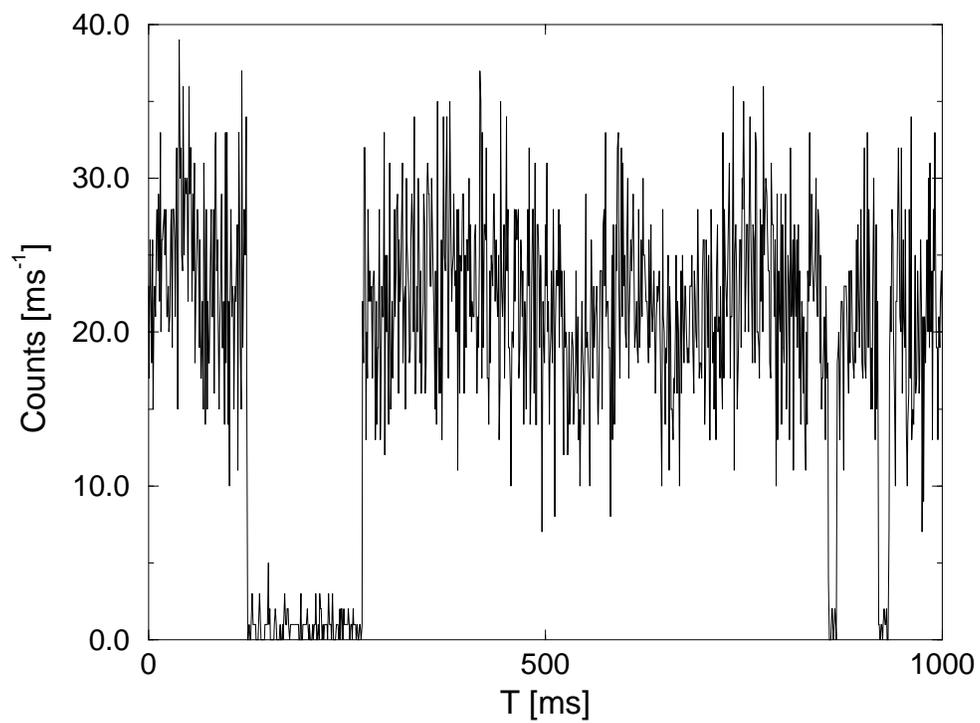}}}
\caption{\label{fig2.2}
Recorded resonance fluorescence signal exhibiting quantumm jumps from a
laser excited ${}^{24}Mg^{+}$ ion \protect\cite{Thompson}. Periods of high
photon count rate are interrupted by periods with negligible count rate (
except for a unavoidable dark count rate).} 
\end{figure}

Instead of observing the photons emitted on the metastable 
two-level system directly, he proposed to use an optical double 
resonance scheme as depicted in Fig. \ref{fig2.1}. One laser drives 
the metastable $0\leftrightarrow 2$ transition 
while a second strong laser saturates the strong 
$0\leftrightarrow 1$; the lifetime of the upper level $1$ is for 
example $10^{-8}s$ while that of level $2$ is of the order of $1s$. 
If the initial state of the system is the 
lower state $0$ then the strong laser will start to excite the 
system to the rapidly decaying level $1$, which will then lead to the 
emission of a photon after a time which is usually very short 
(of the order of the lifetime of level $1$). This emission restores 
the system back to the lower level $0$; the strong laser can 
start to excite the system again to level $1$ which will emit a photon 
on the strong transition again. This procedure repeats until at
some random time the laser on the weak transition manages to 
excite the system into its metastable state $2$ where it remains
shelved for 
a long time, until it jumps back to the ground state either by
spontaneous emission or by stimulated emission due to the laser 
on the $0\leftrightarrow 2$-transition. During the time the 
electron rests in the metastable state $2$, no photons will be 
scattered on the strong transition and only when the electron 
jumps back to state 0 can the fluorescence on the strong transition
start again. Therefore from the switching on and off of the 
resonance fluorescence on the strong transition (which is 
easily observable) we can infer the extremely rare transitions 
on the $0\leftrightarrow 2$ transition. Therefore we have a 
method to monitor rare quantum jumps (transitions) on the 
metastable $0\leftrightarrow 2$ transition by observation of the
fluorescence from the strong $0\leftrightarrow 1$ transition. 
A typical experimental fluorescence signal is depicted in 
Fig. \ref{fig2.2} \cite{Thompson} where the
fluorescence intensity $I(t)$ is plotted. 
However, this scheme only works if we observe a single quantum system, 
because if we observe a large number of systems simultaneously 
the random nature of the transitions between levels $0$ and $2$ 
implies that some systems will be able to scatter photons on the strong
transition while others are not because they are in their 
metastable state at that moment. From a large collection of ions
observed simultaneously one would then obtain 
a more or less constant intensity of photons emitted on the strong 
transition. The calculation of this mean intensity is a straightforward 
task using standard Bloch equations. The 
calculation of single system properties such as the 
distribution of the lengths of the periods of strong fluorescence, 
required some effort which eventually led to the development
of the quantum jump approach. Apart from the interesting theoretical
implications for the study of individual quantum systems, Dehmelt's 
proposal obviously has important practical applications. An often 
cited example is the realization of a new time standard using a 
single atom in a trap. The key idea here is to use either the instantaneous
intensity or the photon statistics of the emitted radiation on the 
strong transition (the statistics of the bright- and dark periods) 
to stabilise the frequency of the laser on the weak transition. 
This is possible because the photon statistics of the strong 
radiation depends on the detuning of the laser on the weak 
transition \cite{Kim1,Kim2,Kim3,Ligare1,Wilser1}. Therefore a change in the
statistics of bright and dark periods indicates that the frequency 
of the weak laser has shifted and has to be adjusted. However, for 
continuously
radiating lasers this frequency shift will also depend on the intensity
of the laser on the strong transition. Therefore in practise
pulsed schemes are preferable for frequency standards
\cite{Arecchi1,Bergquist2}\\
Due to the inability of experimentalists to store, manipulate and
observe single
quantum systems (ions) at the time of Dehmelt's proposal, both the
practical as well as the theoretical implications of his proposal 
were not immediately investigated. It was about ten years later 
that this situation changed. At that time Cook and Kimble published 
a paper \cite{Cook2} in which they made the first attempt to 
analyse the situation described above theoretically. Their 
advance was stimulated by the fact that by that time it had
become possible to actually store {\em single} ions in an ion 
trap (Paul trap) \cite{Paul1,Neuhauser2,Paul2}. 

\begin{figure}[hbt]
\epsfxsize14.cm
\centerline{\epsffile{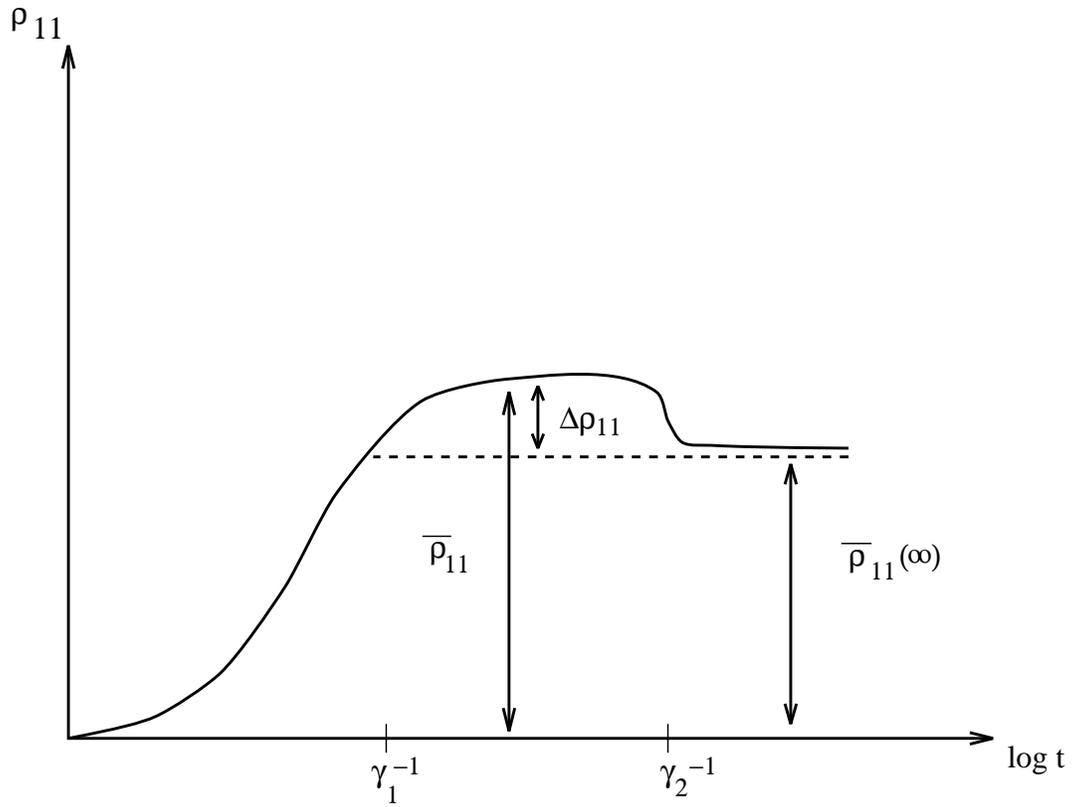}}
\vspace*{1.cm}
\caption{\label{fig2.3}
Time-evolution of the population in the strongly-fluorescing 
level 1 of the three-level ion shown in Fig. (\ref{fig2.1}). The 
lifetimes $\gamma_1^{-1}$ and $\gamma_2^{-1}$ are marked on 
the time axis. What is crucial here is the ``hump'' 
$\Delta \rho_{11}:$ this is a signature of the telegraphic 
nature of the fluorescence.}
\end{figure}

\begin{figure}
\epsfxsize12.cm
\centerline{\epsffile{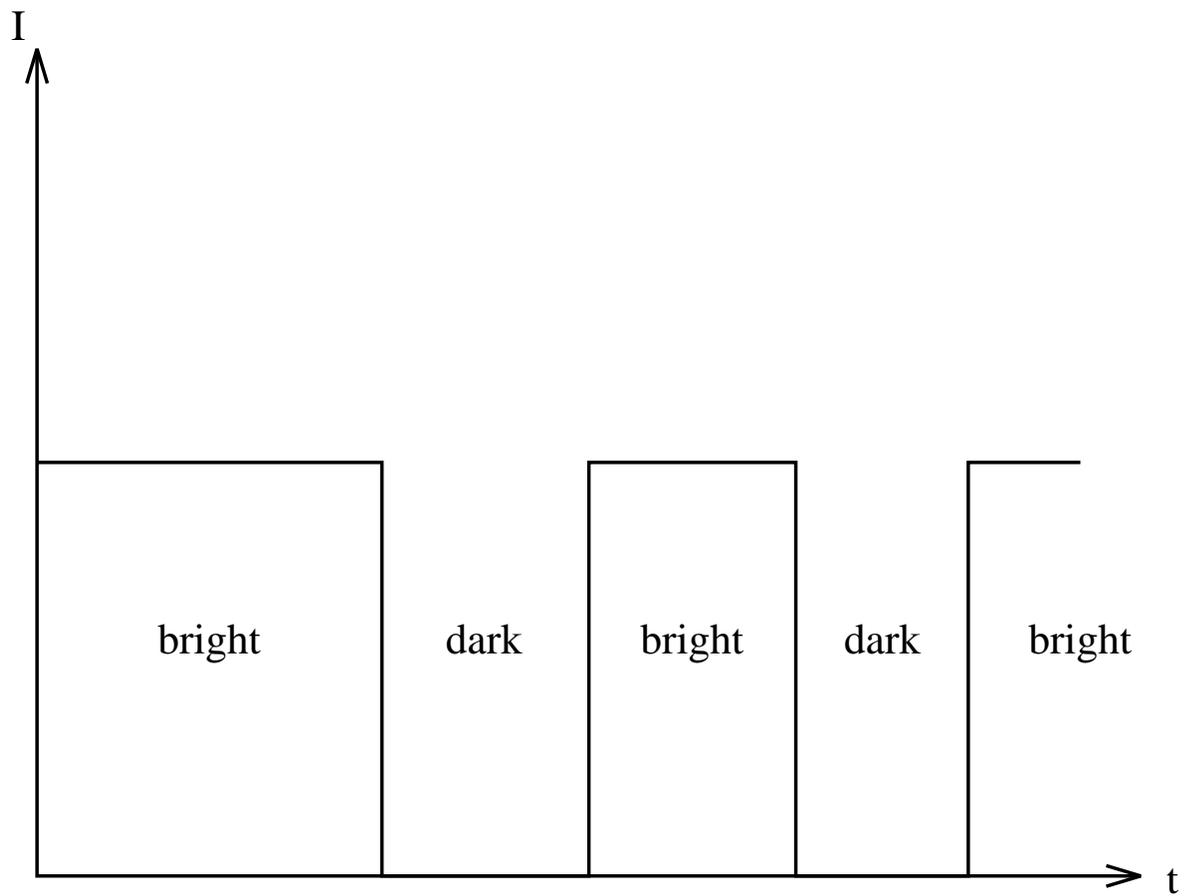}}
  \vspace*{1.cm}
  \caption{\label{fig2.4}
  A few periods of bright and dark sequences in the fluorescence intensity $I$
  from a three-level system. The bright periods last on average $T_B,$ and
  the dark periods $T_D.$
  }
\end{figure}

In their simplified rate equation approach Cook and Kimble started with 
the rate equations for an incoherently driven three level system as shown 
in Fig. \ref{fig2.1} and assumed that the strong $0\leftrightarrow 1$ 
transition is driven to saturation. They consequently simplify their rate
equations introducing the probabilities $P_+$ of being in the metastable 
state and $P_-$ of being in the strongly fluorescing $0\leftrightarrow 1$
transition. This simplification now allows the description of the 
resonance fluorescence to be reduced to that of a two state random telegraph 
process. Either the atomic population is in the levels $0$ and $1$ and
therefore the ion is strongly radiating (on), or the population rests in
the metastable level $2$ and no fluorescence is observed (off). They then 
proceed to calculate the distributions for the lengths of bright and 
dark periods and find that their distribution is Poissonian.
Their analysis that we have outlined very briefly here is of course 
very much simplified in many respects. The most important point is certainly
 the fact that Cook and Kimble assume incoherent driving and therefore adopt
a rate equation model. In a real experiment coherent radiation from 
lasers is used. The complications arising in coherent excitation finally led 
to the development of the quantum jump approach.
Despite of these problems the analysis of Cook and Kimble showed the 
possibility of direct observation of quantum jumps in the fluorescence 
of single ions, a prediction that was confirmed shortly afterwards in a 
number of experiments 
\cite{Bergquist1,Nagourney1,Nagourney3,Sauter1,Sauter2,Dehmelt3}
and triggered off a large number of more detailed 
investigations starting with early works by Javanainen in 
\cite{Javanainen1,Javanainen2,Javanainen3}. 
The following effort of a great number of physicists
eventually culminated in the development of 
the quantum jump approach. Before we present this subsequent development 
in greater detail we would like to study in slightly more detail how 
the dynamics of the system determines the statistics of bright and dark 
periods. Again assume a three-level system as shown in Fig. \ref{fig2.1}.
Provided the $0 \leftrightarrow 1$ and 
$0 \leftrightarrow 2$ Rabi frequencies are small compared with the decay rates, 
one finds for the population in the strongly-fluorescing level 1 
as a function of time something like the behaviour shown in Fig. \ref{fig2.3} 
(we derive this in detail in a later section.). We choose for this figure 
the values $\gamma_1 \gg \gamma_2,$ for the Einstein-coefficients of levels 
$1$ and $2$ reflecting the metastability of level 2.  For times short 
compared with the metastable lifetime
$\gamma_2^{-1},$ then of course the atomic dynamics can hardly be 
aware of level 2 and evolves as a 0 -- 1 {\it two-level} system with the 
''steady state '' population $\bar{\rho}_{11}$ of the upper level. After 
a time $\gamma_2^{-1},$ the metastable state has an effect and the
(ensemble-averaged) population in level 1 reduces to the appropriate 
{\it three-level} equilibrium values. The ``hump'' $\Delta \rho_{11}$ 
shown in Fig. \ref{fig2.3} is actually a signature of the telegraphic 
fluorescence discussed above. To show this, consider
a few sequences of bright and dark periods in the telegraph signal as 
shown in Fig. \ref{fig2.4}. The total rate of emission $R$
is proportional to the rate in a bright period times the fraction of the 
evolution made up of bright periods. This gives
\begin{equation}
	R = \gamma_{1} {\bar \rho}_{11}\left( {T_B \over T_B + T_D} \right)
	\;\; , 
	\label{850}
\end{equation}
but this has to be equal to the true average,
\begin{equation}
 	R = \gamma_{1}\rho_{11} (\infty)\,,
 	\label{860}
\end{equation}
so that 
\begin{equation}
 	{T_D \over T_B} = { {\bar \rho}_{11} -  \rho_{11} (\infty) 
	\over \rho_{11} (\infty) } = 
	{ \Delta \rho_{11} \over \rho_{11} (\infty) }\,,
 	\label{870}
\end{equation}
and the ratio of the period of bright to dark intervals is governed, as
we claimed, by the ``hump'' $\Delta \rho_{11}.$

So far, we have concentrated on situations where the Rabi frequencies are
small (or for incoherent excitation). What happens for {\it coherent}
resonant excitation with larger Rabi frequencies? The answer to this 
question is {\it nothing} \cite{Knight3}: there are essentially {\it no}
quantum jumps, at least at any significant level for coherently-driven
resonantly excited three-level systems!
But this is because of the idea of resonance is tricky: the strong 
Rabi frequency on the $0\leftrightarrow 1$ transition dresses the atom and the 
AC-Stark effect splits \cite{Autler1,Knight2} the transition, forcing
the system substantially out of resonance. If this is recognised and the probe
laser driving the $0\leftrightarrow 2$ transition is detuned from the bare 
resonance until it matches the dressed atom resonance, then the jumps and 
telegraphic fluorescence return. We investigate this phenomenon more closely
in Section V. As far as we know, the dependence of the 
telegraph fluorescence on detuning for coherently excited transitions has 
yet to be confirmed experimentally.

Let us return to the idea of a null measurement. We imagine that we observe
the fluorescence from a driven three-level ion over a time scale which is
long compared with the strongly fluorescing state lifetime $\gamma_1^{-1}$
but very short compared with the shelf state lifetime $\gamma_2^{-1},$
so that $\gamma_1^{-1} \ll \Delta t \ll \gamma_2^{-1}.$ Pegg and Knight 
\cite{Pegg3,Pegg3b} have shown that the average period of brightness and 
darkness
in the telegraphic fluorescence can be obtained very straightforwardly 
from considerations of null
detection. During such an interval $\Delta t,$ the population in the shelf
state, $P_2 (t)$ hardly has time to evolve, but population can be rapidly
cycled from the ground state $| 0 \rangle$ to the strongly-fluorescing 
state $|1 \rangle$ and back. Detection of a photon at the beginning of a
$\Delta t$ interval implies a survival in the 0 -- 1 sector for the whole
interval and a bright period, whereas a {\it null} detection is 
sufficient for us to be confident that the atom is shelved for the whole
$\Delta t$ interval and a dark period ensues.

If we take our origin of time to be after an interval $\Delta t$ in which we
see a photon, then $P_2(0) = 0.$ We can introduce the ``life expectancy''
$T_B$ as the time the atom spends in the 0 -- 1 sector {\it continuously}.
If the atom is still in this sector at a time $t_1$ (known from an 
observation of another fluorescence photon just prior to $t_1$), then the 
life expectancy will be also be $T_B.$ So we can partition the outcomes 
into the 
case where at $t_1$ it has survived in the 0 -- 1 sector with probability 
$P_{10}(t_1)$ and the case where the ion did not survive the whole interval
$t_1$ continuously in the 0 -- 1 sector \cite{Pegg3,Pegg3b}
\begin{equation}
 	T_B = P_{10}(t_1)\, (t_1 + T_B) + (1 - P_{10}(t_1)) f t_1 \,,
 	\label{880}
\end{equation}
where $f$ is a fraction $( <1 ).$ Then for small $t_1$
\begin{equation}
 	T_B = { t_1 P_{10}(t_1) \over 1 - P_{10}(t_1) } =
	{ t_1 \over 1 - P_{10}(t_1) } - t_1 \;\; ,
 	\label{890}
\end{equation}
and if $t_1$ is small so we may neglect the possibility of a {\it return}
from state $|2 \rangle$ back in to the 0 -- 1 sector, $(1 - P_{10}(t_1)) 
\approx P_2(t_1),$ so that
\begin{equation}
	 T_B^{-1} = \left. {dP_2 \over dt} \right|_{t=0}  \ \ \ {\rm given}
	 \ \ \ P_2(0) = 0 \,.
	 \label{900}
\end{equation}
This is finite, so we know that the fluorescence {\it will} 
terminate. To obtain a value for $T_B,$ we merely need to know the evolution
equation (not its solution) for the population in state $|2 \rangle :$ this
would be the Bloch equation for coherent excitation, or the Einstein rate
equation for incoherent excitation.

The calculation of the mean period of darkness proceeds along similar lines:
if {\it no} photons are detected in an interval $\Delta t$ just 
before $t = 0,$ we find
\begin{equation}
 T_D^{-1} = \left. - {dP_2 \over dt} \right|_{t=0}  \ \ \ {\rm given}
 \ \ \ P_2(0) = 1 \,.
 \label{910}
\end{equation}
%
%
%
%
%
The analysis presented here obviously also applies for the density operator 
equations in exactly the same form and we obtain
\begin{eqnarray}
	T_B^{-1} &=& \left(\dot{\rho}_{22}\right)_{t=0} \hspace*{.5cm}
	\mbox{with}\hspace*{.1cm} \rho_{22}(0) = 0\;\; , \label{930}\\
	T_D^{-1} &=& -\left(\dot{\rho}_{22}\right)_{t=0} \hspace*{.5cm}
	\mbox{with}\hspace*{.1cm} \rho_{22}(0) = 1 \label{940}\;\; .
\end{eqnarray}
Here the dot means the {\em average} gradient of the $\rho_{22}$ versus $t$
curve over a range of order $\Delta t$. Because $\Delta t$ is much smaller
than the characteristic change of $\rho_{22}$ it is very close to the normal 
derivative at all points.  

It is straightforward to use this idea of "collapse by non-detection" to 
estimate the characteristic time needed to be sure that a quantum jump 
has occurred \cite{Pegg3b}. There are $T_B/t_d$ times as many short dark 
periods between photon emissions as there are prolonged dark periods of 
average length $T_D$, where $t_d$ ($\approx \gamma_1^{-1}$ for strong
transition saturation) is the average length of the short period. Thus
the probability that an emission will be followed by a long dark period 
is approximately $t_d/T_B$ for $T_B\gg t_d$, and the probability that it 
will be followed by a short dark period is close to unity. 

Immediately following a photon emission, a dark period of length at least 
$\tau$ (with $\tau < T_D$) can exist for two complementary reasons: (a)
the atom goes to state $|2\rangle$ and therefore does not decay for a time 
of the order of $T_D$, or (b) the atom is still in the 
$|0\rangle \leftrightarrow |1\rangle$
plane but has not yet emitted a photon. The probability of (a) occurring 
is $t_d/T_B$ and the probability of (b) is approximately $exp(-\tau/t_d)$. 
Clearly for $\tau<t_d$ it is much more likely that any observed dark period 
of length $\tau$ is due to (b), but this becomes rapidly less likely as
$\tau$ increases. The point where the observation of the dark period is 
just as likely to involve (a) as (b) is found by equating the two expressions 
to give
\begin{equation}
	\frac{\tau}{T_B} = \frac{t_d}{T_B}\, \ln(T_B/t_d) \; \; .
	\label{991}
\end{equation}
It follows that the sampling period $\Delta t$ must be greater than $\tau$ 
given by Eq. (\ref{991}) this 
value in order for the observation of darkness during $\Delta t$ to imply 
(a) with reasonable certainty.

Further, because we know that immediately following the emission the atom
is in $|0\rangle$, so the probability of being in $|2\rangle$ is zero, and 
because Eq. (\ref{991}) gives the order of the time of darkness required 
for the probability of being in $|2\rangle$ to grow to about $\frac{1}{2}$, 
Eq.(\ref{991}) gives the characteristic time for the wave-function 
collapse by non-detection. This characteristic time can be associated with 
the time necessary for us to be certain that a quantum jump from $|0\rangle$ 
to $|2\rangle$ has occurred. For completely coherent excitation Eq. 
(\ref{991}) reduces to an expression similar to that for the "shelving time" 
found in \cite{Porrati1} and in \cite{Zoller1}. 
%
Note that while the collapse by non-detection of the system into the 
metastable state requires a finite time the collapse of the wave function 
due to the detection of a photon has to be viewed as practically 
instantaneous. When we detect a photon our knowledge of the system 
changes suddenly and this sudden change of knowledge is reflected by 
the sudden change of the system state which, after all, represents 
our knowledge of the system.
\section{Ensembles and shelving}
\label{IIIb} 
Before we develop detailed theoretical models to describe individual quantum
trajectories (i.e. state evolution conditioned on a particular sequence
of observed events), it is useful to examine how the entire ensemble 
evolves. This is in line with the historical development as initially 
it was tried to find quantum jump characteristics in the ensemble 
behaviour of the system.We do this in detail for the particular three--level 
V--configuration  (shown in Fig. \ref{fig2.1}) appropriate for Dehmelt's
quantum jump phenomena. For simplicity, we examine the case of 
{\bf incoherent} excitation. Studies for coherent excitation using Bloch 
equations can be found , for example, in 
\cite{Kim3,Kimble1,Ligare1,Nienhuis1,Schenzle2}. The Einstein rate 
equations for the V--system are \cite{Pegg2}
\begin{eqnarray}
	\frac{d}{dt} \rho_{11} &=& -(A_1 +B_1 W_1) \rho_{11} + B_1 W_1 \rho_{00}
	\;\; ,\label{3.1} \\
	\frac{d}{dt} \rho_{22} &=& -(A_2 +B_2 W_2) \rho_{22} + B_2 W_2 \rho_{00}
	\;\; ,\label{3.2} \\
	\frac{d}{dt} \rho_{00} &=& -(B_1 W_1 +B_2 W_2) \rho_{00} 
			+ (A_1 +B_1 W_1) \rho_{11} + (A_2 +B_2 W_2) \rho_{22}
	\;\; ,\label{3.3} 
\end{eqnarray}
where $A_i,B_i$ are the Einstein A- and B-coefficients for the relevant
spontaneous and induced transitions, $W_i$ the applied radiation field energy 
density at the relevant transition frequency, and $\rho_{ii}$ is the relative 
population in state $i$ ($\rho_{00}+\rho_{11}+\rho_{22}=1$ for this {\bf closed}
system). In shelving, we assume that both $B_1W_1$ and $A_1$ are much larger 
than $B_2W_2$ and $A_2$, and furthermore that $B_2W_2\gg A_2$. The 
steady--state solutions of these rate equations are straightforward to obtain, 
and we find
\begin{eqnarray}
	\rho_{11}(t\rightarrow \infty) = \frac{B_1W_1(A_2+B_2W_2)}
	{A_1(A_2+2B_2W_2) + B_1W_1(2A_2 + 3B_2W_2) } \;\; ,
	\label{3.4} \\[.25cm]
	\rho_{22}(t\rightarrow \infty) = \frac{B_2W_2(A_1+B_1W_1)}
	{A_1(A_2+2B_2W_2) + B_1W_1(2A_2 + 3B_2W_2) }\;\; .
	\label{3.5}
\end{eqnarray}
Now if the allowed $0\leftrightarrow 1$ transition is {\bf saturated}
\begin{equation}
	\rho_{11}(t\rightarrow \infty) \approx \frac{A_2 +B_2W_2}
	{2A_2 + 3B_2W_2} \approx \rho_{00}(t\rightarrow \infty) \;\; ,
	\label{3.6}
\end{equation}
and
\begin{equation}
	\rho_{22}(t\rightarrow \infty) \approx \frac{B_2W_2}
	{2A_2 + 3B_2W_2} \approx \frac{1}{3}\;\; .
	\label{3.7}
\end{equation}
Now we see that a small $B_2W_2$ transition rate to the shelf state has a 
major effect on the dynamics. Note that if the induced rates are much larger 
than the spontaneous rates, the steady state populations are
$\rho_{00}=\rho_{11}=\rho_{22}=\frac{1}{3}:$ the populations are evenly 
distributed of course amongst the constituent states of the transition. 

However, the dynamics reveals a different story to that suggested by the 
steady state populations. Again, if the allowed transition is saturated, 
then the time--dependent solutions of the excited state rate equations 
tell us that for $\rho_{00}(0)=1$ we have 
\begin{equation}
	\rho_{11}(t) = \frac{B_2W_2}{2(2A_2+3B_2W_2)} e^{-(A_2+3\,B_2W_2/2)t}
	-\frac{1}{2} e^{-(2B_1W_1 +A_1 + B_2W_2/2) t}
	+ \frac{A_2+B_2W_2}{2A_2+3B_2W_2} \;\; ,
	\label{3.8}
\end{equation}
and
\begin{equation}
	\rho_{22}(t) = \frac{B_2W_2}{2A_2+3B_2W_2} \left\{ 1 -
	e^{-(A_2+3\,B_2W_2/2)t} \right\}\;\; .
	\label{3.9}
\end{equation}
Note that the these expressions are good only for strong driving of the
$0\leftrightarrow 1$ transition. This especially means that for short
times $\rho_{00}$ is of the order of $1/2$ which results in Eq. (\ref{3.9}).
Then for a very long--lived shelf state $2$, we see that for saturated 
transitions ($B_iW_i \gg A_i$)
\begin{equation}
	\rho_{11}(t) \approx \frac{1}{3} \left\{
	1 + \frac{1}{2}\left( e^{-3 B_2W_2t/2} - 3 e^{-2 B_1W_1 t} \right)
	\right\} \;\; ,
	\label{3.10}
\end{equation}
and
\begin{equation}
	\rho_{22}(t) \approx \frac{1}{3} \left\{ 1 - e^{-3 B_2W_2 t/2}
	\right\}\;\; .
	\label{3.11}
\end{equation}
These innocuous--looking expressions contain a lot of physics. We remember 
that state $1$ is the strongly fluorescing state. On a time--scale which 
is short compared with $(B_2W_2)^{-1}$, we see that the populations attain 
a {\bf quasi}--steady state appropriate to the {\bf two-level}
($0\leftrightarrow 1$) dynamics
\begin{equation}
	\rho_{11}(t \,\,\mbox{short})\sim \frac{1}{2}
	\left\{ 1 - e^{-2 B_1W_1 t} 
	\right\} \rightarrow \frac{1}{2}\;\; .
	\label{3.12}
\end{equation}
This can of course be confirmed in an experiment \cite{Finn1,Finn2}.
For truly long times the third, shelving, state makes its effect and
\begin{equation}
	\rho_{11}(t \,\,\mbox{long}) \sim \frac{1}{3} \;\; ,
	\label{3.13}
\end{equation}
as we saw qualitatively in Fig. \ref{fig2.3}. As we saw earlier in the 
discussion of Eq. (\ref{870}), the change from two--level to three--level
dynamics already gives us a signature of quantum jumps and telegraphic
fluorescence provided we are wise enough to recognise the signs. Figure
\ref{fig2.3} illustrates the change from two to three--level dynamics. 

The steady--state populations are sufficient to describe the average level 
of the fluorescent intensity. But how do quantum jumps and shelving show 
up in the intensity correlations? For example, let us examine the second 
order correlation function
\begin{equation}
	g^{(2)}(t,\tau) = \frac{\langle :I(t+\tau) I(t) : \rangle}
	{\langle I(t)\rangle ^2} \;\; ,
	\label{3.14}
\end{equation}
where the colons describe normal ordering \cite{Loudon1}. This correlation 
function is straightforward to compute from the Einstein rate equations 
using the quantum regression theorem \cite{Lax1} which relates one-time 
to two--time correlations given the dynamics is Markovian. Here of course 
there are {\bf two} intensities, that of the $1\rightarrow 0$ and of the 
$2\rightarrow 0$ fluorescence, so we can correlate the two light fields: 
"1" with "1", or 
"1" with "2" and so on, where "1" and "2" represent the fluorescence on the 
$1\rightarrow 0$ and the $2\rightarrow 0$ transitions respectively. So let 
us concentrate on evaluating $g^{(2)}_{ij}(t,\tau)$, which represents the 
joint probability of detecting a fluorescent photon $j$ ($j=1,2$) on the 
transition $j$ at time $t$ and {\bf some} other photon (not necessarily the 
{\bf next} photon) from transition $i$ at time $(t+\tau)$. it is 
straightforward to show \cite{Pegg2} that 
\begin{eqnarray}
	g_{11}^{(2)}(\tau) &=& g_{12}^{(2)}(\tau) \nonumber\\
	&=& 1 + \frac{B_2W_2}{2(A_2+B_2W_2)} 
	e^{-(A_2 + 3 B_2W_2/2) \tau}
	- \frac{2A_2 +3B_2W_2}{2(A_2+B_2W_2)} 
	e^{-(2B_1W_1 + A_1 + B_2W_2/2)\tau} \;\; ,
	\label{3.15}
\end{eqnarray}
and
\begin{equation}
	g_{22}^{(2)}(\tau) = g_{21}^{(2)}(\tau)
	= 1 - e^{-(A_2+3B_2W_2/2)\tau}
	\;\; .
	\label{3.16}
\end{equation}
Using $ B_1W_1\gg B_2W_2$ and that the transitions are saturated, then 
these correlation functions simplify to give
\begin{equation}
	g_{11}^{(2)}(\tau) = g_{12}^{(2)}(\tau) = 1 + \frac{1}{2}
	\left( e^{-3 B_2W_2\tau/2} - 3 e^{-2 B_1W_1 \tau} \right) \;\; ,
	\label{3.17}
\end{equation}
so that as expected, the correlation functions obey the same evolutions as 
the populations. It is worth noting that for short times $\tau$ we expect 
to see anti-bunching \cite{Loudon1} from this {\bf three--level} 
fluorescence, and this has been observed experimentally from trapped ions 
\cite{Itano0,Schubert1}.

\begin{figure}[hbt]
\epsfxsize14.cm
\centerline{\epsffile{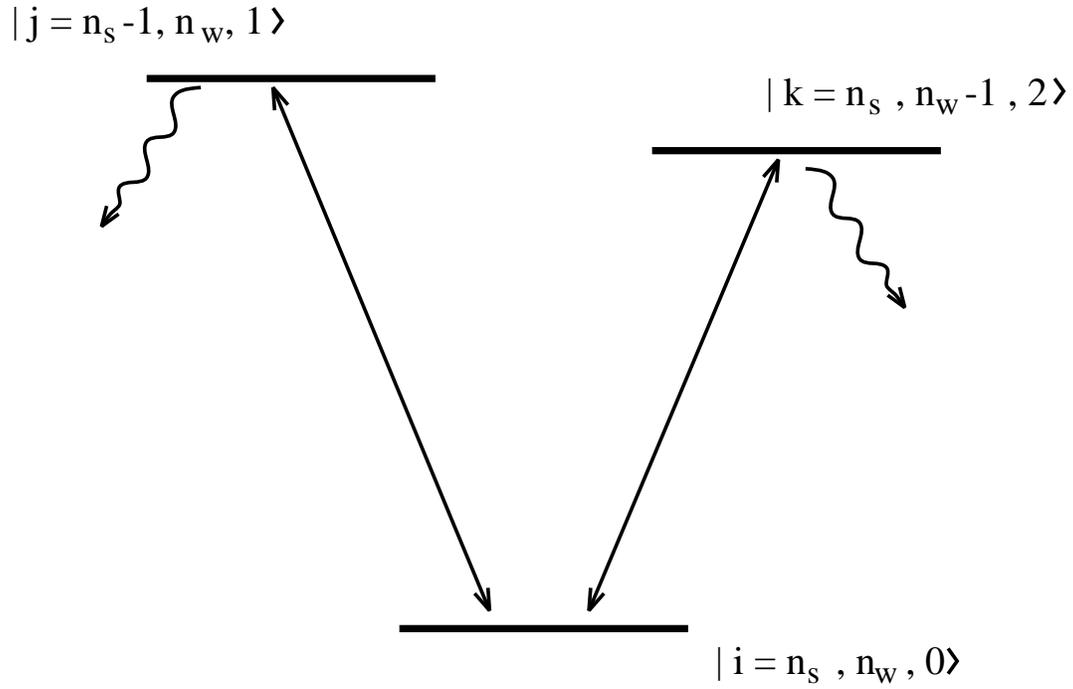}}
\vspace*{1.cm}
\caption{\label{fig2.5}
Atom+field states used to describe the survival of a three-level 
system in an un-decayed state. The numbers $n_s,n_w$ specify photon 
numbers driving the strongly allowed transitions $0\leftrightarrow 1$
and the weak transition $0\leftrightarrow 2$. Note that 
fluorescence takes the system {\em out} of the three atom+field states.}
\end{figure}

In the mid 1980's when studies of quantum jump dynamics of laser--driven 
three--level atoms began in earnest, a great deal of effort was expended 
in determining the relationship between joint probabilities of detection of 
a photon at time $t$ and the {\bf next} or {\bf any} photon a time $\tau$
later. This was addressed in detail by Cohen-Tannoudji and Dalibard 
\cite{Cohen-Tannoudji1,Reynaud1}, by Schenzle and Brewer 
\cite{Schenzle1,Schenzle2} 
and others \cite{Cook1,Lenstra1}.
One attractive approach, advocated by Cohen-Tannoudji and Dalibard, uses a 
dressed 
manifold and from this evaluates the delay function describing
the distribution of delay times before the {\bf next} emission occurs. 
In Fig. \ref{fig2.5} that laser excitation couples together these 
atom+field 
states, but fluorescence does not: spontaneous emission is an irreversible 
loss {\bf out} of this manifold to states with reduced photon number in 
the excitation modes, but with photons created in initially un-occupied 
free-space modes.

We expand our atom+field state vector into the basis states $|l\rangle$ 
with fixed number of fluorescence photons in the radiation field as
shown in Fig. \ref{fig2.5} as 
\begin{equation}
	|\psi(t)\rangle = \sum_{l} a_l(t) e^{-i E_l t/\hbar} |l\rangle
	\;\; ,
	\label{3.18}
\end{equation}
and solve for the probability amplitudes $a_{l}(t)$. The probability of
{\bf remaining} without further emission in the $n$-excitation atom+field 
manifold 
up to time $\tau$ is 
\begin{equation}
	P_0(\tau) = \sum_{i} |a_i(\tau)|^2 \;\; .
	\label{3.19}
\end{equation}
The negative differential of this survival probability describes the
{\bf delay function} $I_1(\tau)$ \cite{Cohen-Tannoudji1}
\begin{equation}
	I_1(\tau) \equiv -\frac{dP_0}{d\tau} \;\; ,
	\label{3.20}
\end{equation}
so that the probability of there being an interval $\tau$ between one
photon being emitted (detected) and the {\bf next} is
\begin{equation}
	P_0(\tau) = 1 - \int_0^{\tau} I_1(\tau') d\tau' \;\; .
	\label{3.21}
\end{equation}
To evaluate this interval distribution function it is sufficient to calculate the 
atom+field dressed state amplitudes. This demonstrates the utility of this approach: 
there is no need to solve the potentially complicated Bloch equations for the driven 
three--level atom, although of course this can be done 
\cite{Schenzle2}. Kim and coworkers \cite{Kim1,Kim2}, 
Grochmalicki and Lewenstein (1989a), Wilser (1991) and later others
have used the delay function to describe the shelving in the V system. 
The 
conditional probability of an atom
emitting {\bf any} photon between time $\tau$ and $\tau+d\tau$ after 
emitting a 
photon at time $\tau=0$ is $I(\tau)d\tau$. The photon emitted at 
time $\tau$
can be the first to be emitted after that at $\tau=0$ or the next 
after any one 
at time $\tau'$ ($0 < \tau' < \tau$), so that
\begin{equation}
	I(\tau) = I_1({\tau}) + \int_{0}^{\tau} d\tau' I(\tau') 
	I_1(\tau-\tau') \;\; ,
	\label{3.22}
\end{equation}
where $I_1(\tau)$ is the interval distribution. If this is expressed in 
terms of its Laplace transform $\tilde{I}(z)$, then
\begin{equation}
	\tilde{I}(z) = \frac{\tilde{I}_1(z)}{1- \tilde{I}_1(z)} \;\; .
	\label{3.23}
\end{equation}
The function $g^{(2)}(\tau)$ is the normalized correlation function for a 
photon detection at $\tau=0$ followed by the detection of any photon
 (not necessarily 
the next) at time $\tau$. It follows that 
\begin{equation}
	g^{(2)}(\tau) = \frac{I(\tau)}{I(\tau\rightarrow \infty)} \;\; ,
	\label{3.24}
\end{equation}
or equivalently in Laplace space
\begin{equation}
	\tilde{g}^{(2)}(z) = \frac{\tilde{I}(z)}{\lim_{z\rightarrow 0}
	z\tilde{I}(z)} \;\; ,
	\label{3.25}
\end{equation}
so that
\begin{equation}
	\tilde{g}^{(2)}(z) = \left[ \lim_{z\rightarrow 0} 
	\frac{1- \tilde{I}_1(z)}{z\tilde{I}_1(z)} \right]
	\frac{\tilde{I}_1(z)}{1- \tilde{I}_1(z)}\;\; .
	\label{3.26}
\end{equation}
For the case of incoherent, rate-equation excitation of a three--level 
V--system atom, it is straightforward to calculate the atom+field survival
 probability, differentiate this to generate the delay function and from 
Eq. (\ref{3.26}) deduce $g^{(2)}(\tau)$. If this is done, precisely the same 
form is obtained as that from the quantum regression theorem. The merit of 
this approach is easier to appreciate for the case of {\em coherent} excitation, 
where the regression theorem requires the solution of the three--level Bloch 
equations and the solution of 8th order polynomial characteristic equations, 
compared with the need to solve three coupled equations, using the delay 
function route. In Section V we further illustrate the connection between the
''next'' photon probability density and the "any' photon rate in 
Eqs. (\ref{508})-(\ref{519}).

\begin{figure}[hbt]
\epsfxsize16.cm
\centerline{\epsffile{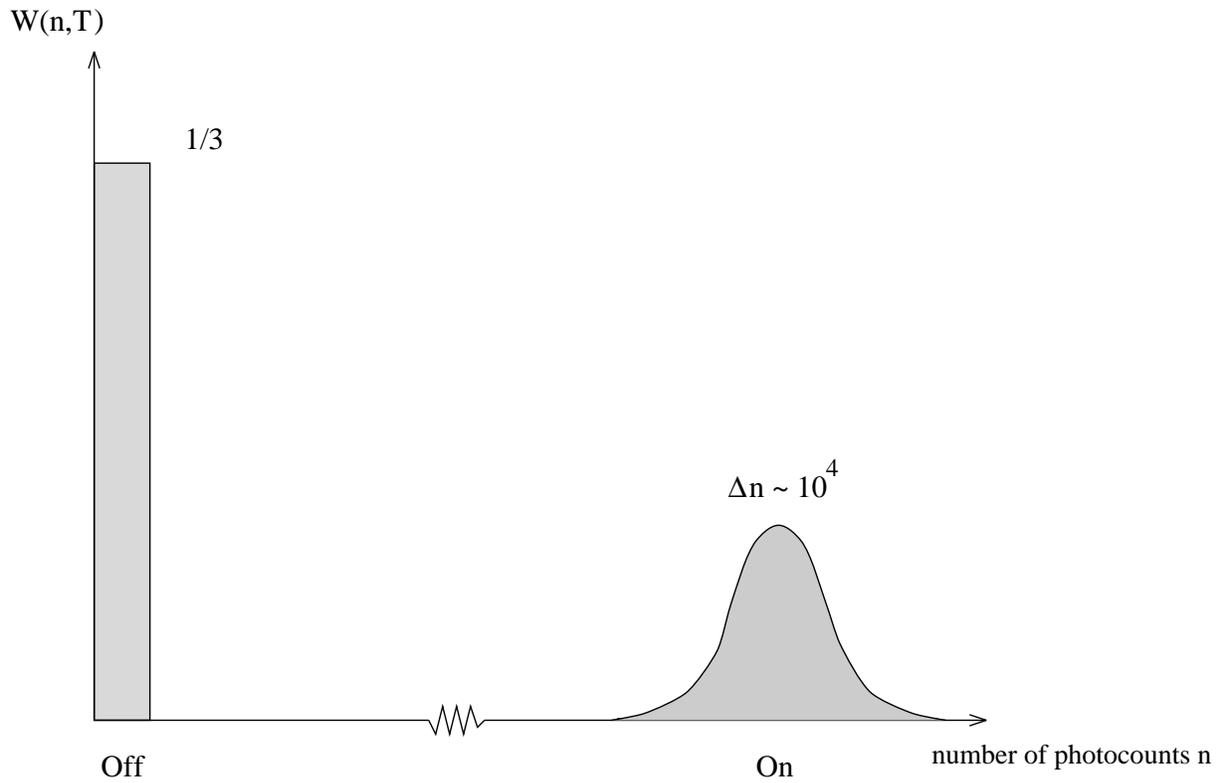}}
\vspace*{1.cm}
\caption{\label{fig2.6} Macroscopic photocount fluctuations revealed in 
photo count distribution $W(n,T)$ of counts of the allowed 
$0\leftrightarrow 1$ transition fluorescence from a saturated 
V --  configuration. One either registers a large number of counts in the 
time interval $[0,T]$ (On) which is a sign of a bright period, or one
registers no counts at all (Off) which shows that one is in a dark period.
After Schenzle and Brewer (1986).}
\end{figure}

Rather than examine the correlation functions, it may be useful to examine 
other properties of the photon statistics, and in particular the variance 
in the photon numbers in the detected fluorescence 
\cite{Kim1,Kim2,Kim3,Jayarao1}. For a Poissonian field, 
$(\Delta n)^2=\langle n \rangle$, but for a sub-Poissonian field 
$(\Delta n)^2<\langle n \rangle$, and for a super-Poissonian
field, $(\Delta n)^2=\langle n \rangle$. To characterise the deviation from pure
Poissonian fluctuations, Mandel \cite{Mandel1} defined
the parameter 
\begin{equation}
	Q_M(\tau) \equiv \frac{(\Delta n)^2-\langle n \rangle}
	{\langle n\rangle} \;\; ,
	\label{3.38}
\end{equation}
which can be written in terms of the mean intensity $\langle I\rangle$ and
the second order correlation function $g^{(2)}(t)$ as
\begin{equation}
	Q_M(\tau) = \frac{\langle I\rangle}{\tau} \left\{\int_{0}^{\tau} dt_2
	\int_{0}^{t_2} dt_1 g^{(2)}(t_1) \right\}- \langle I \rangle \tau
	\;\; .
	\label{3.39}
\end{equation}
If this is used to describe the fluorescence from a three--level atom 
involving shelving \cite{Kim1,Kim2,Kim3} then as $\tau\rightarrow \infty$ 
for {\bf saturated} transitions
\begin{equation}
	Q_M(\tau\rightarrow \infty) = \frac{T_D^2}{T_B} \langle I \rangle\;\; ,
	\label{3.40}
\end{equation}
where $T_B,T_D$ are the mean times of bright and dark periods in the 
telegraphic fluorescence signal from the three-level atom shown in 
Fig. \ref{fig2.1}. If a dark period does not occur, 
$Q_M(\tau\rightarrow \infty)\rightarrow 0$
whereas the larger $T_D$ becomes the larger the Mandel parameter becomes, 
reflecting the large fluctuations implicit in jumps from dark to bright 
periods. These macroscopic fluctuations are manifested in the photon counting
distributions studied in detail by Schenzle and Brewer \cite{Schenzle2}
using Bloch equations. They showed that the count distribution of photons 
detected from the strongly 
allowed transition were 
Poissonian {\bf except} for an excess of zero counts. In an interval $T$ 
of order of the lifetime of the shelving state one either counts a large number 
of photons (a bright period) {\bf or} one counts nothing (a dark period). 
The probability of counting $n$ photons in time $T$ is Poissonian except 
again for an excess of zeros \cite{Schenzle2} and is evaluated from the 
Mandel counting formula (or its quantum equivalent derived in \cite{Kelley1})
\begin{equation}
	W(n,T) = \frac{1}{n!} \left(\frac{\gamma_1 \eta T}{2} \right)^n 
	e^{-\frac{1}{2}\gamma_1 \eta T} \;\; ,
	\label{3.41}
\end{equation}
{\em except} for $n=0$. Here $\gamma_1$ is the decay rate of the 
strongly-fluorescing 
state and $\eta$ is the detector efficiency. So in an 
interval $T\sim\gamma_2^{-1}$ where $\gamma_2^{-1}$ is the lifetime of the
shelving state, we count {\em either} 
$n\sim \frac{1}{2}\gamma_1\eta\gamma_{2}^{-1}\approx 10^8$ for typical 
transitions {\em or} we find $n=0$. Note we essentially do {\em not} see
$n=1,2,3,\ldots$ as $W(n=1,T=\gamma_2^{-1})\sim 10^8 e^{-10^8}\cong 0$, as 
shown in Fig. \ref{fig2.6}.
Schenzle and Brewer \cite{Schenzle2} interpret these macroscopic intensity 
fluctuations 
in terms of quantum jumps. Imagine the fluorescent intensity 
to be jumping from dark (off) states to bright (on) states with probability
distribution 
\begin{equation}
	P(I) = A\delta(I) + (1-A) \delta(I-I_0) \;\; ,
	\label{3.42}
\end{equation}
then we find for the probability of counting $n$ photons in a time $T$
\begin{equation}
	W(n,T) = A\delta_{n,0} + (1-A) \frac{(H I_0 T)^n}{n!} e^{-H I_0 T} \;\; ,
	\label{3.43}
\end{equation}
where $H$ is the counting efficiency. The zero count is $W(n=0,T)=A\cong 1/3$ 
for saturated transitions. The behaviour of Eq. (\ref{3.43}) is schematically
shown in Fig. \ref{fig2.6} where we see the excess probability for no counts 
(dark period) together with a high probability for a large number of jumps 
(bright period).

In the past two sections we have discussed the initial attempts towards a
theoretical description of single ion resonance fluorescence.
However, these attempts did not yet yield a satisfying approach to the 
problem as the single system properties described, e.g., by the delay 
function, were deduced from equations of motion describing the whole 
ensemble. In the next section we will now explain and summarize a number 
of approaches all giving the quantum jump approach that allows the most 
natural description of many properties of resonance fluorescence and 
time evolution of single quantum systems. 
\section{Discussion of different derivations of the quantum 
jump approach}
\label{IV}
\subsection{Quantum jumps}
Prior to the development of quantum jump methods, all investigations 
of the photon statistics started out from the ensemble description 
via optical Bloch equations, or rate equations as presented above,
which were used to calculate {\em nonexclusive} 'probability densities' 
for the emission of one or several photons at time $t_1,\ldots,t_n$ in 
the time interval $[0,t]$. It is important to note that 
only the probability of emission of {\em any}
photon was asked for. Therefore many more photons might have been
emitted in between the times $t_i$. An example for 
such a function which we discussed in section \ref{IIIb} is the intensity 
correlation function $g^{(2)}(t)$ 
which gives the normalized rate at which one can expect to detect photons 
(any photon rather than the next) at time $t$ when one has been 
found at $t=0$. \\
Efforts were made to use nonexclusive ''probability densities''
to deduce the 
photon statistics of the single three-level ion and the aim was 
to show that a single ion exhibits bright and dark periods in 
its resonance fluorescence on the strong transition 
\cite{Pegg3,Schenzle2}. This approach is, however, not very satisfying 
as it requires the solution of the full master equation and the inversion 
of a Laplace transformation. Furthermore this approach is very indirect as 
we first calculate the ensemble properties and then try to derive the 
single particle properties. It would be much more elegant 
to have a method which enables us to calculate the photon statistics 
of the single ion directly. This was discussed widely at a workshop 
at NORDITA in Copenhagen in December 1985, following a paper at that 
meeting by Javanainen \cite{Javanainen3}. This intention was finally 
realized with the development of the quantum jump approach. Its development 
essentially started when Cohen-Tannoudji and Dalibard (1986) 
and much at the same time Zoller, Marte and Walls (1987) derived 
the {\em exclusive} probability $P_0(t)$ that, 
after an emission at time $t=0$, no other photon has been emitted 
in the time interval $[0,t]$ \cite{Cohen-Tannoudji1} or the 
exclusive n-photon probability density $p_{[0,t]}(t|t_1,\ldots,t_n)$ 
that in $[0,t]$ n photons are emitted exactly at the times
$t_1,\ldots,t_n$ \cite{Zoller1} without going back to the master equation
of the full ensemble. Both quantities are intimately related, 
as the
probability density $I_1(t)$ for the emission of the first photon 
after a time $t$ is given by 
\begin{equation}
	I_1(t) = P_0(t) \;\; ,
	\label{1}
\end{equation}
and because it turns out that the exclusive n-photon probability 
density essentially factorizes into next photon probability densities
$I_1(t)$
\begin{equation}
	p_{[0,t]}(t|t_1,\ldots,t_n) = P_0(t-t_n)I_1(t_n-t_{n-1})
	\cdot\ldots\cdot I_1(t_1)\; .
	\label{2}
\end{equation}
This factorization property was initially assumed and then
justified by physical arguments by Cohen-Tannoudji and Dalibard
in \cite{Cohen-Tannoudji1} 
while in \cite{Zoller1} first the exclusive n-photon probability
$p_{[0,t]}(t|t_1,\ldots,t_n)$ was calculated and subsequently 
its decomposition into next photon probabilities was derived. Before 
we discuss the approaches of Cohen-Tannoudji and Dalibard
\cite{Cohen-Tannoudji1} and Zoller, Marte and Walls \cite{Zoller1} we 
point out that although the exclusive 
(next photon at time $t$) $I_1(t)$ and nonexclusive distributions 
(a photon at time $t$) $I(t)$ are very different, they are related 
by a simple integral equation \cite{Kim1}. We have, as discussed 
in Eq. (\ref{3.22}) in Section \ref{IIIb}
\begin{equation}
	I(t) = I_1(t) + \int_{0}^{t} dt' I_{1}(t-t')I(t') \;\; ,
	\label{3}
\end{equation} 
which becomes especially simple as we saw when one considers the Laplace 
transform of that equation as we have a convolution on the 
right hand side of Eq. (\ref{3}). This relationship enables us in 
principle to obtain the exclusive probability density $I_1(t)$ 
from the nonexclusive quantity $I(t)$. In practice however this 
is exceedingly difficult to do as one has to know all the 
eigenvalues of the corresponding Bloch equations. Therefore a 
direct approach is needed.\\ 
The idea put forward in \cite{Zoller1} was to calculate not the 
complete density operator $\rho$ irrespective of the number of 
photons that have been emitted but to discriminate between 
density operators corresponding to different numbers of emitted 
photons in the quantized radiation field. The quantity of interest 
is therefore 
\begin{equation}
	\rho^{(n)}_A(t) = tr_{F}\{\pr_n \rho(t)\} \;\; ,
	\label{4}
\end{equation}
where $\rho(t)$ is the density operator of atom and quantized 
radiation field, $tr_{F}\{.\}$ the partial trace over the 
modes of the quantized radiation field and $\pr_{n}$ the 
projection operator onto the state of 
the quantized radiation field that contains $n$ photons. This 
projector is given by
\begin{equation}
	\pr_n = \frac{1}{n!} \sum_{{\bf k}_1\lambda_1}\ldots 
		\sum_{{\bf k}_n\lambda_n}
		a^{\dagger}_{{\bf k}_1\lambda_1}\ldots 
		a^{\dagger}_{{\bf k}_n\lambda_n}
		|0\rangle\langle 0| a_{{\bf k}_n\lambda_n}\ldots 
		a_{{\bf k}_1\lambda_1}\; .
	\label{5}
\end{equation}
This method to calculate the density operator for a given number of
photons in the quantized radiation field was first used by Mollow (1975)
to investigate the resonance fluorescence
spectrum of two-level systems. However, as at that time the
investigation of single ions was completely beyond then-current
experimental possibilities, so that he did not draw further conclusions 
from his approach concerning single quantum systems. This was only
triggered later by the experimental realization of single ions in ion 
traps. 

In the following we discuss the approach of Zoller {\em et al} (1987)
for a three-level system in V configuration (see Fig. \ref{fig2.1}) case 
rather than as in the original paper for the two-level 
case. Following \cite{Mollow1,Blatt1} one obtains for 
Eq. (\ref{4}) the equations of motion
\begin{equation}
	\frac{d}{dt} \rho_{A}^{(0)}(t) = 
	-i ( H_{eff}\rho_{A}^{(0)}(t) - 
	\rho_{A}^{(0)}(t)H_{eff}^{\dagger} )
	\label{6a}
\end{equation}
and
\begin{equation}
	\frac{d}{dt} \rho_{A}^{(n)}(t) = 
	-i ( H_{eff}\rho_{A}^{(n)}(t) - 
	\rho_{A}^{(n)}(t)H_{eff}^{\dagger} )
	+ \sum_{i=1}^{2} 2\Gamma_{ii} |0\rangle\langle i|
	 \rho_{A}^{(n-1)}(t)|i\rangle\langle 0| \;\; ,
	\label{6b}
\end{equation}
where the effective Hamilton operator $H_{eff}$ is given by 
\begin{equation}
	H_{eff} = -\sum_{i=1}^{2} \left\{\hbar(\Delta_i +
	 i\Gamma_{ii})|i\rangle\langle i|
	+ \frac{\hbar\Omega_{i}}{2} (|0\rangle\langle i| +
	 |i\rangle\langle 0|)
	\right\} \;\; ,
	\label{7}
\end{equation}
with the detunings $\Delta_i=\tilde{\omega}_i-\omega_{i1}$ and,
$\tilde{\omega}_i$ the laser frequency, $\Omega_i$ the Rabi frequency 
and $2\,\Gamma_{ii}$ the Einstein coefficient on the 
$i\leftrightarrow 1$ transition. It is now important to note that the 
effective Hamilton operator $H_{eff}$
is a {\em non-Hermitean} operator. The real part of $-i\,H_{eff}$ 
is negative which implies that the trace of the density operator 
$\rho_{A}^{(0)}(t)$ decreases in time. This is not surprising 
because $\rho_{A}^{(0)}(t)$ describes the conditional time 
evolution under the assumption that no photon has been emitted 
into the quantized radiation field. The probability that 
an excited atom has not emitted a photon decreases in time and
therefore the trace of $\rho_{A}^{(0)}(t)$ describing this probability
should decrease in time. 
This decrease is necessary for the trace of the density
operator $\rho(t)$ disregarding the number of emitted photons, 
\begin{equation}
	T(t,0)\rho(0) = \rho(t) = \sum_{n=0}^{\infty}
	 \rho_{A}^{(n)}(t)\; .
	\label{8}
\end{equation}
to be preserved under the time evolution.
The equations of motion Eqs. (\ref{6a}) and (\ref{6b}) have the solution
\begin{equation}
	\rho_{A}^{(0)}(t) = S(t,t_0) \rho_{A}^{(0)}(t_0) \;\; ,
	\label{9}
\end{equation}
where
\begin{equation}
	S(t,t_0) \rho_{A}^{(0)}(t_0) = 
	e^{-i\,H_{eff}(t-t_0)}\rho_{A}^{(0)}(t_0)
	 e^{i\,H_{eff}^{\dagger}(t-t_0)}
	\label{10}
\end{equation}
and
\begin{equation}
	\rho_{A}^{(n)}(t) = \int_{0}^{t} dt' S(t,t') R
	 \rho_{A}^{(n-1)}(t')
	\label{11}
\end{equation}
with 
\begin{equation}
	R \rho_{A}^{(n)}(t) = \sum_{i=1}^{2} 2\Gamma_{ii} 
	|0\rangle\langle i| \rho_{A}^{(n)}(t) |i\rangle\langle 0|\; .
	\label{12}
\end{equation}
From this result it is now possible to deduce the probability that
exactly $n$ photons have been emitted in the time interval $[0,t]$.
The probability that no photon has been found should then be given by
\begin{equation}
	P_{0}(t) = tr_{A}\{ S(t,0)\rho_{A}(0) \}\;\; ,
	\label{13}
\end{equation}
and
\begin{equation}
	P_{n}(t) = \int_{0}^{t} dt_n \ldots \int_{0}^{t_2} dt_1 
	tr_{A}\{ S(t,t_n) R \ldots R S(t_1,0) \rho_{A}(0) \}
	\label{14}
\end{equation}
is the probability that exactly $n$ photons have been emitted. 
Zoller, Marte and Walls then realized that the structure of these
expressions coincides with that derived from an abstract theory of
continuous measurement constructed by Srinivas and Davies 
\cite{Davies1,Davies2,Davies3,Davies4,Srinivas1,Srinivas2}. This
theory supports the interpretation of
\begin{equation}
	p_n(t_1,\ldots,t_n) = 
	tr_{A}\{ S(t,t_n) R \ldots R S(t_1,0) \rho_{A}(0) \}
	\label{15}
\end{equation}
as the probability density that exactly $n$ photons have been 
emitted at times $t_1,\ldots,t_n$ and no photons in between. 
From the general theory of measurement, they interpreted the
quantity
\begin{equation}
	P_0(t_1-t_0) = \frac{tr_{A}\{R S(t_1,t_0) R T(t_0,0) \rho_{A}(0) \} }
	{tr_{A}\{R T(t_0,0) \rho_{A}(0) \} }
	\label{16}
\end{equation} 
as the probability density that after an emission of a photon at 
time $t_0$ the {\em next} photon will be emitted at $t_1$. It 
should be stressed that although in \cite{Zoller1} the super--operator 
$T(t,0)$ is identified with the time evolution of 
the ensemble irrespective of how many photons have been emitted in 
$[0,t]$ for the following it should always be chosen to be 
the time evolution if a given number of emissions have taken 
place at the times $t_1,\ldots,t_n$. Assuming this (as is also
implicitly done later in \cite{Zoller1}) one obtains for the 
probability density that photons are emitted exactly at times 
$t_1,\ldots, t_n$
\begin{eqnarray}
	p_{[0,t]}(t_1,\ldots,t_n) &=& 
	tr_{A}\{ S(t,t_n) R \ldots R S(t_1,0) \rho_{A}(0) \}
	\nonumber\\[.25cm]
	&=& \frac{tr_{A}\{ S(t,t_n) R \ldots R S(t_1,0) \rho_{A}(0) \}}
	{tr_{A}\{ R S(t,t_{n-1})\ldots R S(t_1,0) \rho_{A}(0) \}} 
	\ldots
	\frac{ tr_{A}\{ R S(t_{1},0)\rho_{A}(0) \} }{tr_{A}\{\rho_{A}(0) \} }
	\nonumber\\[.25cm]
	&=& P_0(t,t_n)I_1(t_n,t_{n-1})\ldots I_1(t_1,0) \;\; .
	\label{17} 
\end{eqnarray}
Here we have factorized $p_{[0,t]}(t_1,\ldots,t_n)$ into products of
$I_1(t_l,t_{l-1})$. In principle these functions can depend on the atomic 
state at time $t_{l-1}$ (after the emission). However, in most cases 
this state will be the ground state of the system and will be the same after
each emission.
Having found that the knowledge of $P_0(t)$ is sufficient 
($I_1(t)$ can be obtained via Eq. (\ref{1})), Zoller {\em et al} then continue 
to discuss the photon statistics of the three-level V system. The results 
found in \cite{Zoller1} may also be used to implement 
a simulation approach for the time evolution of a single three-level
system \cite{Dalibard1,Dum1,Hegerfeldt1}. However, the application of the 
quantum jump approach in numerical simulations will be discussed later in 
this section.\\[1ex]
The approach of Zoller, Marte and Walls already reveals many features 
of the quantum jump approach. However there is a slight complication 
in their approach as they rely on the abstract theory of
continuous measurement of Srinivas and Davies to give interpretations 
to their expressions Eqs. (\ref{15})-(\ref{16}). The reason that 
they need the support of the theory of Srinivas and Davies is that they 
never talk about the way the photons are measured. 
In fact only the emission of photons is mentioned and not the detection 
of photons. From a quantum mechanical point of view,
however, one has to be very careful, as the emission of 
a photon is {\em not} well defined. It is the detection of a photon 
in the radiation field which is a real event. Of course the treatment 
in \cite{Zoller1} already
implies some properties of the measurement process, e.g., they implicitly 
assume time resolved photon counting. However, no explicit treatment  
of such measurements was given in \cite{Zoller1}. This problem was then
addressed in greater detail by several authors and in the following
we discuss these ideas. 

The first approach to include the result of quantum mechanical
measurements into their calculation explicitly was given by Porrati 
and Putterman (1989). They \cite{Porrati1} as well as others 
\cite{Pegg3} noticed that the failure to detect a photon in a measurement 
leads to a state reduction, as information is gained through that null
measurement. Essentially we can be increasingly confident that the ion 
is in a non radiating state (examples of this will be shown in Section V).
Porrati and Putterman assume that at some large time $t$ a measurement 
on the whole quantized radiation field is performed. Assuming the result 
of this measurement is the detection of no photons, they calculate all
Heisenberg operators at that time projected onto the null-photon subspace 
of the complete Hilbert space, i.e., operators of the form
\begin{equation}
	(\pr_{0}\hat{O}_A \pr_{0})(t)\; .
	\label{24a}
\end{equation}  
Although not mentioned explicitly in 
\cite{Porrati1,Porrati2}, the calculation of this operator turns 
out to be closely related to the projector formalism 
\cite{Agarwal1,Haake1,Nakajima1,Zwanzig1},
a connection that was elaborated on by Reibold \cite{Reibold1}. 
Although their approach can in principle lead to the quantum jump method, 
there are some conceptual problems in the actual execution 
of the use of the null-measurements idea. The main problem is that 
Porrati and Putterman only talk about a single measurement at a large 
time $t$ performed 
on the complete quantized radiation field. This does not seem to be 
a very realistic model of measurements performed by a broadband
counter, which informs us immediately whether he has detected 
a photon or not. Also the calculation of the state after the detection of 
a photon was not elaborated on in  \cite{Porrati1,Porrati2}, 
where it was merely stated that the system is reset back to its ground 
state on photodetection which is of course a physically correct picture. 
These conceptual concerns to this approach were later addressed in the 
work of Hegerfeldt and Wilser \cite{Hegerfeldt1,Wilser1}, of Carmichael
\cite{Carmichael2} and of
Dalibard, Castin and M{\o}lmer\cite{Dalibard1,Molmer1,Molmer2} 
and in the following we give a more detailed account of their approach. 

We will 
follow closely the presentation given by the Hegerfeldt group  
\cite{Hegerfeldt1,Wilser1} 
as it directly leads to the delay function that was also found in 
earlier papers \cite{Cohen-Tannoudji1,Zoller1}. The
physical ideas, however, are very similar to those presented elsewhere
\cite{Dalibard1,Molmer1,Molmer2}. We treat the same three level system as
in the discussion of Zoller {\em et al} (1987). 

In  Hegerfeldt and Wilser (1991) and Wilser (1991) (as well as in 
\cite{Dalibard1,Molmer1,Molmer2}), the following simple model of how 
the photons are detected was proposed. It was assumed that the 
radiating ion is surrounded by a $4\pi$--photodetector that 
detects photons irrespectively of their frequency and that the
efficiency of the detector is unity. Efficiencies less than unity 
may be treated in a mathematically slightly different way using the 
same physical ideas \cite{Plenio1,Hegerfeldt8} and leads to a natural 
connection between the 
{\em next} photon probability density $I_1(t)$ and the {\em any} 
photon rate (intensity correlation function) $g^{(2)}(t)$ 
\cite{Plenio1,Kim1}. We will return to this point later.
As truly continuous measurements in quantum mechanics are not 
possible without freezing the time evolution of the system
through the Zeno effect \cite{Mahler1,Misra1,Reibold2}, it was instead 
assumed that measurements 
are performed in rapid succession where the time difference 
$\Delta t$ between successive measurements should be much larger 
than the correlation time of the quantized radiation field. This 
means that 
\begin{equation}
	\Delta t \gg \omega_{10}^{-1}\; .
	\label{18}
\end{equation}
If successive measurements are more frequent than $\omega_{10}^{-1}$ we 
enter the regime of the quantum Zeno effect and we significantly 
inhibit the possibility of spontaneous emissions \cite{Reibold2}. 
On the other hand $\Delta t$ should be much smaller than all time
constants of the atomic time evolution to ensure that one finds the 
photons one by one and because we want determine the time evolution 
using perturbation theory. Therefore
\begin{equation}
	\Gamma_{ii}^{-1},\Delta_i^{-1},\Omega_i^{-1} \ll \Delta t\;\; .
	\label{19}
\end{equation} 
For optical transitions it is easy to satisfy both inequalities
Eqs. (\ref{18}) and (\ref{19}) simultaneously. 

Now the density operator at time $t$ under the condition that 
no photons have been detected in all measurements which took 
place at time $s_1,\ldots,s_n$ has to be calculated. Although 
the result of each measurement was negative, in the sense that no 
photon was found, this still has an impact on the 
wavefunction of the system, as it represents an increase of
knowledge about the system \cite{Dicke1,Porrati1,Pegg3}. Using the 
projection operator $\pr_0$ onto the vacuum state of the quantized
radiation field and the time evolution operator $U(t,t')$ of 
system and radiation field in a suitable interaction picture, we find
\begin{equation}
	|\psi(s_n)\rangle = \pr_0 U(s_n,s_{n-1}) \pr_0 \ldots \pr_0 U(s_1,0) 
	|\psi(0)\rangle\; .
	\label{20}
\end{equation}
as, after each measurement which has failed to detect a photon, we 
have to project the quantized radiation field onto the vacuum 
state according to the von Neumann-L{\"u}ders postulate
\cite{Lueders1,Neumann1}. As $s_j-s_{j-1}$ obeys Eqs. (\ref{18}) and 
(\ref{19}), it is possible to calculate the time evolution operator
$U(s_j,s_{j-1})$ in second order perturbation theory to obtain
\begin{equation}
	\pr_0 U(s_j,s_{j-1}) \pr_0 \approx 
	(\id - \frac{i}{\hbar}H_{eff}(s_{i-1})(s_i-s_{i-1}) ) \pr_0
	\label{21}
\end{equation}
with the effective Hamilton operator Eq. (\ref{7}). In the quantum jump 
method presented in \cite{Dalibard1,Molmer1,Molmer2} 
the Weisskopf-Wigner approach \cite{Weisskopf1} was used to find a 
formula equivalent to Eq. (\ref{21}). Inserting into Eq. (\ref{20}) and 
going over from a coarse grained time scale to a continuous time, 
we obtain for the atomic part of the wavefunction (the radiation 
field is in its vacuum state) where no photons have been detected
in all measurements in the interval $[0,t]$ 
\begin{equation}
	|\psi_{A}^{(0)}(t)\rangle = e^{-i\,H_{eff} t/\hbar}
	 |\psi_{A}(0)\rangle\; .
	\label{22}
\end{equation}
One should note that the effective time evolution does not preserve 
the norm of the state and that it maps pure states onto pure states. In fact 
the square of the norm of Eq. (\ref{22}) is just the delay function 
\begin{equation}
	P_{0}(t) = \langle\psi_{A}^{(0)}(t)|\psi_{A}^{(0)}(t)\rangle
	\label{22a}
\end{equation}
and coincides with $P_0(t)$ given in Eq. (\ref{13}). The delay function
$P_0(t)$ will become important in applications of the method in 
simulations \cite{Dalibard1,Dum1}. It should be noted here in passing that 
if we consider the normalized version of the time evolution Eq. (\ref{22}) 
for a two-level system, then one finds that it is identical to the
time evolution according to the neoclassical radiation theory
of Jaynes \cite{Bouwmeester1}. The reason for this is claimed by
Bouwmeester et al to rest on 
the fact that $H_{eff}$ contains all contributions from virtual 
photons (i.e. all radiation reaction terms) but does not include 
the real photons as their detection leads to state reduction 
according to the projection postulate. However, neoclassical theory
predicts quantum beats from a three-level system in $\Lambda$ 
configuration while it is easily seen that an analysis of the 
problem using the quantum jump approach does not predict quantum 
beats; a result in accordance with experiment Milonni (1976)
and references therein.

Eventually the photo detector will find a photon and the state 
after this detection can be determined by the projection postulate. 
We write the state after the detection of a photon as a density 
operator, as the state after the emission can be a mixture (e.g. as in 
the three-level $\Lambda$ configuration 
\cite{Javanainen4,Hegerfeldt3,Hegerfeldt6,Hegerfeldt9}) 
although in the case of the three-level V system it is not:
\begin{equation}
	\tilde{\rho}_{R}(s_n) = (\id-\pr_0) U(s_n,s_{n-1})\pr_0
 	\rho_{A}^{(0)}(s_{n-1})
	\pr_0 U(s_n,s_{n-1})(\id-\pr_0) \;\; .
	\label{23}
\end{equation}
At this point the additional assumption is made that the photo 
detector absorbs the photon (which it does in reality) and that 
the state after the detection is simply obtained by removing the 
photons from the radiation field. This can be done by tracing over 
the quantized radiation field and multiplying with $\pr_0$
\begin{eqnarray}
	\rho_{R} &=& tr_{F}\{\tilde{\rho}_{R}(s_n)\} \otimes \pr_0
	\nonumber\\
	&=& \sum_{i=1}^{2} 2\Gamma_{ii} |0\rangle\langle i| 
	\rho_{A}^{(0)}(s_{n-1}) |i\rangle\langle 0| (s_n -s_{n-1})\;\; .
	\label{24}
\end{eqnarray}
It should be noted that the {\em assumption} that the state after the 
detection of a photon in the counter is given by Eq. (\ref{24}) 
is not included in the
projection postulate but enters as a physically justified 
additional assumption. However using slightly different 
mathematical methods it is possible to show that the procedure
Eq. (\ref{24}) is not really necessary. Ideal quantum mechanical
measurements where the photon is not absorbed lead to the same 
results \cite{Plenio1,Hegerfeldt8}. This is a consequence of the intuitively 
obvious fact that in free space photons emitted by the system will 
never return to it and is implicit in the treatment of 
Zoller {\em et al} (1987).\\
A different approach towards the quantum jump method was presented earlier
by Carmichael and coworkers \cite{Carmichael1,Carmichael2}. 
They derive the quantum jump method from a discussion 
of photoelectron counting distributions that are found in
experiments. A quantum mechanical theory for photoelectron 
counting distributions was developed in 1964 by Kelley and Kleiner
\cite{Kelley1} who derived the quantum mechanical expressions 
for nonexclusive multicoincidence rates. For the probability to 
have $n$ photoelectron counts in the time interval $[t,t+T]$, they 
find
\begin{equation}
	p(n,t,T) = \langle :
	\left\{ \frac{\xi\int_{t}^{t+T} dt' \hat{E}^{(-)}(t')
	\hat{E}^{(+)}(t')}{n!}
	\right\}
	exp\{ -\xi\int_{t}^{t+T} dt' \hat{E}^{(-)}(t')\hat{E}^{(+)}(t') \}
	:\rangle \;\; ,
	\label{25}
\end{equation}
where $\xi$ is the product of detector efficiency and a factor 
to convert field intensity into photon flux. The notation
 $\langle:\ldots :\rangle$ means that all operators have to be 
normally ordered {\em} and time ordered in such a way that times
decrease from the centre towards the left and right. Expanding
Eq. (\ref{25}) one can write $p(n,t,T)$ as a complicated series of 
integrals over the {\em nonexclusive} multicoincidence rates
 \begin{equation}
	I(t_1,\ldots,t_m) = \xi^{m} \langle 
	\hat{E}^{(-)}(t_1)\ldots \hat{E}^{(-)}(t_m)
	\hat{E}^{(+)}(t_m)\ldots \hat{E}^{(+)}(t_1) \rangle \;\; ,
	\label{26}
\end{equation} 
which gives the rate for the joint detection of photons at times
$t_1,\ldots,t_m$. It is a nonexclusive rate as there may be more
detections in between the times $t_1,\ldots,t_m$. That these 
possible events are included in Eq. (\ref{26}) is obvious, as the 
Heisenberg operators are calculated with respect to the total
Hamiltonian of the system which describes a time evolution in 
which arbitrarily many photons may be created. The analysis of 
the photon statistics by means of the coincidence rate Eq. (\ref{26}) 
has long been the standard way of investigation. It was, however, 
realized that this is not the only possibility, and for 
certain problems it is not even the most natural way. In fact Eq. (\ref{25}) 
, for example, can be expressed very easily by the {\em exclusive}
probability density to find photon counts at exactly the times 
$t_1,\ldots,t_n$ and at no other time in $[t,t+T]$. One finds for this the 
expression
\begin{equation}
	p(n,t,T) = \int_{t}^{t+T} dt_n \int_{t}^{t_n} dt_{n-1} 
	\ldots \int_{t}^{t_2} dt_1
	p_{[t,t+T]}(t_1,\ldots,t_n)\;\; .
	\label{27}
\end{equation}
Carmichael and coworkers \cite{Carmichael1,Carmichael2} then undertook 
the step to express the 
exclusive probability density in terms of the intensity operators. 
They find
\begin{equation}
	p_{[t,t+T]}(t_1,\ldots,t_n) = \sum_{r=0}^{\infty}
	 \frac{(-1)^r}{r!} 
	\int_{t}^{t+T} dt'_{r} \ldots \int_{t}^{t+T} dt'_{1}
	\langle : \hat{I}(t'_r)\ldots\hat{I}(t'_1)\hat{I}(t_m)
	\ldots \hat{I}(t_1) 
	:\rangle \;\; ,
	\label{28}
\end{equation}
where 
\begin{equation}
	\hat{I}(t) = \xi\hat{E}^{(-)}(t)\hat{E}^{(+)}(t)\; .
	\label{29}
\end{equation}
This can be checked by inserting  Eqs. (\ref{28}) and (\ref{29}) into 
Eq. (\ref{27}) and showing that the result coincides with Eq. (\ref{25}) 
\cite{Saleh1,Stratonovitch1}. The aim now is to rewrite Eq. (\ref{28})
in terms of (super)-operators that only act in the atomic space, as 
these are much easier to handle than the Heisenberg-operators
$\hat{I}(t)$. It turns out that the resulting equations are quite
simple. To this end it is important to note that the electric field
operator in the Heisenberg picture can be decomposed into a 
free-field part and a source field part
\begin{equation}
	\hat{E}^{(-)}(t) = \hat{E}^{(-)}_{f}(t_i) 
	+ \hat{E}^{(-)}_{s}(t_i)\; .
	\label{30}
\end{equation} 
In the Markov approximation the free field commutes with all
electric field operators at earlier times Cohen-Tannoudji, 
J. Dupont-Roc, and G. Grynberg (1992) 
and when acting onto the vacuum state it vanishes. It is therefore
possible to replace the intensity operators $\hat{I}(t_i)$ for
$i=1,\ldots,m$ by $\hat{E}^{(-)}_{s}(t_i)\hat{E}^{(+)}_{s}(t_i)$ 
in Eq. (\ref{28}). All other intensity operators remain unchanged. 
Carrying out the time ordering in Eq. (\ref{28}) explicitly and 
after tedious calculations (for details we refer the reader to
\cite{Carmichael1,Carmichael2}) one obtains for an overall counter
efficiency $\xi$
\begin{equation}
	p_{[t,t+T]}(t_1,\ldots,t_n) = \xi^n 
	tr_A\{ e^{({\cal L}-\xi {\cal R})(t+T-t_m)}
	{\cal R}\ldots {\cal R} e^{({\cal L}-\xi 
	{\cal R})(t_1-t)}\rho(t)\} \;\; ,
	\label{31}
\end{equation}
where ${\cal L}$ is the superoperator given by
\begin{equation}
	{\cal L}\rho = -\frac{i}{\hbar} (H_{eff}(\xi)\rho - 
	\rho H_{eff}(\xi) ) \;\; ,
	\label{32}
\end{equation}
where $H_{eff}(\xi)$ is obtained from $H_{eff}$ for perfect efficiency 
$\xi=1$ by substituting
$\Gamma\rightarrow(1-\xi)\Gamma$. ${\cal R}$ is the reset operator
giving the state after a photon has been detected. Assuming unit
efficiency ($\xi=1$) of the detection process we recover 
Eq. (\ref{7}).\\[1ex]
So far we have discussed a number of approaches to the quantum 
jump description of dissipation. These approaches can be formulated 
somewhat 
differently in the language of quantum stochastic differential 
equations \cite{Gardiner4}. This formulation is certainly rather 
formal at first glance, but it has the advantage that certain 
operations where one uses the Markov approximation become simpler. 
On the other hand one has to use the somewhat unintuitive
Ito formalism \cite{Gardiner4,Gardiner1} and a more physically oriented 
derivation would sometimes be helpful for the interpretation 
of the occurring equations.\\
To illustrate the idea of this formalism we consider a laser-driven 
two-level atom in a quantized radiation field which is in the vacuum 
state. We follow the description in Sondermann (1995b).
The Hamilton operator in a suitable interaction picture is given by
\begin{eqnarray}
	H &=& -\hbar\Delta_1|1\rangle\langle 1| + \frac{\hbar\Omega}{1}
	(|0\rangle\langle 1| + |1\rangle\langle 0|) + 
	\sum_{{\bf k}\lambda} (i\hbar g_{{\bf k}\lambda} 
	\sigma_{10} a_{{\bf k}\lambda} 
	e^{-i(\omega_{{\bf k}\lambda}-\omega_{10})t} + h.c.)\nonumber\\
	&=& H_A + \sigma_{10}{\bf D_{10}}{\bf E}^{(+)}(t) + 
	{\bf D_{10}^{\dagger}}{\bf E}^{(-)}(t)\sigma_{01} \;\; ,
	\label{96}
\end{eqnarray}
where $\sigma_{ij}=|i\rangle\langle j|$ is an operator annihilating 
an electron in level $j$ and creating an electron in level $i$ and
where 
\begin{equation}
	{\bf D_{10}}{\bf E}^{(+)}(t) = {\bf D_{10}}
	\sum_{{\bf k}\lambda} i\hbar 
	\left(\frac{e^2 \omega_{{\bf k}\lambda})}
	{2\epsilon_0\hbar V}\right)^{(1/2)}
	\biggreec{\epsilon}_{{\bf k}\lambda} \sigma_{10} 
	a_{{\bf k}\lambda} 
	e^{-i(\omega_{{\bf k}\lambda}-\omega_{10})t}
	\label{97}
\end{equation}
is the interaction energy between the electric-field operator 
$E^{\dagger}(t)$ in the Schr\"odinger picture (or more precisely in 
the chosen interaction picture) and the atomic dipole moment 
${\bf D}_{21}$ of the transition. The time discretized Schr\"odinger 
equation then reads
\begin{eqnarray}
	i\hbar\Delta|\psi(t)\rangle &=& H|\psi(t)\rangle \Delta t 
	\nonumber \\
	&=& \{ H_A \Delta t + \sigma_{10}\Delta A^{\dagger}(t) 
	- \sigma_{01} \Delta A(t)\}
	|\psi(t)\rangle \;\; ,
	\label{98}
\end{eqnarray}
where
\begin{equation}
	\Delta A(t) = \int_{t}^{t+\Delta t} dA(t) = 
	\int_{t}^{t+\Delta t} dt \, {\bf D_{10}}
	\sum_{{\bf k}\lambda} i\hbar 
	\left(\frac{e^2 \omega_{{\bf k}\lambda})}
	{2\epsilon_0\hbar V}\right)^{(1/2)}
	\biggreec{\epsilon}_{{\bf k}\lambda} \sigma_{10} 
	a_{{\bf k}\lambda} 
	e^{-i(\omega_{{\bf k}\lambda}-\omega_{10})t}\;\; .
	\label{99}
\end{equation}
We assume in the following that $\Delta t\gg\omega_{10}^{-1}$, 
which is crucial for us to be able to perform the Markov approximation. 
The idea is now to perform the Markovian limit directly
in the Schr\"odinger equation instead of performing this limit on the 
results. This is the step where we have to introduce the notion 
of quantum stochastic differential equations, as in performing 
this limit we cannot subsequently interpret the resultant Schr\"odinger 
equation as an ordinary differential equation anymore
\cite{Gardiner4,Gardiner1}. Under the Markov assumption we have 
\begin{equation}
	[\Delta A(t),\Delta A^{\dagger}(s)] = \left\{\begin{array}{cl}
	0              & \mbox{if}\, |t-s|\geq\Delta t\\
	\Gamma\Delta t & else 
	\end{array}\right. \;\; ,\label{100}
\end{equation}
which in the limit $\Delta t\rightarrow 0$ results in
\begin{equation}
 	[dA(t),dA^{\dagger}(s)] = \left\{\begin{array}{cl}
	0 & \mbox{if}\; t\neq s\\
	\Gamma dt & else 
	\end{array}\right.\; .\label{101}
\end{equation}
We also need to know that if we assume that the initial state of 
the quantized radiation field is the vacuum, then
\begin{equation}
	dA(t)dA(t) = 0 = dA^{\dagger}(t)dA^{\dagger}(t) \;\; ,
	\label{102}
\end{equation}
where this equation is defined in the mean-square topology sense, i.e.,
in brief one applies both sides on an initial vector and takes the
absolute square of the result afterwards \cite{Gardiner4,Gardiner1}. Taking 
the limit $\Delta t\rightarrow 0$ in Eq. (\ref{99}) we have assumed 
the ordinary rules of calculus, and therefore generated a stochastic 
differential equation in the sense of Stratonovitch
\begin{eqnarray}
	i\hbar d|\psi(t)\rangle|_{S} &=& 
	\{ H_A dt + \sigma_{10} dA^{\dagger}(t) 
	- \sigma_{01} dA(t)\} |\psi(t)\rangle \;\; .
	\label{103}
\end{eqnarray}
As a Stratonovitch equation is not easy to integrate, we would 
like to transform it to an Ito-form using the rules
Eqs. (\ref{100})-(\ref{102}) \cite{Gardiner4,Gardiner1}. We then find 
\begin{equation}
	i\hbar d|\psi(t)\rangle|_{I} 
	= \{ H_A dt + dA^{\dagger}(t)\sigma_{01} - 
	\Gamma\sigma_{11} dt\}|\psi(t)\rangle \;\; ,
	\label{104}
\end{equation}
where the $\Gamma\sigma_{11}dt$ arises from a 
$dA(t)dA^{\dagger}(t)$ contribution. As the $dA(t)$ commute with 
all earlier $dA(s)$ upon which $|\psi(t)\rangle$ depends, it can
be commuted to the right until it operates on the initial state 
and therefore the vacuum. Therefore the contribution of the 
$dA(t)$ vanishes and only the $dA^{\dagger}(t)$ contribution survives. \\
What we are in fact interested in here is the rederivation of the 
quantum jump approach. Therefore we are interested in the time 
evolution when no photon is present in the field, i.e., we are interested
in the state vector
\begin{equation}
	\pro|\psi (t)\rangle = |\psi(t)\rangle_0 \;\; ,
	\label{105}
\end{equation}
where $\pro$ is the projector onto the vacuum state of the 
quantized radiation field. We find
\begin{eqnarray}
	d|\psi(t)\rangle_0 &=& \pro d|\psi(t)\rangle\nonumber\\
	&=& (-i H_A/\hbar - \Gamma\sigma_{11}) \pro |\psi(t)\rangle 
	dt\; .
	\label{106}
\end{eqnarray}
The $dA^{\dagger}(t)$ now vanishes because acting on a vacuum state to 
its left it gives zero contribution.
The norm of the conditional state vector $|\psi(t)\rangle_0$ is 
just the probability to find no photon until $t$ if there was no photon
at $t=0$. This is just the reduced time evolution found in the previously
discussed derivations other approaches, too. The probability density 
$I_1(t)$ for an 
emission at time t is just the rate of decrease of the norm of 
the emission free time evolution, i.e.,
\begin{eqnarray}
	d_0\langle\psi(t)|\psi(t)\rangle_o|_I &=& 
	\left( d\, {}_0\langle\psi(t)|\right)|\psi(t)\rangle_0 + 
	{}_0\langle\psi(t)|\left(d|\psi(t)\rangle_0\right) + 
	\left(d\,{}_0\langle\psi(t)|\right)
	\left(d|\psi(t)\rangle_0\right)
	\nonumber\\
	&=& {}_0\langle\psi(t)|i H_A/\hbar - \Gamma\sigma_{11}
	 |\psi(t)\rangle_0
	+ {}_0\langle\psi(t)| -i H_A/\hbar - \Gamma\sigma_{11}
	|\psi(t)\rangle_0
	\nonumber\\
	&=& -2\Gamma\,\, {}_0\langle\psi(t)|\sigma_{11}|\psi(t)
	\rangle_{0}\;\; .
	\label{107}
\end{eqnarray}
Therefore using this formulation we have recovered the quantum 
jump approach; we observe that this formalism, although not
delivering new insights into physics different from those from previous
derivations of the quantum jump approach, is very elegant from a formal 
point of view. To 
understand the formalism a little better, we now show how one may 
obtain Eq. (\ref{106}) without referring to the formalism of 
stochastic differential equations \cite{Zoller2}. We consider a 
finite time step for the state vector $\pro|\psi(t)\rangle$ using 
the Hamilton operator Eq. (\ref{96}) and first order perturbation theory. 
We obtain
\begin{eqnarray}
	\Delta\pro |\psi(t)\rangle &=& 
	\pro (-i H_A/\hbar \Delta t - 
	\sigma_{10}\Delta A(t))|\psi(t)\rangle
	\nonumber\\
	&=& \pro U(t,0)U^{(\dagger)}(t,0) 
	(-i H_A/\hbar \Delta t - \sigma_{10}\Delta A(t)) 
	U(t,0) U^{(\dagger)}(t,0)|\psi(t)\rangle
	\nonumber\\
	&=& -i H_A/\hbar \Delta t\pro|\psi(t)\rangle + 
	\pro U(t,0) \sigma_{10}\left(\Delta A(t) - 
	\Gamma \sigma_{01}(t) \right) |\psi(0)\rangle \;\; ,
	\label{108}
\end{eqnarray}
where in the last line we have used the well known expression for  
the Heisenberg operator of the electric field operator which can 
be written as the free field contribution and a source term (the 
dipole of the atom radiates the outgoing field) \cite{Loudon1}. 
Note that $\sigma_{01}(t)$ is now a Heisenberg operator.
Eliminating $A(t)$ in the last row of Eq. (\ref{108}), as it 
operates on the initial vacuum state, we obtain
\begin{eqnarray}
	\Delta\pro |\psi(t)\rangle &=& 
	\Delta t \left( -i H_A/\hbar - \Gamma \sigma_{11} \right)
	\pro|\psi(t)\rangle \;\; .
	\label{109}
\end{eqnarray}
Now we may easily perform the limit $\Delta t\rightarrow 0$ to 
obtain the same result as in Eq. (\ref{106}), however, without the 
explicit use of the quantum stochastic differential calculus. 
\subsection{Quantum state diffusion and other approaches to single system
dynamics}
So far we have discussed the quantum jump approach for the 
description of single radiating quantum systems. The main ingredient 
in the derivation was the assumption of time resolved photon counting
measurements on the quantized radiation field. The resulting time 
evolution could be divided into a coherent time evolution governed by a 
non Hermitean Hamilton operator which is interrupted by instantaneous 
jumps caused by the detection of a photon and the 
consequent gain in knowledge about the system.  
One could ask whether this description is unique, that is, it 
represents the only possibility. From the emphasis we put on the
importance of the measurement process in the derivation of the 
quantum jump approach one can already guess that other measurement 
prescriptions will yield different kind of quantum trajectories.
In the following we will discuss an important example, quantum state 
diffusion, \cite{Gisin3} of a different
kind of quantum trajectories which in fact can be derived from a 
very important measurement method in Quantum Optics, namely, the balanced 
heterodyne detection. Before we show the connection of quantum state 
diffusion to balanced heterodyne detection let us point out that quantum state
diffusion was originally derived independently from a measurement 
context. Steps in this direction 
were made when several authors became interested in alternative 
versions of quantum mechanics 
\cite{Pearle1,Ghirardi1,Ghirardi2,Diosi1,Diosi3} and the
investigation of the wavefunction function collapse, i.e., the 
projection postulate \cite{Gisin1,Gisin2}. In these investigations 
stochastic differential equations for the time evolution of the 
state vector of the system were studied. Again there is a 
multitude of possible equations; however, Gisin and Percival (1992a)
provided a natural symmetry condition under which 
it is possible to derive a unique diffusion equation which is
referred to as the quantum state diffusion model (QSD).
Given a Bloch equation in Linblad form \cite{Lindblad1} 
\begin{equation}
	\dot{\rho} = -\,\frac{i}{\hbar} \left[ H_{sys},\rho \right] 
	+ \sum_{m} 
	\left( 2\,L_m\rho L_m^{\dagger} - L_m^{\dagger} L_m \rho - 
	\rho L_m^{\dagger} L_m \right) \;\; ,
	\label{110}
\end{equation}
with the system Hamiltonian $H_{sys}$, and the Lindblad operators 
$L_m$, the quantum state diffusion equation for the state vector is
\begin{eqnarray}
	|d\psi\rangle &=& 
	-\,\frac{i}{\hbar}\left(H_{sys} - i\hbar 
	L_m^{\dagger} L_m\right)|\psi\rangle
	dt + \sum_m
	\left( 2\langle L_m \rangle L_m - 
	\langle L_m^{\dagger} \rangle
	\langle L_m \rangle \right) |\psi\rangle dt
	\nonumber\\
	&& +\sum_m \left( L_m - \langle L_m \rangle \right) 
	|\psi\rangle d\xi_m\;\; .
	\label{111}
\end{eqnarray}
The $d\xi_m$ represent independent complex normalized Wiener 
processes whose averages, denoted by M(\ldots), satisfy
\begin{eqnarray}
	M(d\xi_m) &=& 0 \;\; ,\nonumber\\
	M(Re(d\xi_m)Re(d\xi_n)) = M(Im(d\xi_m)Im(d\xi_n)) &=& \delta_{mn}dt 
	\;\; ,
	\nonumber\\
	M(Re(d\xi_m)Im(d\xi_n)) &=& 0 \; .
	\label{112}
\end{eqnarray}
Equation (\ref{111}) has to be interpreted as an Ito stochastic
differential equation, see , for example,
\cite{Stratonovitch1,Gardiner4}. It is easy to check 
that averaging Eq. (\ref{111}) over the stochastic Wiener process yields 
the density operator equation Eq. (\ref{110}) and that therefore (in the 
mean) normalization is preserved. For numerical studies often a 
somewhat simpler equation is used that does not 
preserve normalization even under the mean. This is given by
\begin{equation}
	|d\psi\rangle = -\,\frac{i}{\hbar} H |\psi\rangle dt + 
	\sum_m \left( 2\langle L_m^{\dagger}\rangle L_m - 
	L_m^{\dagger}L_m \right)
	|\psi\rangle dt + \sum_m L_m|\psi\rangle d\xi_m\; .
	\label{111a}
\end{equation}
It should be noted that Eq. (\ref{112}) is a nonlinear equation as it 
also depends on the expectation values of the Lindblad operators 
$L_m$. This makes the analytical treatment of this equation very
difficult and there are only a few cases for which an analytical
solution is known \cite{Gisin1,Gisin2,Salama1,Wiseman3,Carmichael4}. However 
it was found by Goetsch, Graham and Haake \cite{Goetsch1,Goetsch2,Goetsch3} 
that it is possible to find linear stochastic differential equations
which also reproduce the ensemble average. Stochastic differential
equations for the wavefunction have also been derived by Barchielli
\cite{Barchielli1,Barchielli2,Barchielli3} (see also \cite{Zoller2} for 
good summary of these approaches) from a more abstract mathematical point 
of view. The approach of Barchielli also gives a common mathematical 
basis for both diffusion and jump processes. 

\begin{figure}[hbt]
\epsfxsize14.cm
\centerline{\epsffile{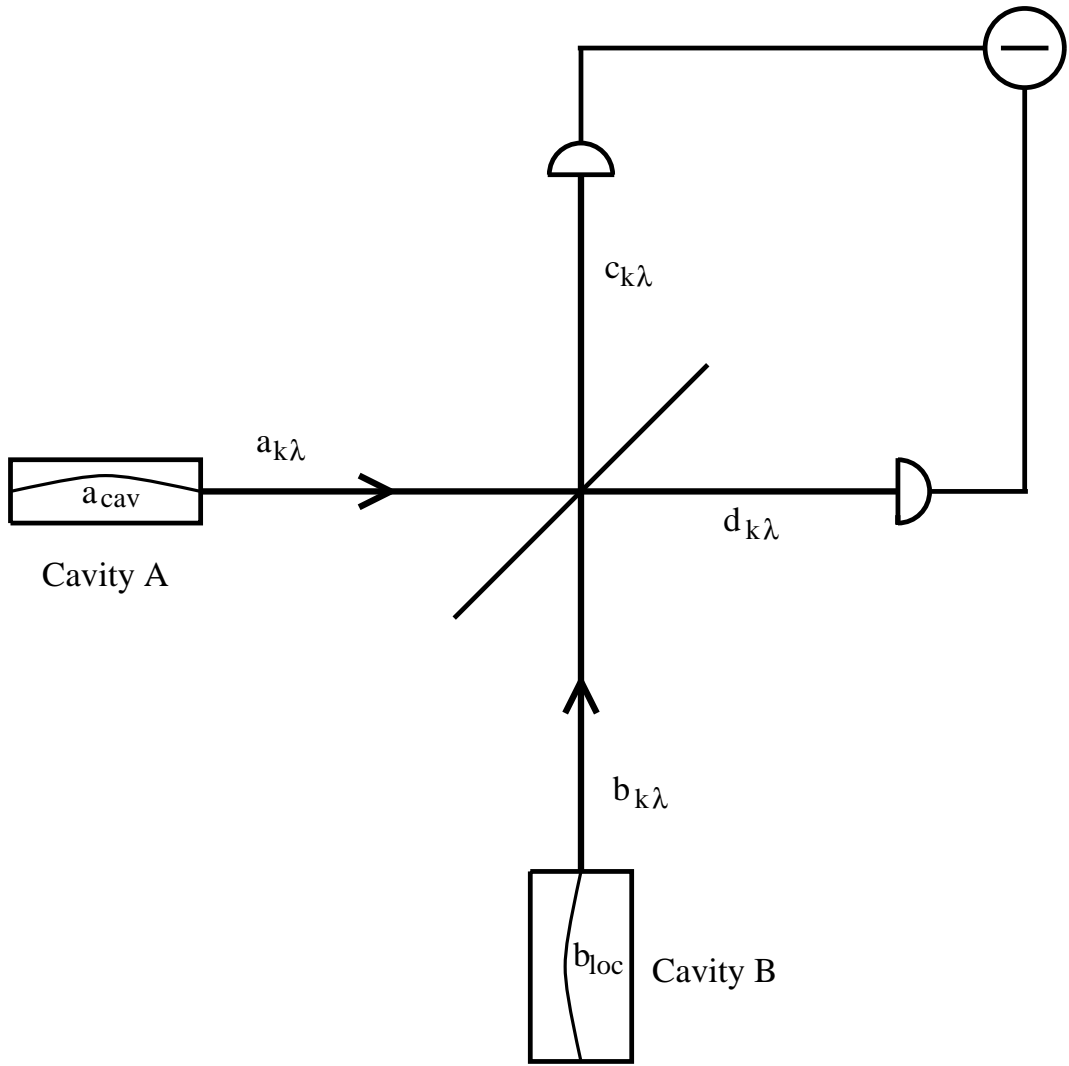}}
\vspace*{1.cm}
\caption{\label{fig3.1} Schematic picture of the heterodyne detection scheme.
Cavity A emits a weak signal that we mix with the signal from the local
oscillator cavity B. We measure the difference in the counts in the two
counters that detect the photons that leave the beamsplitter.}
\end{figure}

However, we do not intend to elaborate further on the mathematical side 
of the theory. Instead we would like to show that it is possible 
to derive QSD from the quantum jump approach in a certain limiting 
case, i.e., the case of infinitely many jumps where each jump has 
an infinitesimal impact on the wavefunction. In fact it turns out
that QSD can be related to an explicit and well known physical
measurement process in quantum optics, namely, the method of balanced 
heterodyne detection, see , for example, 
\cite{Castin92,Wiseman1,Wiseman2,Wiseman4,Carmichael2,Molmer2,Knight1}.
In the following we would like to show this explicitly for the 
specific example of a decaying cavity and we  follow a similar 
path to that used in the approach of Garraway and Knight
\cite{Castin92,Garraway3,Knight1}. To be specific we will illustrate 
the method for 
the case of balanced heterodyne detection of the output of an undriven 
optical cavity. We have in mind the situation given in Fig. \ref{fig3.1}. 

The left hand cavity A (with mode operators $a_{cav}$) is the source a weak 
output field (mode operators 
$a_{{\bf k}\lambda}$) which we want to analyze, while the lower cavity 
(mode operators $b_{loc}$) is assumed to be in a coherent state 
$|\beta\rangle$ with a very large amplitude $\beta$ for all times so 
that the radiated field (with mode
operators $b_{{\bf k}\lambda}$) of that cavity is very large. The 
Hamilton operator describing this situation is
\begin{eqnarray}
	H &=& \hbar\omega_{cav}a^{\dagger}_{cav}a_{cav} +
	    \sum_{{\bf k}\lambda} \hbar\omega_{{\bf k}\lambda} 
	    a_{{\bf k}\lambda}^{\dagger}a_{{\bf k}\lambda} +
	    \sum_{{\bf k}\lambda} \left\{ i\hbar g_{{\bf k}\lambda} 
	    a^{\dagger}_{cav} a_{{\bf k}\lambda}^{} + h.c.
	    \right\}
	    \nonumber\\
	  && + \hbar\omega_{loc}b^{\dagger}_{loc}b_{loc} +
	    \sum_{{\bf k}\lambda} \hbar\omega_{{\bf k}\lambda} 
	    b_{{\bf k}\lambda}^{\dagger}b_{{\bf k}\lambda} +
	    \sum_{{\bf k}\lambda} \left\{ i\hbar f_{{\bf k}\lambda} 
	    b^{\dagger}_{loc} b_{{\bf k}\lambda}^{} + h.c.
	    \right\}\; .
	    \label{113}
\end{eqnarray}
where the $g_{{\bf k}\lambda}$ and $f_{{\bf k}\lambda}$ are the coupling
constants between the cavity and the outside world and $\omega_{cav}$
and $\omega_{loc}$ the frequencies of the cavity A and the local oscillator
cavity respectively.
The action of the beamsplitter is to mix the two incoming modes.
Assuming a 50\% beamsplitter we find for the new mode operators
\cite{Loudon2}
\begin{eqnarray}
	c_{{\bf k}\lambda} &=& \frac{1}{\sqrt{2}} 
	( a_{{\bf k}\lambda} + b_{{\bf k}\lambda} ) \;\; ,
	\nonumber\\
	d_{{\bf k}\lambda} &=& \frac{1}{\sqrt{2}} 
	( -a_{{\bf k}\lambda} + b_{{\bf k}\lambda} )\; .
	\label{114}
\end{eqnarray}
Now going over to an interaction picture with respect to
\begin{equation}
	H_0 =  \hbar\omega_{cav}a^{\dagger}_{cav}a_{cav} +
	       \sum_{{\bf k}\lambda} \hbar\omega_{{\bf k}\lambda} 
	       a_{{\bf k}\lambda}^{\dagger}a_{{\bf k}\lambda} +
	       \hbar\omega_{loc}b^{\dagger}_{loc}b_{loc} +
	       \sum_{{\bf k}\lambda} \hbar\omega_{{\bf k}\lambda} 
	       b_{{\bf k}\lambda}^{\dagger}b_{{\bf k}\lambda}
	\label{115}
\end{equation}
and subsequently changing the basis via a displacement operator such
that the initial state of the local oscillator is the vacuum 
\cite{Mollow1,Pegg4} we obtain using $\Omega=\omega_{loc}-\omega_{cav}$
\begin{eqnarray}
	H &=& \sum_{{\bf k}\lambda} \frac{i\hbar}{\sqrt{2}} \left(
	g_{{\bf k}\lambda}a^{\dagger}_{cav} e^{-i\Omega t} + 
	f_{{\bf k}\lambda}(b^{\dagger}_{loc}+\beta^{*} )
	\right) c_{{\bf k}\lambda} 
	e^{-i(\omega_{{\bf k}\lambda}-\omega_{loc})t} + h.c.
	\nonumber\\
	&& + \sum_{{\bf k}\lambda} \frac{i\hbar}{\sqrt{2}} \left(
	- g_{{\bf k}\lambda}a^{\dagger}_{cav} e^{-i\Omega t} + 
	f_{{\bf k}\lambda}(b^{\dagger}_{loc}+\beta^{*} ) 
	\right) d_{{\bf k}\lambda} 
	e^{-i(\omega_{{\bf k}\lambda}-\omega_{loc})t} + h.c.
	\; .
	\label{116}
\end{eqnarray}
Now applying the methods that we used to derive the quantum jump approach, we 
easily obtain the two jump operators
\begin{eqnarray}
	J_c &=& \frac{1}{\sqrt{2}} \left( \sqrt{\gamma_{cav}}a_{cav} 
	e^{i \Omega t} + \sqrt{\gamma_{loc}}\beta \right)\;\; ,
	\nonumber\\
	J_d &=& \frac{1}{\sqrt{2}} \left( -\sqrt{\gamma_{cav}}a_{cav} 
	e^{i \Omega t} + \sqrt{\gamma_{loc}}\beta \right)\;\; ,
	\label{117}
\end{eqnarray}
where $\gamma_{cav}$ and $\gamma_{loc}$ are the decay rates of 
the cavity and the local oscillator. Within a short time interval
$\Delta t$, i.e., such that 
$(\omega_{cav}-\omega_{loc})\Delta t \ll 1$, we will count on 
average 
\begin{equation}
	\langle J^{\dagger}_c J_c\rangle = \frac{\beta^2 \gamma_{loc}}{2} 
	\left( 1 + \sqrt{\frac{4\gamma_{cav}}{\gamma_{loc}\beta^2}}
	\langle x_{\Omega t} \rangle \right)
	\label{118}
\end{equation}
counts in mode $c$ and 
\begin{equation}
	\langle J^{\dagger}_d J_d\rangle = \frac{\beta^2 \gamma_{loc}}{2} 
	\left( 1 - \sqrt{\frac{4\gamma_{cav}}{\gamma_{loc}\beta^2}}
	\langle x_{\Omega t} \rangle \right)\;\; ,
	\label{119}
\end{equation}
counts in mode $d$ where 
\begin{equation}
	\langle x_{\phi} \rangle := 
	\langle a^{\dagger}_{cav} e^{i\phi} + a_{cav} e^{-i\phi}
	\rangle    \; .
	\label{120}
\end{equation}
These are average values around which the actual number of counts 
in the two counters fluctuates. We can approximate this number of 
counts $m(t)$ by the stochastic process
\begin{equation}
	m(t) = \langle J^{\dagger} J \rangle \Delta t + 
	\langle J^{\dagger} J \rangle^{1/2} \Delta W
	\label{121}
\end{equation}
such that $\langle (\Delta W)^2 \rangle = \Delta t$. For the powers 
of the jump operators we find
\begin{eqnarray}
	J_c^{m_1} &=& \beta^{m_1}
	 \left(\frac{\gamma_{loc}}{2}\right)^{m_1/2} 
	\left( 1 +
	\frac{m_1}{\beta}\sqrt{\frac{\gamma_{cav}}{\gamma_{loc}}}
	a_{cav} e^{i\Omega t} \right)
	\nonumber\\
	J_d^{m_2} &=& \beta^{m_2}
	\left(\frac{\gamma_{loc}}{2}\right)^{m_2/2} 
	\left( 1 -
	\frac{m_2}{\beta}\sqrt{\frac{\gamma_{cav}}{\gamma_{loc}}}
	a_{cav} e^{i\Omega t} \right)\; .
	\label{122}
\end{eqnarray}
As we normalize after each emission the prefactors are not 
really important and we can divide the jump operators by these. 
One should note that the phase of these prefactors is fixed due to the 
fact that 
in the limit of infinite $\beta$ the jump operators have to become the 
unit operator. Therefore there is no freedom in the choice of the sign
of the prefactors. It is now easy to derive the effective Hamilton 
operator $H_{eff}$ for which we find
\begin{equation}
	H_{eff} = -i\hbar\frac{\gamma_{cav}}{2} 
	( a^{\dagger}a + \frac{\gamma_{loc}}{\gamma_{cav}}\beta^2 )\; .
	\label{123}
\end{equation}
Using this together with Eq. (\ref{122}) we obtain
\begin{equation}
	|\tilde{\psi}(t+\Delta t)\rangle = 
	(\id - \frac{i}{\hbar} H_{eff} \Delta t
	+ e^{-i \Omega t} a_{cav} ( 2\gamma_{cav}\langle x_{\Omega t} \rangle 
	+ \sqrt{\frac{\gamma_{cav}}{2}}(\Delta W_1 - \Delta W_2) )
	 |\tilde{\psi}(t)\rangle \;\; .
	\label{124}
\end{equation}
Adding together the two Wiener noises 
$\Delta W = (\Delta W_1 - \Delta W_2)/\sqrt{2}$,
taking the limits $\Delta t\rightarrow dt$ and 
$\Delta W \rightarrow dW$ and defining 
\begin{equation}
	d\xi = e^{-i\Omega t} dW \;\; ,
	\label{125}
\end{equation} 
we obtain after dropping a counter-rotating term of the form $e^{-2i\Omega t}$ 
\begin{equation}
	|d\tilde{\psi}\rangle = \left[ -\frac{\gamma_{cav}}{2}(a^{\dagger}_{cav}
	a_{cav} + \frac{\gamma_{loc}}{\gamma_{cav}}\beta^2) dt + 
	\gamma_{cav} a_{cav}\langle a^{\dagger}_{cav}\rangle dt
	+\sqrt{\gamma_{cav}} a_{cav} d\xi \right] |\tilde{\psi}\rangle
	\; .
	\label{126}
\end{equation}
This is the unnormalized diffusion equation given , for example, by 
Gisin and Percival \cite{Gisin3}. One should note that if we had
considered homodyne detection, i.e., the case $\Omega=0$, then we 
would have found a different diffusion equation, as there would not 
have been a counter-rotating term which we could have dropped. 
Therefore an additional term in Eq. (\ref{126}) would appear 
\cite{Carmichael2,Molmer2}.

To yield the normalized equations for QSD as they are given by
Gisin and Percival (1992a) we have to normalize the wavefunction 
{\em and} we have 
to include a stochastic phase factor $\alpha(t)$ into the 
wavefunction \cite{Garraway3}
, i.e., we look for a diffusion equation for
\begin{equation}
	|\psi(t)\rangle = \frac{ e^{i\alpha(t)} |\tilde{\psi}\rangle} 
	{\langle\tilde{\psi}|\tilde{\psi}\rangle} \;\; .
	\label{127}
\end{equation} 
The reason we have to include this seemingly unmotivated phase 
factor is that in the derivation of the QSD equation in 
\cite{Gisin3}, a term is added to the diffusion equation to give 
it the simplest possible form. This term in fact gives rise to a 
random phase change. To yield QSD we choose $\alpha(t)$ as
\begin{equation}
	\alpha(t) = \frac{i\gamma}{2}\langle a_{cav} \rangle d\xi -
	\frac{i\gamma}{2}
	\langle a^{\dagger}_{cav} \rangle d\xi^{*}\;\; .
	\label{128}
\end{equation}   
This choice has the effect of removing $d\xi^{*}$ that would 
appear in the diffusion equation of the normalized wavefunction 
without the additional phase factor. Using Eq. (\ref{128}) in 
Eq. (\ref{126}) and assuming $d\xi d\xi^{*}=dt$ we finally obtain
\begin{eqnarray}
	|d\psi\rangle &=& -\frac{\gamma}{2}a^{\dagger}_{cav}a_{cav} + 
	\gamma (a_{cav}\langle a^{\dagger}_{cav}\rangle + \frac{1}{2}
	\langle a^{\dagger}_{cav}a_{cav} \rangle - \frac{1}{2}
	\langle a^{\dagger}_{cav}\rangle\langle a_{cav}\rangle )
	|\psi\rangle dt
	\nonumber\\
	&& + \sqrt{\gamma} (a_{cav} - \langle a_{cav} \rangle ) d\xi\;\; .
	\label{129}
\end{eqnarray} 
We have therefore shown that the quantum state diffusion equation Eq. 
(\ref{129}) (or Eq. (\ref{111}) ) can be regarded as a limiting case of 
the quantum jump approach. In Section V we will illustrate the
transition from the quantum jump behaviour to the quantum state diffusion 
behaviour which takes place when we increase the amplitude of the local 
oscillator \cite{Granzow96}. It should be noted that it is also possible 
to obtain a jump--like behaviour from quantum 
state diffusion equations. However, this procedure is much less 
satisfying than the above derivation of quantum state diffusion from
quantum jumps. The reason is that one has to modify the quantum state 
diffusion equation by adding an additional operator, the localisation 
operator. The amplitude with which this localisation operator appears
in the equations is arbitrary and has to be adjusted according to the 
experimental situation \cite{Gisin9}. This is not particularly satisfying 
at least in cases in which we deal with single ion resonance fluorescence. 
Here the quantum jump approach appears to be much more natural. In fact it
can be shown that one can {\em not} associate the jumps occurring in the 
quantum state diffusion picture with photon emissions as such an 
interpretation can lead to more than one emission from an undriven 
two-level system \cite{Granzow96}.
Taking these considerations into account one could be tempted to say that 
the quantum jump approach is more fundamental then quantum state diffusion. 
However, both approaches have the same justification as they were both
derived from a particular measurement situation. Depending on the 
experimental situation and the measurement scheme employed we have to
choose either the quantum jump approach or the quantum state diffusion
model to obtain the correct description of the experimental situation.
Quantum state diffusion was, as mentioned before, not originally introduced 
to describe a specific experimental situation. It was rather seen as
an attempt to formulate alternative versions of quantum mechanics and there
are attempts to derive diffusion equations from fundamental ideas such as
, for example, decoherence induced by gravitational fluctuations 
\cite{Percival2,Percival3,Percival4}. Although the quantum state diffusion 
model can not be regarded as the proper description of quantum jumps in 
single photon counting experiments but rather as the 
description of heterodyne detection it is nevertheless useful in the 
investigation of single system behaviour. Interesting phenomena such as 
localisation in phase and position space
\cite{Gisin6,Gisin7,Herkommer96,Percival1}
are found. These can be used to improve the performance of simulation 
procedures using a ''moving basis'' approach \cite{Schack0,Schack1} where only
time dependent subset of all basis states is used in the simulation. A
similar method is also possible for a variant of the quantum jump approach
\cite{Holland2}.

So far, we have discussed the evolution of {\it open} systems,
that is of microsystems in contact with Markovian reservoirs such as
the bath of vacuum field modes responsible for spontaneous emission.
The quantum jump concept within an open system context has to do with
the gain in information about the microsystem which is accessible from
the record available in the dissipative environment. Such jump processes
do {\it not} require an extension or modification of conventional 
quantum mechanics, and we refer to these as ``extrinsic'' jumps.
A very different jump mechanism has been studied by a number of authors
\cite{Diosi3,Ghirardi1,Ghirardi2,Percival2,Percival3}. In these approaches 
the Schr\"odinger equation is modified in such a way that quantum coherences
are automatically destroyed in a {\it closed} system by an 
{\it intrinsic}  stochastic jump mechanism. This should be distinguished
from the extrinsic mechanisms we are concerned with in the bulk of this
review. 

To see how an intrinsic jump mechanism works, we need a concrete
realisation which we can apply to a specific time evolution. Milburn
has proposed just such a realisation \cite{Milburn3}, in which standard
quantum mechanics is modified in a simple way to generate intrinsic 
decoherence. He assumes that on sufficiently short time steps, the 
system does not evolve continuously under normal unitary evolution, but
rather in a {\it stochastic} sequence of identical unitary 
transformations. This assumption leads to a modification of the 
Schr\"odinger equation which contains a term responsible for the 
decay of quantum coherence in the energy eigenstate basis, without the
intervention of a reservoir and therefore without the usual energy 
dissipation associated with normal decay \cite{Moya-Cessa1}. The decay 
is entirely of phase-dependence only, akin to the dephasing decay of
coherences produced by impact-theory collisions or by fluctuations in the
phase of a laser in laser spectroscopy, but here of intrinsic origin.

It is interesting to apply Milburn's model of intrinsic decoherence to 
a problem of dynamical evolution: that is, the interaction of two
subsystems and the coherences which establish themselves as a consequence
of their interaction. In \cite{Moya-Cessa1} the interaction between a 
single two-level atom and a quantized cavity mode was considered and shown 
how the intrinsic decoherence affects the long--time coherence characteristics 
of the entangled atom--field system. In particular it could be shown how the
revivals (a signature of long--time coherence) are removed by this intrinsic
decoherence.
The quantum jump approach as we have discussed it so far only 
treats systems (atoms) interacting with a Markovian bath 
(the quantized multimode radiation field). However, one might 
be interested to apply the quantum jump approach to non 
Markovian interactions. Examples are electrons interacting 
with phonons or, in quantum optics, an atom in a cavity interacting 
with a mode which loses photons to the outside world
\cite{Garraway4}. The second example already suggests a possible
way one could model such systems. Here the atom sees a 
cavity mode with finite width, i.e., a spectral function which is 
not flat but a Lorentzian (Piraux {\em et al}, 1990 and references 
therein. However, one does not 
need to solve a non Markovian master equation, as the width of 
the mode is produced by its coupling to the outside world. Taking 
this coupling explicitly into account, by describing a coupled 
atom-cavity field mode with a dissipative field coupling to the
environment, one again obtains a 
Markovian master equation. This is also the recipe for the treatment 
of an interaction with a bath with a general spectral function 
$R(\omega)$ \cite{Imamoglu2}. One has to decompose $R(\omega)$ 
into a sum (or integral) of Lorentzians with positive weights. 
Each Lorentzian can then be modelled by a mode interacting with 
both the system and a Markovian reservoir. This method is 
practical only if the number of additional modes that one has to 
take into account is not too large. One should also note that 
in this case the meaning of a jump in the simulation can become 
obscure, as the excitation of the system is transferred to the 
Markovian bath in two steps via the additional mode \cite{Garraway4}.
However, if one is only interested in a simulation method to obtain 
the master equation for non Markovian interactions this is not
important.

We have discussed a number of derivations of the quantum jump approach 
so far. A different approach towards the description of single system
dynamics has been proposed in \cite{Teich1,Teich2}. In their method the 
dynamics described by the master equation is split into two distinct
parts. One part changes smoothly the instantaneous basis of the density 
operator (coherent evolution) while the other part causes jumps between
the basis states according to a rate equation. The instantaneous basis
can be viewed as a generalisation of the dressed state basis. For a
stationary state the basis states are fixed so that only jump 
processes occur. However, the approach is analytically quite complicated
for nonstationary processes and in addition there are interpretational
problems \cite{Wiseman1}. 

At this point we would like to explain briefly a recently proven 
connection \cite{Brun1,Yu1} between the quantum jump approach and 
a totally different concept, the Decoherent Histories formulation of quantum
mechanics. A similar connection, although mathematically 
more involved, between the quantum
state diffusion model and the Decoherent Histories approach has also been
established \cite{Diosi5}. The Decoherent Histories formulation of quantum 
mechanics was introduced by Griffiths, Omn\`es, and Gell-Mann and Hartle
\cite{Griffiths1,Omnes1,Omnes2,Omnes3,Gell-Mann1,Gell-Mann2}. In
this formalism, one describes a quantum system in terms of an exhaustive
set of possible histories, which must obey a {\it decoherence criterion}
which prevents them from interfering, so that these histories may be
assigned classical probabilities. 

In ordinary nonrelativistic quantum mechanics, a set of histories for a
system can be specified by choosing a sequence of times $t_1,\ldots,t_N$
and a complete set of projections $\{\pr^j_{\alpha_j}(t_j)\}$ at each time
$t_j$, which represent different exclusive possibilities, i.e., they obey
\begin{eqnarray}
	\sum_{\alpha_j} \pr^j_{\alpha_j}(t_j) &=& \id \;\; ,
	\label{dec1}\\ 
 	\pr^j_{\alpha_j}(t_j) \pr^j_{\alpha_j'}(t_j) &=&
 	\delta_{\alpha_j \alpha_j'} \pr^j_{\alpha_j}(t_j)\label{dec2}
	\;\; .
\end{eqnarray}
Note that the projection operators $\pr^j_{\alpha_j}$ are Heisenberg 
operators; one could represent them
in the Schr\"odinger picture by
\begin{equation}
\pr^j_{\alpha_j}(t_j) = e^{-i H t} \pr^j_{\alpha_j} e^{i H t}.
\label{schrodinger}
\end{equation}
The Schr{\"o}dinger picture projection operators are assumed to be 
operators in the system space.

A particular history is given by choosing one $\pr^j_{\alpha_j}$ at each point 
in time, specified by the sequence of indices $\{\alpha_j\}$, denoted
$\alpha$ for short.  The {\it decoherence functional} on a pair of 
histories $\alpha$ and $\alpha'$ is then given by
\begin{equation}
  D[\alpha,\alpha'] = Tr \biggl\{ \pr^N_{\alpha_N}(t_N) \cdots
  \pr^1_{\alpha_1}(t_1) \rho(t_0) \pr^1_{\alpha_1'}(t_1) \cdots
  \pr^N_{\alpha_N'} \biggr\},
\end{equation}
where $\rho(t_0)$ is the initial density matrix of the system.
The decoherence criterion is now given by this
decoherence functional $D[h,h']$. Two histories $h$ and $h'$ are 
said to {\it decohere} if they satisfy the relationship
\begin{equation}
	D[h,h'] = p(h) \delta_{hh'},
	\label{decoherence}
\end{equation}
where $p(h)$ is the probability of history $h$.  A set of 
histories $\{h\}$ is said to be exhaustive and decoherent if 
all pairs of histories satisfy the criterion Eq. (\ref{decoherence}) 
and the probabilities of all the histories sum to $1$.

To establish a connection between quantum jumps and Decoherent 
Histories the idea is to use a system that interacts with the outside
world in one direction. An example of such a system is a cavity.
The counter outside the cavity is now modelled by a two level system
that is strongly coupled to a bath so that both its coherence as well
as its excitation is damped much faster than all time constants of
the evolution of the system. One then defines the two projection 
operators 
\begin{equation}
	\pr_0 =\id \otimes |0\rangle\langle 0|
	\hspace{1.cm}
	\pr_1 = \id \otimes |1\rangle\langle 1| \;\; .\label{dec3}
\end{equation}
These projections model the presence or absence of photons outside
the system. It would be more general to consider more than one mode
of the radiation field and the the proof can be generalised to that case.
We now space these projections a short time $\delta t$ 
apart, and each history is composed of $N$ projections representing a 
total time $T=N \delta t$. A single history is a string 
$\{\alpha_1,\alpha_2,\ldots,\alpha_N\}$, where $\alpha_j=0,1$ represents
whether or not a photon has been emitted at time $t_j=(j-1)\delta t$.
Using this, it is possible to write  the decoherence functional as
\begin{equation}
	D[h,h'] = Tr\biggl\{ \pr_{\alpha_N} e^{{\cal L}\delta t}
	\bigl(\pr_{\alpha_{N-1}}e^{{\cal L}\delta t}\bigl(\ldots
	e^{{\cal L}\delta t}\bigl(\pr_{\alpha_{1}} |\psi\rangle\langle
	\psi|\pr_{\alpha'_{1}}\bigr)\ldots \bigr)\pr_{\alpha'_{N}}
	\biggr\} \;\; ,
\end{equation}
where ${\cal L}$ is the superoperator describing the time evolution
according to the Bloch equations for the system (cavity) coupled to 
the two-level system. It is now possible to show that the decoherence
functional in fact obeys Eq. (\ref{decoherence}) to a very good approximation.
It should be noted that the construction of the decoherent histories 
using the two operators in Eq. (\ref{dec3}) closely resembles the 
derivation of the quantum jump approach as given by
Hegerfeldt and Wilser (1991) and Wilser (1991).

The crucial point in the quantum jump approach is the fact that 
we assume that we perform time resolved measurements on the photons 
that are emitted by the atom. These photons may be mixed with a 
local oscillator in a heterodyne detection as we did for the 
derivation of QSD but even there we assume time resolved measurements. 
A nice feature of the quantum 
jump approach has been that it allows us to describe the radiating 
system by a wavefunction instead of a density matrix. However, one 
may ask the question whether this is the only possible way to reach 
a wavefunction description of radiating systems. In fact it turns 
out that in some sense there is a complementary way to the quantum 
jump approach that also yields a wavefunction description. This 
method was proposed in \cite{Holland1} and 
uses frequency resolved measurements instead of time resolved
measurements. It turns out that again it is possible to decompose
the density operator into pure states \cite{Mollow1} which, however, 
are now characterized by the number of photons that have been detected 
and for which the frequency instead of their precise emission time 
is known.
\subsection{Simulation of single trajectories}
After we have introduced and discussed different derivations of 
the quantum jump approach we will now briefly explain how the quantum
jump approach is used to simulate single quantum systems.
We will describe the simulation approach for a decaying undriven cavity.
The generalization to an arbitrary system should then be obvious.  
Carmichael (1993a) has given a precise relationship  between the
conditioned density operator contingent on a precise sequence of 
detection events (a ``record'') and the ensemble averaged density operator. 
He shows
, for example, that if the zero temperature boson damping master equation
is written in Liouvillian form
\begin{equation}
	\frac{d\rho}{dt} \; = \; {\cal L}\rho
	\label{610}
\end{equation}
and we split the Liouvillian action ${\cal L}$ as
a sum of two terms, an anticommutator and a ``sandwich'' term, 
\begin{eqnarray}
	{\cal L}\rho \;&=& \; -\frac{\gamma}{2}
	\left[\hat a^\dagger\hat a\;,\;\rho\right]_+ \;+\; 
	\gamma\hat a \rho \hat a^\dagger \nonumber\\
	&=&\; ({\cal L-S}) + {\cal S} \;\; ,
	\label{620}
\end{eqnarray}
then we may identify the ``sandwich'' term $\cal S$ as a jump operator.
Equation (\ref{610}) can be integrated formally as
\begin{eqnarray}
	\rho(t) \;&=&\; \exp\left\{\left[({\cal L-S})+
	{\cal S}\right]t\right\}\rho(0)
	\nonumber\\
	&=& \sum_{m=0}^\infty \int_0^t d t_m \int_0^{t_m} 
	d t_{m-1} \; ... \;\int_0^{t_2} d t_1
    	\nonumber\\
	&&\times \left\{
	e^{ ({\cal L-S})(t-t_m)}  {\cal S} 
	e^{ ({\cal L-S})(t_m-t_{m-1})}  {\cal S} 
	...  {\cal S} 
	e^{ ({\cal L-S})t_1}  \rho(0) \right\} \;\; ,
	\label{630}
\end{eqnarray}
where the quantity in curly brackets in \eqn{630} is labelled
$\overline{\rho_c}(t)$ by Carmichael and is the conditioned density operator
describing a specific ``trajectory'' or detection sequence.
We can write $\overline{\rho_c}(t)$ in terms of the conditioned pure state
projectors
\begin{equation}
	\overline{\rho_c}(t) = | \overline{\Psi_c}(t) \rangle\langle
 	\overline{\Psi_c}(t)| \;.
	\label{640}
\end{equation}
The component $\exp[({\cal L-S})\Delta t]$ propagates $\overline{\rho_c}(t)$
for a time $\Delta t$ without a decay being recorded: for the conditioned
state $|\overline{\Psi_c}(t)\rangle$
\begin{equation}
	|\overline{\Psi_c}(t+\Delta t)\rangle \;=\;
	\exp\left[ -i \heff \Delta t/\hbar\right]\;|\overline{\Psi_c}(t)\rangle
	\;\; ,
	\label{650}
\end{equation}
where the non-Hermitian effective Hamiltonian
\begin{equation}
	H_{eff} = H \;-\; i \hbar \frac{\gamma}{2} \hat a^{\dagger} \hat a
	\label{660}
\end{equation}
derives from the anticommutator in \eqn{620}. Once a decay is registered,
the gain in information is responsible for the jump
\begin{equation}
	|\overline{\Psi_c}(t)\rangle \longrightarrow \hat a
 	\;|\overline{\Psi_c}(t)\rangle
	\;.
	\label{670}
\end{equation}
The procedure adopted in quantum jump simulations can then be summarised 
as follows \cite{Dalibard1,Dum1}:
\begin{enumerate}
 \item
 Determine the current probability of an emission:
 \begin{equation}
  	\Delta P = \gamma \, \Delta t\, \langle \Psi |
  	{\hat a}^\dagger \hat a | \Psi \rangle \,.
 	 \label{680}
 \end{equation}
 \item
 Obtain a random number $r$ between zero and one, compare with 
  $\Delta P$ and decide on emission as follows:
 \item
 Emit if $r < \Delta P,$ so that the system jumps to the 
 renormalised form:
 \begin{equation}
  	| \Psi \rangle \longrightarrow { \hat a | \Psi \rangle
  	\over \sqrt{ \langle \Psi | {\hat a}^\dagger \hat a 
 	| \Psi \rangle} }\,.
 	\label{690}
 \end{equation}
 \item
 Or no emission if $r > \Delta P,$ so the system evolves under 
 the influence of the non-Hermitian form
 \begin{equation}
  	| \Psi \rangle \longrightarrow { \{1 - (i/\hbar) \, H \Delta t -
 	(\gamma/2)\, \Delta t\,{\hat a}^\dagger \hat a \} | \Psi \rangle
 	 \over (1 - \Delta P)^{1 \over 2} }\,.
 	 \label{700}
 \end{equation}
 \item
 Repeat to obtain an individual trajectory, or history.
 \item
 Average observables over many such trajectories.
\end{enumerate}
To reassure ourselves that this is all true, we note the history 
splits into two alternatives in a time $\Delta t:$
\begin{equation}
	|\Psi\rangle = \left\{ \begin{array}{cc}
	|\Psi_{emit}\rangle         & \mbox{with probability} \, \Delta P\\
	|\Psi_{no emission}\rangle  & \mbox{with probability} \, 1 - \Delta P
	\end{array} \right.
\end{equation}
Then in terms of the density matrix, the evolution for a step
$\Delta t$ becomes a sum of the two possible outcomes,
\begin{eqnarray}
 | \Psi \rangle \langle \Psi | 
 &\longrightarrow& \Delta P\, | \Psi_{\rm emit} \rangle \langle
 \Psi_{\rm emit} | 
 \nonumber \\
 & &\ +\ (1 - \Delta P) \, | \Psi_{\rm no\ emit} \rangle \langle
 \Psi_{\rm no\ emit} |
 \label{710} \\[2ex]
 &=& \gamma \Delta t\, \hat a | \Psi \rangle \langle \Psi |
 {\hat a}^\dagger
 \nonumber \\
 & &\ +\  \{1 - {i \over \hbar} \, H \Delta t -
 {\gamma \over 2}\, \Delta t\,{\hat a}^\dagger \hat a \} | \Psi \rangle
 \langle \Psi | \{1 + {i \over \hbar} \, H \Delta t -
  {\gamma \over 2}\, \Delta t\,{\hat a}^\dagger \hat a \}
 \nonumber \\[2ex]
 &\sim& | \Psi \rangle \langle \Psi | 
 -{i \over \hbar} \, \Delta t\,[H,| \Psi \rangle \langle \Psi |]
 \nonumber \\
 & &\ +\ {\gamma \over 2}\, \Delta t \,
 \{2\, \hat a | \Psi \rangle \langle \Psi | {\hat a}^\dagger -
 {\hat a}^\dagger \hat a | \Psi \rangle \langle \Psi |  -
 | \Psi \rangle \langle \Psi | {\hat a}^\dagger \hat a \}
 \label{720}
\end{eqnarray}
so that
\begin{equation}
 {\Delta \rho \over \Delta t} = -{i \over \hbar}\,[H,\rho] +
 {\gamma \over 2} \,\{ 2\, \hat a \rho {\hat a}^\dagger -
 {\hat a}^\dagger \hat a \rho - \rho {\hat a}^\dagger \hat a\}
 \label{730}
\end{equation}
as in the original master equation Eq. (\ref{610}). We have now seen how the 
quantum jump approach can be used to simulate a master equation. In Section 
\ref{V} we will see some examples of such simulations and also of single
realizations of quantum trajectories. After this simple approach to simulating
the master equation we now give a brief exposition of the idea of higher 
order unravellings 
of the master-equation \cite{Steinbach1}. To see the motivation we have 
to realize that the quantum jump approach is based on the simulation of 
the conditioned evolution of either a density operator or a state vector.
However, at one point it is not a rigorous
implementation of the trajectory concept. Because this method discretises
time into small steps $\delta t\,,$ a quantum jump in
the simulation takes a finite time $\delta t\,,$ whereas in a simulation 
of quantum trajectories the information gained 
from detection should instantaneously be used in conditioning
the quantum state of the system. This pinpoints the subtle difference
between conditioned trajectories and the slightly simpler idea of
evaluating the probability of decay quanta at discrete timesteps.
The simplest way to remedy the fact that conditioning takes time 
in the simulation is to add evolution with the effective Hamiltonian to the
projection step that has to be performed when a photon is detected. Having 
said this, the question arises
at what point during the time interval $\delta t$ we need to condition
the quantum state according to the result of the detection process.
First, it is worth noting that wherever we decide to do this, it would
not change the accuracy of integrating the master equation in first order.
Second, we may try to increase the accuracy by choosing
a specific point in the interval $\delta t.$ Let us integrate the
master equation to second order in $\delta t\,:$
\begin{equation}
 \rho_S (t+\delta t) = \rho_S (t) + {1 \over 2}\, \delta t\,
 (\,[{\cal L}\rho_S]_t + 
 [{\cal L}\rho_S]_{t+ \delta t}\,) +
 O({\delta t}^3)\,.
 \label{550}
\end{equation}
Here ${\cal L}$ is the Liouville operator describing the evolution of 
the complete master equation. The terms that result from evaluating 
the right hand side of this equation can be cast into the following 
form (for details see \cite{Steinbach1}). By $C$ we denote the reset 
operator that has to be applied after the detection of a photon.

$$
   \begin{array}{crlr@{\rho_S (t)}l}
     \displaystyle \rho_S (t+\delta t) =
     &
     &
     & \displaystyle U 
     & \displaystyle U^{\dagger }
     \\[2ex]
     & +
     & {1 \over 2} \delta t
     & U C 
     & C^{\dagger} U^{\dagger}
     \\[1ex]
     & +
     & {1 \over 2} \delta t
     & C U
     & U^{\dagger} C^{\dagger}
     \\[2ex]
     & +
     & {1 \over 2} {\delta t}^2
     & U C C
     & C^{\dagger} C^{\dagger} U^{\dagger} \ + \ O({\delta t}^3)\,.
   \end{array}
$$
\begin{equation}
 \label{560}
\end{equation}
Here $U$ denotes evolution under the influence of the effective Hamiltonian
\begin{equation}
 U = \exp{(-{i \over \hbar}\, H_{\rm eff} \,\delta t)}\,,
 \label{565}
\end{equation}
which we call the ``no-jump'' evolution.
The four terms on the right hand side of Eq. (\ref{560}) represent
four specific conditioned evolutions or {\it mini-trajectories} 
that the system might follow. An expansion into mini-trajectories is 
important because only then can the density matrix evolution Eq. (\ref{560})
be simulated with pure states.
The first mini-trajectory in Eq. (\ref{560}) represents evolution without
any jump, the second and the third represent a jump followed by
evolution without jumps and vice versa respectively and the fourth includes
two successive jumps followed by no-jump evolution. 

\noindent
We see that it is not sufficient to specify one point in the interval
$\delta t$ at which to condition the density operator due to the quantum
jump. We have to consider two points, at the beginning and at the
end of $\delta t,$ and also the possibility of two immediately
successive quantum jumps in order to increase the accuracy in
$\delta t$ by one order.

\noindent
One can pursue this idea to obtain results which are accurate up to
fourth order \mbox{(in $\delta t$).}
The master equation has to be integrated along the lines of a
fourth-order Runge-Kutta method for ordinary differential equations.
The result in fourth order then contains thirteen mini-trajectories
(including the no-jump evolution) as follows

$$
\begin{array}{crlr@{\rho_S (t)}lrlr@{\rho_S (t)}l}
 \rho_S (t+\delta t) =
 &
 &
 &
 \multicolumn{6}{l}{U_1 \rho_S (t) U^{\dagger}_1}
 \\[2ex]
 & +
 & {1 \over 8} \delta t
 & \multicolumn{2}{l}{U_1 C \rho_S (t) C^{\dagger} U^{\dagger}_1}
 & +
 & {1 \over 8} \delta t
 & \multicolumn{2}{l}{C U_1 \rho_S (t) U^{\dagger}_1 C^{\dagger}}
 \\[1ex]
 & +
 & {3 \over 8} \delta t
 & \multicolumn{2}{l}{U_{1 \over 3} C U_{2 \over 3} \rho_S (t)
   U^\dagger_{2 \over 3} C^\dagger U^\dagger_{1 \over 3}}
 & +
 & {3 \over 8} \delta t
 & \multicolumn{2}{l}{U_{2 \over 3} C U_{1 \over 3} \rho_S (t)
   U^\dagger_{1 \over 3} C^\dagger U^\dagger_{2 \over 3}}
 \\[2ex]
 & +
 & {1 \over 6} {\delta t}^2
 & \multicolumn{6}{l}
 {U_{1 \over 2} C U_{1 \over 2} C \rho_S (t) 
 C^\dagger U^\dagger_{1 \over 2} C^\dagger U^\dagger_{1 \over 2}}  
 \\[1ex]
 & +
 & {1 \over 6} {\delta t}^2
 & \multicolumn{6}{l}
 {C U_{1 \over 2} C U_{1 \over 2} \rho_S (t) 
 U^\dagger_{1 \over 2} C^\dagger
 U^\dagger_{1 \over 2} C^\dagger}
 \\[1ex] 
 & +
 & {1 \over 6} {\delta t}^2
 & \multicolumn{6}{l}
 {U_{1 \over 2} C C U_{1 \over 2} \rho_S (t) 
 U^\dagger_{1 \over 2} C^\dagger C^\dagger U^\dagger_{1 \over 2}}
 \\[2ex] 
 & +
 & {1 \over 24} {\delta t}^3
 & U_1 C C C
 & C^\dagger C^\dagger C^\dagger U^\dagger_1
 & +
 & {1 \over 24} {\delta t}^3
 & C U_1 C C
 & C^\dagger C^\dagger U^\dagger_1 C^\dagger
 \\[1ex]
 & +
 & {1 \over 24} {\delta t}^3
 & C C U_1 C
 & C^\dagger U^\dagger_1 C^\dagger C^\dagger
 & +
 & {1 \over 24} {\delta t}^3
 & C C C U_1
 & U^\dagger_1 C^\dagger C^\dagger C^\dagger
 \\[2ex]
 & +
 & {1 \over 24} {\delta t}^4
 & \multicolumn{6}{l}{
 U_1 C C C C \rho_S (t) C^\dagger C^\dagger C^\dagger
 C^\dagger U^\dagger_1 \ \ + \ O({\delta t}^5)\,.} 
 \end{array}
$$ 
\begin{equation}
 \label{570}
\end{equation}
\noindent
The subscripts on the non-Hermitian evolution $U$ indicate the fraction
of the time interval $\delta t$ for which each particular $U$ evolves the 
density operator, e.g., $U_{1/3} = \exp{(-i\,H_{\rm eff}\,
\delta t/3 \hbar )}.$ The way in which Eqs. (\ref{560}) and (\ref{570}) are
turned into a Monte-Carlo simulation is clear: each mini-trajectory
defines the conditioned evolution of the system and is assigned a 
specific probability with which it occurs, analogous to the jump
and no-jump probabilities in the standard method.
Just as in the standard procedure, a random number uniformly 
distributed between $0$ and $1$ is drawn to choose at random which
of the mini-trajectories will govern the system evolution in the
next timestep $\delta t.$ The no-jump evolution is tested first as this,
for small $\delta t,$ is the most likely mini-trajectory.
We note that the probability for evolution without detection remains
unchanged as compared with the standard method and because the no-jump 
evolution is most likely the diversity of the mini-trajectories hardly
influences the necessary computing time. However, if the no-jump
mini-trajectory is not selected then one of the alternative trajectories
in Eq. (\ref{570}) must be chosen. For example, if the normalized state
of the system at time $t$ is $|\Psi (t)\rangle$ then the state of the
system after evolution corresponding to the fourth mini-trajectory in
Eq. (\ref{570}) is,
\begin{equation}
 |\Psi (t+\delta t)\rangle = {1 \over {\cal N}}\,
 e^{-i\, H_{\rm eff} \,
 \delta t/3 \hbar } \ C \ e^{-i\, H_{\rm eff} \,
 2 \delta t/3 \hbar } \ |\Psi (t) \rangle\,,
 \label{75}
\end{equation}
which includes a renormalization factor ${\cal N.}$ The state $|\Psi \rangle$
is evolved with the effective Hamiltonian over two thirds of the timestep
$\delta t$ using a fourth-order Runge-Kutta integration step
\cite{Molmer1}. After projection with the Lindblad operator $C$ the
evolution is continued with the non-Hermitian Hamiltonian for the remaining
third of the timestep $\delta t.$ Only then is the resulting state vector
renormalized. The probability for this mini-trajectory to occur is given
by the product of the factor $3 \delta t/8$ and the renormalization
${\cal N}.$

To illustrate the improvement the method of higher order unravellings 
presents \cite{Steinbach1} one can simulate a laser driven two-level 
system using the
ordinary first-order (fourth order integration of the effective time 
evolution operator) and compare the result to the same simulation 
using the method of higher order unravellings. In Fig. \ref{fig4.0} the 
simulation
results of the inversion 
$\langle \sigma_3\rangle=\langle |1\rangle\langle 1| - 
|0\rangle\langle 0|\rangle/2$ of the two-level system
are plotted after $250000$ runs and for a Rabi frequency $\Omega=A$ equal 
to the Einstein coefficient
of the transition, zero detuning and a time step $\delta t = 0.1 A^{-1}$.
We clearly see that the first-order quantum jump approach (dashed line)
deviates from the exact result (solid line) while the dotted line obtained
from a fourth-order unravelling is much closer to the exact result. 
\subsection{A quantum system driven by another quantum system}
\begin{figure}[hbt]
\epsfxsize14.cm
\centerline{\epsffile{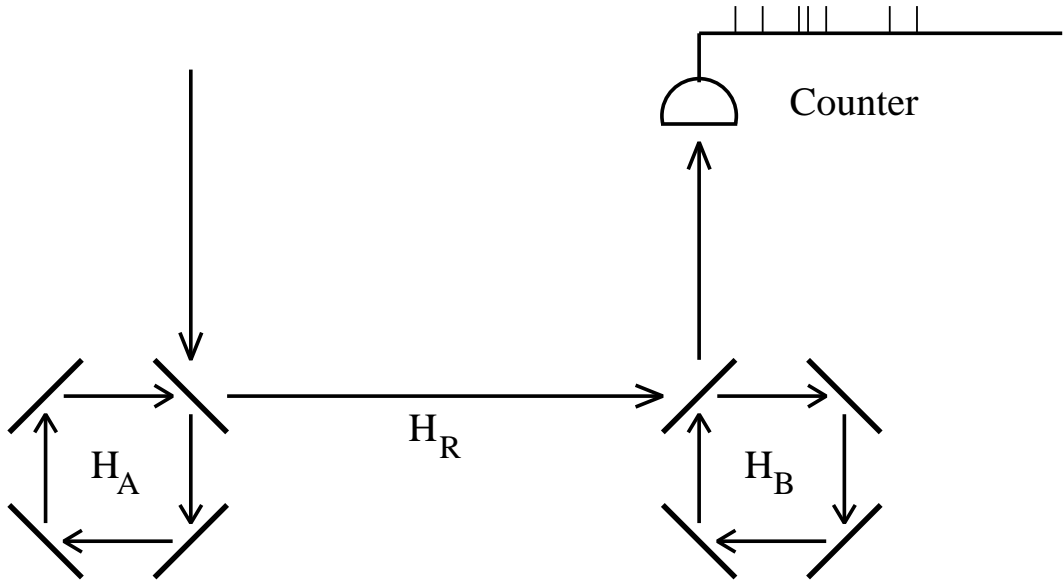}}
\vspace*{1.cm}
\caption{\label{fig4.1a} Schematic picture of the experimental situation
envisaged in the theory of ''cascaded'' quantum systems. 
System $B$ is driven by the quantum system $A$. The counter registers a
superposition of both fields, one emitted from system $A$ and the other
emitted from system $B$. After Carmichael (1993b.)}
\end{figure}

\begin{figure}[hbt]
\epsfxsize14.cm
\centerline{\epsffile{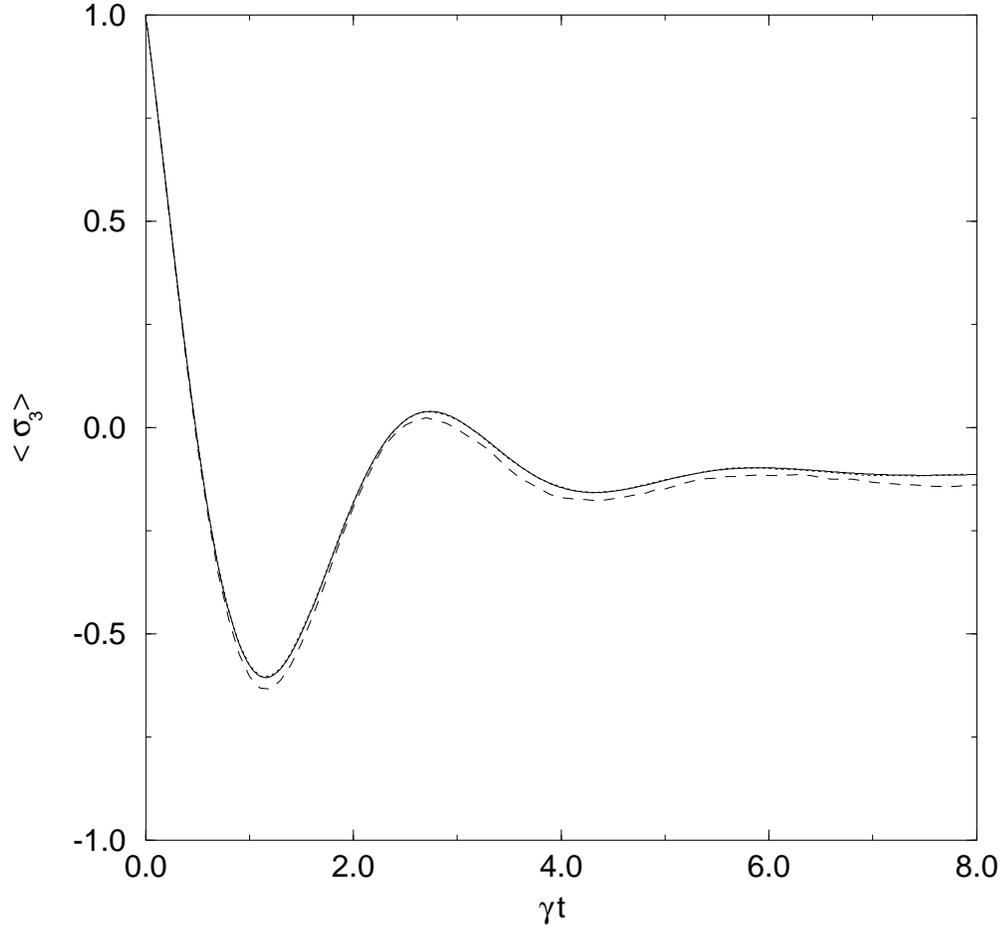}}
\caption{\label{fig4.0} Ensemble averaged time evolution for the
expectation value $\langle \sigma_3 \rangle = 
\langle |1\rangle\langle 1| - |0\rangle\langle 0| \rangle/2$
(inversion in the two-level system). 
The dotted line shows a sample of $250000$ trajectories obtained
by the fourth-order Monte-Carlo method ($\Omega= A,\,
\gamma \delta t = 0.1 A^{-1}$ and zero detuning). It is hard to distinguish
the dotted line from the solid line showing the analytical result. 
The dashed line shows a sample of $250000$ trajectories obtained
by the first-order Monte-Carlo method for the same parameters. From
\protect\cite{Steinbach1}.}
\end{figure}

The next problem we want to investigate is that of a quantum system $B$
driven by the radiation emitted from another quantum system $A$
\cite{Gardiner2,Carmichael3}. In the following we closely follow
\cite{Carmichael3}. One could
try to solve the problem by determining the dynamics of the driving
system $A$ first and from that the statistics of the emitted light. However,
in general an infinite number of correlation functions is required to 
characterise the state of the light emitted from $A$. In semiclassical 
theory one could instead simulate the properties of the light by 
implementing a suitable stochastic process, unfortunately this is not
possible in the quantum case. Therefore it is better not to divide the 
problem into two but to determine the dynamics of the composite system
$A\oplus B$. To obtain the broken time symmetry one uses an interaction
between A and B that is mediated by a reservoir R and one assumes the
Born-Markoff approximation. A simplified version of the problem is
illustrated in Fig. \ref{fig4.1a} where it is assumed that only
one mode of each cavity needs to be assumed. The cavities have three
perfectly reflecting mirrors and one with transmission coefficient $T\ll 1$.
The Hamiltonians $H_A$ and $H_B$ describe the free cavity modes and any 
interactions that take place inside the cavities. $H_R$ is the free 
Hamiltonian of a travelling wave reservoir R which couples the cavities 
in one direction only. The fields ${\cal E}(0)$ and ${\cal E}(l)$ that 
couple to the cavities are written in photon flux units. The complete
Hamiltonian for $A\oplus B \oplus R$ is
\begin{equation}
	H = H_A + H_B + H_R + H_{AR} + H_{BR}
	\label{5.1a}
\end{equation}
with 
\begin{eqnarray}
	H_{AR} &=& i\hbar \left( 2\kappa_A \right)^{1/2} \left[
	a {\cal E}^{\dagger}(0) - {\cal E}(0) a^{\dagger} \right]\;\; ,
	\nonumber\\
	H_{BR} &=& i\hbar \left( 2\kappa_B \right)^{1/2} \left[
	b {\cal E}^{\dagger}(l) - {\cal E}(l) b^{\dagger} \right]\;\; ,
	\label{5.2a}
\end{eqnarray}
where $\kappa_A$ and $\kappa_B$ are the cavity linewidths, and $a$ and $b$ 
are annihilation operators for the cavity modes. $H$ describes two systems 
interacting with the same reservoir. It should be noted that $A$ and $B$ 
couple to that reservoir at different positions in space. Usually spatially 
separated reservoir fields are treated as independent an assumption that 
cannot be valid for the geometry shown in Fig. \ref{fig4.1a} where the output 
from cavity $A$ appears a time $\tau=l/c$ later at the input of cavity $B$.
The spatial separation of the two cavities can in fact be eliminated using the 
Born-Markoff approximation in the Heisenberg picture to relate 
${\cal E}^{\dagger}(0)$ and ${\cal E}^{\dagger}(l)$. One obtains
\begin{equation}
	U_A(\tau){\cal E}^{\dagger}(l)U_A^{\dagger}(\tau) = 
	{\cal E}^{\dagger}(0) + \frac{1}{2} \left( 2\kappa_A\right)^{1/2} a
	\;\; ,
	\label{5.3a}
\end{equation}
where
\begin{equation}
	U_A(\tau) = e^{i\,(H_A + H_R + H_{AR})\tau/\hbar}\;\; .
	\label{5.4a1}
\end{equation}
If $\chi(t)$ is the density operator of system $A\oplus B\oplus R$ we may 
define the retarded density operator of the system $A\oplus B\oplus R$ as
\begin{equation}
	\chi'(t) = U_A(\tau) \chi(t) U_A^{\dagger}(\tau)
	\label{5.5a}
\end{equation}
and easily shows that $\chi'(t)$ satisfies the Liouville equation with the 
Hamiltonian
\begin{equation}
	H' = H_S + H_R + H_{SR},
	\label{5.6a}
\end{equation}
where
\begin{eqnarray}
	H_S &=& H_A +H_B + i\hbar \left( \kappa_A\kappa_B \right)^{1/2}
	\left( a^{\dagger} b - a b^{\dagger} \right) \;\; , 
	\nonumber\\
	H_{SR} &=& i\hbar \left\{ 
	\left[ (2\kappa_A)^{1/2} a + (2\kappa_B)^{1/2} b \right] 
	{\cal E}^{\dagger}(0) - H.c. \right\}\;\; .
	\label{5.7a}
\end{eqnarray}
Now $a$ and $B$ couple to the reservoir at the same spatial location and 
in addition they also couple directly with coupling constant
$\left(\kappa_A\kappa_B\right)^{1/2}$. Now one can easily derive the master
equation for the density operator of $A\oplus B$ $\rho'=tr_R\{\chi'\}$
and obtains
\begin{equation}
	\dot{\rho}' = \frac{1}{i\hbar} \left[ H_S,\rho'\right]
	+ C\rho' C^{\dagger} - \frac{1}{2} C^{\dagger}C \rho'
	- \frac{1}{2} \rho' C^{\dagger}C,
	\label{5.8a}
\end{equation}
with
\begin{equation}
	C =\left(2\kappa_A\right)^{1/2} a + \left(2\kappa_B\right)^{1/2} b\;\; .
	\label{5.9a}
\end{equation}
Having found the master equation one can easily find the conditional time
evolution that then allows to unravel the dynamics of the composite
system $A\oplus B$. The time evolution of the conditional wave function
$|\psi_c(t)\rangle$ between photodetections is governed by the Hamilton 
operator 
\begin{equation}
	H = H_A + H_B -i\hbar \left[ \kappa_A a^{\dagger} a +
	\kappa_B b^{\dagger} b + 
	2\left(\kappa_A\kappa_B\right)^{1/2} a b^{\dagger} \right]\;\; .
	\label{5.10a}
\end{equation}
After a photodetection we have to reset the system using the operator $C$
, i.e., $|\psi_C(t)\rangle \rightarrow C|\psi_C(t)\rangle$. Now we are 
in a position to simulate individual trajectories of two coupled quantum
systems. For applications of the theory , for example, to an atom driven by
squeezed light or by antibunched light emitted from another atom see 
\cite{Carmichael3,Gardiner2,Kochan1,Gardiner3}. An early example of an
investigation of atoms driven by antibunched light was discussed by
Knight and Pegg (1982). 
\subsection{Spectral information and correlation functions}
Until now we have discussed the quantum jump approach only in connection
with quantities of the system or its resonance fluorescence that require a 
temporal resolution; no frequency information has been obtained as only 
broadband photon counting has been assumed. However, it would be nice to
be able to use the quantum jump approach also for spectral properties of 
the resonance fluorescence of the system. Of special interest are , for 
example, 
the power spectrum of resonance fluorescence or the absorption spectrum 
of a weak probe laser. It turns out, that it is in fact possible to use
the quantum jump approach
\cite{Gardiner1,Dum2,Molmer1,Mu1,Plenio1,Plenio2,Hegerfeldt8} and also 
quantum state diffusion \cite{Gisin8,Sondermann1,Brun2,Schack1}, to calculate 
those spectra. There are several different ways to derive quantum jump 
equations that enable us to calculate spectral information and in the 
following we will discuss three of them. 

One possible approach can be made via the method of quantum stochastic 
differential equations (QSDE) \cite{Gardiner1,Dum2}. We already outlined 
the spirit of their approach in Eqs. (\ref{96}) - (\ref{109}). Following
essentially Gardiner {\em et al} (1992) but using the notation used in 
Eqs. (\ref{96}) - (\ref{109}) one defines an output mode operator
\begin{equation}
	dM(t) = \int_{t}^{t+dt} ds \, {\bf D_{21}}
	\sum_{{\bf k}\lambda} i\hbar 
	\left(\frac{e^2 \omega_{{\bf k}\lambda})}
	{2\epsilon_0\hbar V}\right)^{(1/2)}
	\biggreec{\epsilon}_{{\bf k}\lambda} \sigma_{21} 
	a_{{\bf k}\lambda} 
	e^{-i(\omega_{{\bf k}\lambda}-\omega_{21})s}
	\label{5.20a}
\end{equation}
and then the spectrum in the Schr{\"o}dinger picture as
\begin{equation}
	S(\omega) = \lim_{t\rightarrow\infty} 
	\frac{\langle \phi(t) | r^{\dagger}(\omega,t) r(\omega,t) 
	|\phi(t)\rangle}
	{t-t_0}\;\; ,
	\label{5.21a}
\end{equation}
where we defined
\begin{equation}
	r(\omega,t) = \int_{t_0}^{t} dM(s)\, e^{-i\omega (t-s)} \;\; .
	\label{5.22a}
\end{equation}
We can now introduce the auxiliary wave function
\begin{equation}
	|\beta(t)\rangle = r(\omega,t) |\phi(t)\rangle\;\; .
	\label{5.23a}
\end{equation}
For these two wavefunctions one then obtains a Stratonovitch stochastic 
differential equation which can then be transformed into the Ito form. 
One then obtains
\begin{equation}
	d\left(\begin{array}{c}
	|\phi(t)\rangle \\
	|\beta(t)\rangle
	\end{array}\right)
	=
	\left( \begin{array}{cc}
	-i\,H_{eff}/\hbar    &                0               \\
	\sqrt{A} \sigma_{12} & -i\,H_{eff}/\hbar -i\,\omega
	\end{array}\right)
	\left(\begin{array}{c}
	|\phi(t)\rangle \\
	|\beta(t)\rangle
	\end{array}\right)\;\; .
	\label{5.24a}
\end{equation}  
The spectrum is now by averaging 
\begin{equation}
	S(\omega) = \lim_{t\rightarrow\infty} 
	\frac{\langle \beta(t) |\beta(t)\rangle}
	{t-t_0}
	\label{5.25a}
\end{equation}
over many realization. Note that Eq. (\ref{5.24a}) for $|\phi(t)\rangle$
is just the conditional time evolution when no photons have been emitted.
This equation is used to determine the jump times. The equation for 
$|\beta (t)\rangle$ has a free evolution similar to $|\phi(t)\rangle$
except for an additional rotation with frequency $\omega$. In addition
it is driven by $|\phi(t)\rangle$. This driving can be interpreted as a 
process where a photon is emitted into the modes described by $dM(t)$ 
and where therefore the atom is deexcited. The approach given here can 
be used to simulate spectra in a number of situations 
\cite{Dum1,Marte1,Marte2}. This approach is in fact similar in spirit
to the approach of Schack {\em et al} (1996) who also assume the interaction 
of the system with one mode of the quantized radiation field. Schack 
{\em et al} then derive the quantum state diffusion equations that 
allow the calculation of the spectrum. It should be noted that using
Eq. (\ref{5.24a}) and then following our derivation of quantum state 
diffusion from quantum jumps would have led to the same result.

\begin{figure}[hbt]
\epsfxsize14.cm
\centerline{\epsffile{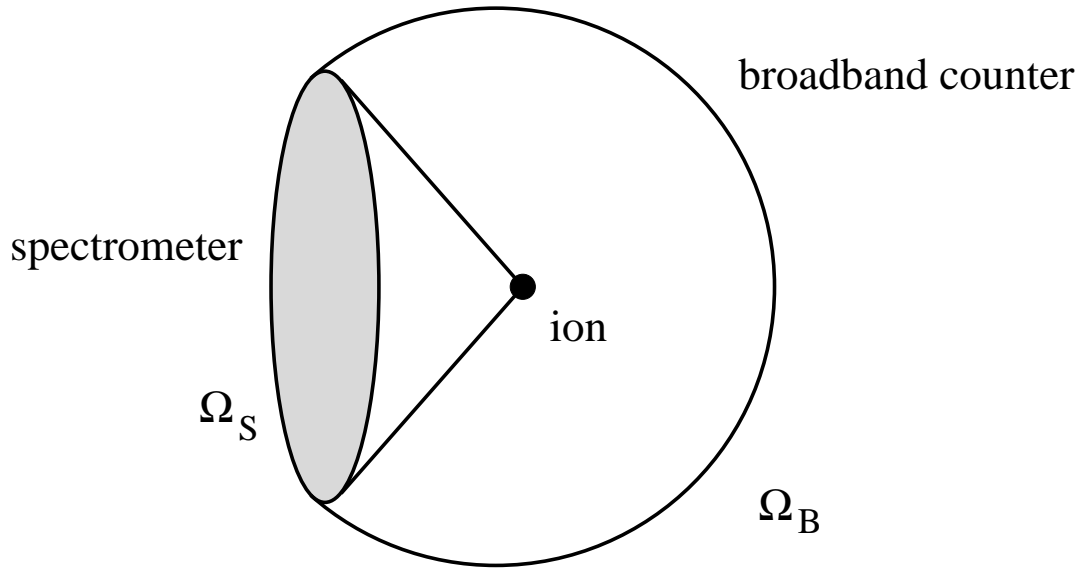}}
\vspace*{1.5cm}
\caption{\label{fig4.1} A schematic representation of a possible experimental 
setup for the measurement of conditional spectra. The spectrometer occupies
a solid angle $\Omega_S$ while the broadband counter occupies $\Omega_B$.
The broadband counter performs frequent measurements while in the spectrometer
only one measurement at a late time $T$ is performed.}
\end{figure}

A slightly different approach to the problem more in the spirit of the 
derivation of the quantum jump approach of \cite{Hegerfeldt1,Wilser1}
was undertaken in \cite{Plenio1,Hegerfeldt8}. Their starting 
point is the photon number operator 
and they define the spectrum via the number of photons that have 
been emitted into certain narrow frequency intervals \cite{Agarwal1} 
or to be more precise, the number of photons in a certain mode of 
the quantized radiation field (an approach using the electric field operator
following similar lines is also possible and is essentially equivalent
to the approach of \cite{Gardiner1}). To remain close to a physical 
picture, they envisaged an experimental situation as depicted in 
Fig. \ref{fig4.1}. A part of the quantized radiation field 
(in a solid angle $\Omega_B$) is observed by a broadband counter 
while in the rest of the space (solid angle $\Omega_S$) a 
spectrometer (for example a Fabry-Perot) is situated.

The broadband counter performs time resolved observations on 
the quantized radiation field. It is again assumed that the 
time resolved measurements can be modelled by a sequence of 
gedanken measurements that are performed in rapid succession as
in the derivation of the quantum jump approach
\cite{Hegerfeldt1,Wilser1}. At a large time $T$, we then perform 
a spectrally resolved measurement of those photons that have 
entered the spectrometer. In that way one has measured a 
spectrum with a spectral resolution $\sim 1/T$, which is conditioned 
on the particular 
detection sequence one has found in the broadband counter. 
The derivation of the relevant set of differential equations 
to calculate the spectrum follows similar ideas as they were 
developed in the work of Hegerfeldt and Wilser (1991) and Wilser (1991). 
The aim is to calculate differential equations for the conditional photon 
number in the mode ${\bf k}\lambda$ which is given by
\begin{equation} 
        tr\{a_{{\bf k}\lambda}^{\dagger}(0)
	a_{{\bf k}\lambda}(0)\rho(t|t_1, t_2, \ldots,t_n)\}~,
	\label{derivation1}
\end{equation}
where $\omega=|{\bf k}|/c,\hat{\bf k}={\bf k}/|{\bf k}|$, and we 
assume that ${\bf k}$ is a vector pointing into the solid angle
$\Omega_S$ of the spectrometer. The conditional spectrum of 
resonance fluorescence is now obtained by summing over all vectors 
${\bf \hat k}$ in $\Omega_S$. For the calculation of Eq. (\ref{8}) we 
need the state 
$\rho(t|t_1, t_2, \ldots,t_n)$ where photons were detected at times 
$t_1,\ldots,t_n$ only. With ${{\,\sf P \hspace{-1.45ex}
 \rule{0.1ex}{1.54ex}\hspace{1.25ex}}_{0\Omega_B}}$ 
the projection operator onto the vacuum state for all modes with 
${\bf \hat k}\in\Omega_B$ and with the abbreviation
\begin{equation}
   	{\cal A} \equiv {{\,\sf P \hspace{-1.45ex} 
	\rule{0.1ex}{1.54ex}
	\hspace{1.25ex}}_{0\Omega_B}}
	 U(t,s^n_{m_n})\prod_{k=1}^{m_n}{{\,\sf P 
	\hspace{-1.45ex}
 	\rule{0.1ex}{1.54ex}\hspace{1.25ex}}_{0\Omega_B}} 
	U(s_k^n,s_{k-1}^n) \;\; ,
	\label{derivation2}
\end{equation}
where $s^{n}_{k}$ are the times of those measurements where no photon 
was found,
we find, for $t_n < t < t_{n+1}$,
\begin{equation}
	\rho(t|t_1, t_2, \ldots) = 
	{\cal A}\rho(t_n+0|t_1,\ldots, t_{n-1}){\cal A}^{\dagger} \;\; ,
	\label{derivation3}
\end{equation}
where $\rho(t_n+0|t_1,\ldots, t_{n-1})$ is the state right after 
the detection of a photon in $\Omega_B$; it  is recursively given by 
\begin{eqnarray}
 	\rho(t_n+0|t_1,\ldots t_{n-1}) & \equiv & \nonumber\\
	& &\hspace*{-4.8cm} tr_{\Omega_B}\left\{
	({{\sf 1 \hspace{-0.3ex} 	
	\rule{0.1ex}{1.52ex} \rule[-.01ex]{0.3ex}{0.1ex}}}-
	{{\,\sf P \hspace{-1.45ex} 
	\rule{0.1ex}{1.54ex}\hspace{1.25ex}}_{0\Omega_B}}) 
	U(t_n,s^{n-1}_{m_{n-1}})\rho(s^{n-1}_{m_{n-1}}|t_1,...,t_{n-1})
	U^{\dagger}(t_n,s^{n-1}_{m_{n-1}})
	({{\sf 1 \hspace{-0.3ex} \rule{0.1ex}{1.52ex}
 	\rule[-.01ex]{0.3ex}{0.1ex}}}-
	{{\,\sf P \hspace{-1.45ex}
 	\rule{0.1ex}{1.54ex}\hspace{1.25ex}}_{0\Omega_B}})
	\right\}\otimes{{\,\sf P \hspace{-1.45ex} \rule{0.1ex}{1.54ex}
	\hspace{1.25ex}}_{0\Omega_B}}
 	 \: .\nonumber\\[-.1cm]
	\label{derivation4}
\end{eqnarray}
Here $tr_{\Omega_B}\{.\}$ denotes the partial trace over all modes 
with a ${\bf k}$ vector that points into the solid angle $\Omega_B$. 
As in the usual quantum jump approach, at this point the assumption
enters that the photons detected in the broadband  detector are 
absorbed during the measurement as in a real counter. One can show,
however, that this assumption is not necessary for obtaining the
equations of motion used in this section \cite{Plenio1}. In fact 
one can just as well assume ideal quantum mechanical measurements 
instead and obtain the same results \cite{Plenio1}. This is of 
course a manifestation of the intuitive physical idea that 
photons emitted by the atom will not be reabsorbed as long as there 
are no reflecting mirrors close to the atom. Mathematically it is
essentially a consequence of the Markov approximation which is
in the approach of Gardiner {\em et al} (1992) incorporated elegantly in the 
formalism.\\ 
The  general 
principle is to find first the Heisenberg-Langevin equations 
for operators of the form
\begin{equation}
	Q^{0}(t) \equiv U_I^{\dagger}(t,0){{\,\sf P 
	\hspace{-1.45ex} 
	\rule{0.1ex}{1.54ex}\hspace{1.25ex}}_{0\Omega_B}} Q 
	{{\,\sf P \hspace{-1.45ex} 
	\rule{0.1ex}{1.54ex}\hspace{1.25ex}}_{0\Omega_B}} U_I(t,0) \;\; ,
	\label{derivation5}
\end{equation}
where $U_I(t,0)$ is the interaction picture time evolution operator. 
With the abbreviation 
\begin{equation}
   	{\cal P}_n := \prod_{j=0}^{m_n} {{\,\sf P \hspace{-1.45ex} 
	\rule{0.1ex}{1.54ex}\hspace{1.25ex}}_{0\Omega_B}}(s_j^n) 
   	\prod_{k=0}^{n-1} \left\{ C(s_0^{k+1}) \prod_{i=0}^{m_k} {{\,\sf P 
	\hspace{-1.45ex} 
	\rule{0.1ex}{1.54ex}\hspace{1.25ex}}_{0\Omega_B}}(s_i^k)
 	\right\} \;\; ,
	\label{derivation6}
\end{equation}
where ${{\,\sf P \hspace{-1.45ex} \rule{0.1ex}{1.54ex}
\hspace{1.25ex}}_{0\Omega_B}}(t)$
and $C(t)$ are defined similar to Eq. (\ref{30})
without ${{\,\sf P \hspace{-1.45ex} \rule{0.1ex}{1.54ex}
\hspace{1.25ex}}_{0\Omega_B}}$, 
we then find for Eq. (\ref{12}) 
\begin{equation}
   	tr\{{a_{M}^{\dagger}}(0){\mbox{\boldmath $\sigma$\unboldmath}}(0)
	{a_{M}^{ }}(0) 
	\rho(t|t_1, t_2, \ldots)\} =
 	\langle {\cal P}_n^{\dagger} ({a_{M}^{\dagger}}{\mbox{\boldmath 
	$\sigma$\unboldmath}}
	{a_{M}^{ }})^{0}(t)
	{\cal P}_n \rangle\; .
	\label{derivation7}
\end{equation}  
Having found the Heisenberg-Langevin equations the next step is 
to show that the occurring Langevin noise operators give a 
negligible contribution. Having shown this one arrives at a set 
of differential equations. As long as the solid angle covered by 
the broadband counter is not equal to $4\,\pi$ these equation 
still describe mixtures as the photons emitted into the 
spectrometer are not observed in a time resolved way. For the 
explicit form of the equations of motion see \cite{Plenio1,Hegerfeldt8}.
Going over to the limit of a $4\,\pi$ broadband counter
yields equations that maps pure states into pure states and which 
are actually the same as those derived by Gardiner et al \cite{Gardiner1}. 
The main ingredients in this derivation are the fact that the 
initial state of the quantized radiation field is the vacuum 
state and the Markov approximation. \\
Using this approach it is straightforward to derive the quantum 
jump approach for the calculation of the absorption spectrum of 
a weak probe beam by an atom. The idea is to assume that the 
probe beam consists of a mode which is in a coherent state. 
\begin{equation}
	|\alpha_{{\bf k}\lambda}\rangle = 
	D(\alpha_{{\bf k}\lambda})|0\rangle \;\; ,
\end{equation}
where 
\begin{equation}
        D(\alpha_{{\bf k}\lambda}) = 
        e^{\alpha_{{\bf k}\lambda}a^{\dagger}_{{\bf k}\lambda} - 
           \alpha_{{\bf k}\lambda}^{*}a_{{\bf k}\lambda}}\;\; .
\end{equation}
One then performs a unitary transformation that maps
this coherent state onto the vacuum state. This leads to 
additional oscillating fields in the Hamilton operator and to an 
initial state in the probe mode which is now the vacuum. Therefore the 
approach 
outlined above may be used again. Without further complicated
calculations one obtains the spectrum as the change in the photon 
number in the probe mode. It is possible to show that this 
definition of the absorption spectrum reduces to the stationary
absorption spectrum for sufficiently long times \cite{Plenio1,Plenio2}. \\
The approaches exhibited so far enable us to simulate spectral
information (e.g. the spectrum of resonance fluorescence). They were 
derived for both a definition of the spectrum via the electric field operator 
\cite{Gardiner1} and via the photon number operator \cite{Plenio1,Hegerfeldt8}.
The physically motivated derivation of the formalism by Hegerfeldt and
Plenio \cite{Hegerfeldt7,Plenio1} with a finite size spectrometer yields as a
byproduct equations of motion for a system that is observed by a counter that 
does not cover the whole solid angle and/or has below unit efficiency.
These equations of motion will later be used to illustrate the connection 
between
the next photon and the any photon probability and to show that an inefficient
photon counter will not measure the next photon probability but the any photon
probability which is proportional to the intensity correlation function
$g^{(2)}(t)$.\\
The approaches discussed so far gave equations that enabled 
us to calculate the spectrum of resonance fluorescence directly 
from the norm of a component of a propagated 
'wave'function. However, there is a third way to obtain the spectrum of 
resonance fluorescence which is via the simulation of the correlation function
$\langle\sigma_{10}(t+\tau)\sigma_{01}(t)
\rangle$ (where $\sigma_{01}(t)$ is the Heisenberg operator corresponding
to $|0\rangle\langle 1|$)
and subsequent Fourier transformation of the 
simulation results. Indeed this is the way in which
Dalibard, Castin and M{\o}lmer \cite{Molmer1,Molmer2} and others
\cite{Garraway5,Mu1} employed the quantum jump approach to obtain simulations
of the spectrum of resonance fluorescence. The simulation procedure for a 
correlation function of the form $C(t+\tau)=\langle A(t+\tau)B(t)\rangle$ 
runs as follows. First one evolves using 
the MCWF approach a wave function $|\phi(0)\rangle$ towards $|\phi(t)\rangle$. 
Then one forms the auxiliary wave-functions
\begin{eqnarray}
	|\chi_{\pm}(0)\rangle &=& \frac{1}{\sqrt{\mu_{\pm}}}(1\pm B)
                                   |\phi(t)\rangle \;\; ,\\
        |\chi'_{\pm}(0)\rangle &=& \frac{1}{\sqrt{\mu'_{\pm}}}(1\pm iB)
                                   |\phi(t)\rangle \;\; ,
\end{eqnarray}
where the $\mu_{\pm},\mu'_{\pm}$ are normalization constants. 
Now one has to evolve each of these four wave functions according 
to the MCWF procedure and then to form
\begin{eqnarray}
	c_{\pm}(\tau) &=&
 	\langle\chi_{\pm}(\tau)|A|\chi_{\pm}(\tau)\rangle \;\; ,\\
        c_{\pm}'(\tau) &=& 
	\langle \chi_{\pm}'(\tau)|A|\chi_{\pm}'(\tau)\rangle \;\; ,
\end{eqnarray}
from which on obtains
\begin{equation}
	C(t,\tau) = \frac{1}{4} \left[ \mu_{+} \bar{c}_{+}(\tau)
                    - \mu_{-} \bar{c}_{-}(\tau) - 
		      i \mu_{+}' \bar{c}_{+}'(\tau) 
                    + i \mu_{-}' \bar{c}_{-}'(\tau) \right]\;\; .
\end{equation}
It can be shown that this procedure produces the correct 
ensemble averages \cite{Dalibard1}. A subsequent Fourier 
transform of the simulation results yields a spectrum. This 
procedure has been used , for example, \cite{Garraway5} to
simulate the spectrum of a three level system in a V configuration.
However care has to be taken in this procedure because if one
only performs a single Fourier transform to obtain the stationary
spectrum
\begin{equation}
	S_1(\omega) := 2\,Re\int_{0}^{\infty} d\tau\, e^{i\Delta\tau} 
        \langle\sigma_{10}(\tau)\sigma_{01}(0)\rangle_{ss}\;\; ,
\end{equation}
one inevitably runs into difficulties as one can only simulate finite times.
Then, however, the Fourier transformation yields spurious negativities in
in the power spectrum which are due to finite time effects.
This problem can be circumvented by using a finite time double 
integral of the correlation function as required in the definition 
of the time dependent spectrum.
\begin{equation}
	S_T(\omega) := \frac{1}{T}
	\int_{0}^{T} dt' \int_{0}^{T} dt'' e^{i\Delta(t'-t'')}
        \langle \sigma_{10}(t') \sigma_{01}(t'')\rangle \;\; .
\end{equation}
This quantity has the advantage that it is manifestly positive 
for each realization as it is of the form 
$\langle A^{\dagger}A\rangle$. However this considerably slows 
down the simulation procedure because the integral transform is 
now much more complicated. 

A different point of criticism should be mentioned concerning 
the interpretation of a single run of this simulation procedure. 
While the derivation of the simulation procedure by Hegerfeldt
and Plenio \cite{Hegerfeldt8} clearly shows that a physical 
interpretation of a single run in the schemes derived
in  Hegerfeldt and Plenio (1996), Plenio (1994) and Gardiner {\em et al} 
(1992) is possible, although this is not obviously the case for the 
simulation procedure of M{\o}lmer {\em et al} (1993) . A problem lies in 
the fact that four 
different wave functions have to be propagated in the 
procedure, each of which might follow different jump sequences. 
Furthermore even if only one jump sequence is followed,
it is not clear what the interpretation of the correlation 
function is. This may be illustrated by the following example. 
We now would like to simulate a trivial correlation function of 
the form $\langle 1(t+\tau)A(t)\rangle$. We would expect that in 
each realization the simulation procedure would give the same 
result as for the single time expectation value 
$\langle A(t) \rangle$. This, however, is not the case as is 
easily seen in the following example. We assume a spontaneously 
decaying TLS and the initial state 
\begin{equation}
	|\phi(0)\rangle = \left( \frac{1}{1+e^2} \right)^{1/2} 
                          (|0\rangle + e|1\rangle)\, .
\end{equation}
The emission--free time evolution is given by
\begin{equation}
	U_0(t+t',t) = e^{\bf{M} t'} \;\; ,
\end{equation}
where
\begin{equation}
	{\bf M} = \left(
	\begin{array}{ll}
	0 & 0\\
	0 & -\Gamma_{22}
	\end{array}
	\right)\, .
\end{equation}
If we assume that no jump at all occurs until $t+\tau$ we 
obtain according to the procedure of M{\o}lmer {\em et al} (1993) the single 
run result for the correlation function 
$\langle 1(t+\tau)\sigma_{01}(t)\rangle$ to be 
\begin{equation}
	\langle 1(t+\tau)\sigma_{01}(t)\rangle = \frac{1}{1+e^{-2\Gamma_{11}\tau}}
\end{equation}
while for the expectation value of the operator $\sigma_{01}(t)$ at time 
$t=1/\Gamma_{11}$ is given by
we obtain
\begin{equation}
	\langle \sigma_{01}(t=1/\Gamma_{11}) \rangle = \frac{1}{2}\;\; .
\end{equation}
These two expressions obviously differ. Therefore we can conclude
that the simulation procedure of M{\o}lmer {\em et al} (1993) yields the 
correct
ensemble results; however if we are interested in questions
concerning single runs or spectra conditioned on a given jump 
statistics care has to be taken and it is safer to resort to the approaches 
in \cite{Gardiner1,Dum2} and
Hegerfeldt and Plenio \cite{Hegerfeldt8,Plenio1}. 
Applications to conditioned spectra will be given later in Section \ref{V}.

Here we have only described in detail quantum jump approaches to calculate
spectral information. Of course one can also obtain a quantum state diffusion
simulation of spectral information. One can derive the equations either by
starting out from the quantum jump equations given above and then follow our
derivation of the quantum state diffusion equations as a result of heterodyne
detection. Otherwise one can deduce the equations by assuming that the system
couples to an additional output mode (i.e. similar to \cite{Gardiner1} but here 
literally one mode of the quantized radiation field is chosen) and then
writes down the master equation for that enlarged system. From that one can 
then easily obtain quantum state diffusion equations \cite{Brun2}. However, 
if one tries to apply the equations to the calculation of time correlation
functions of the form $\langle A(t+\tau) B(t)\rangle$ one again encounters
interpretational problems similar to those explained for the approach of
M{\o}lmer described above. It should be noted that the approach to calculate
time correlation functions given by Gisin (1993) does not lead to 
$\langle A(t+\tau) B(t)\rangle$ as usually defined in quantum optics but 
to a different correlation quantity which is difficult to interpret
\cite{Sondermann1}.
\section{Applications of the quantum jump approach}
\label{V}
\subsection{Photon statistics}
In the previous sections we have introduced and discussed the 
quantum jump approach illuminating many different approaches 
to it. After these sometimes formal considerations we would like 
to give a number of examples to give a better understanding of the 
formalism and its physical implications. The examples will 
be drawn mainly from two physical situations; single trapped 
ions driven by lasers and electromagnetic fields in cavities, 
i.e., cavity QED.

We start by illustrating the difference between single realizations 
and ensemble averages by investigating a laser driven two--level system 
(TLS) damped by a zero temperature quantized radiation field. The
master equation is then given by 
\begin{equation}
	\dot{\rho} = -\;\frac{i}{\hbar}[H_{A},\rho] - 
	\Gamma\{ \sigma_{11}\rho + \rho\sigma_{11} \} 
	+ 2\,\Gamma \sigma_{01}\rho\sigma_{10}\;\; ,
	\label{500}
\end{equation}
with
\begin{equation}
	H_A = -\hbar\Delta |1\rangle\langle 1 | + 
	\frac{\hbar\Omega}{2} \left( |0\rangle\langle 1| + 
	|1\rangle\langle 0| \right)
	\label{501}
\end{equation}
where $2\Gamma$ equals the Einstein coefficient of the observed TLS,
$\Omega$ the Rabi frequency and $\Delta$ the detuning of the laser.

\begin{figure}[hbt]
\epsfxsize14.cm
\centerline{{\epsffile{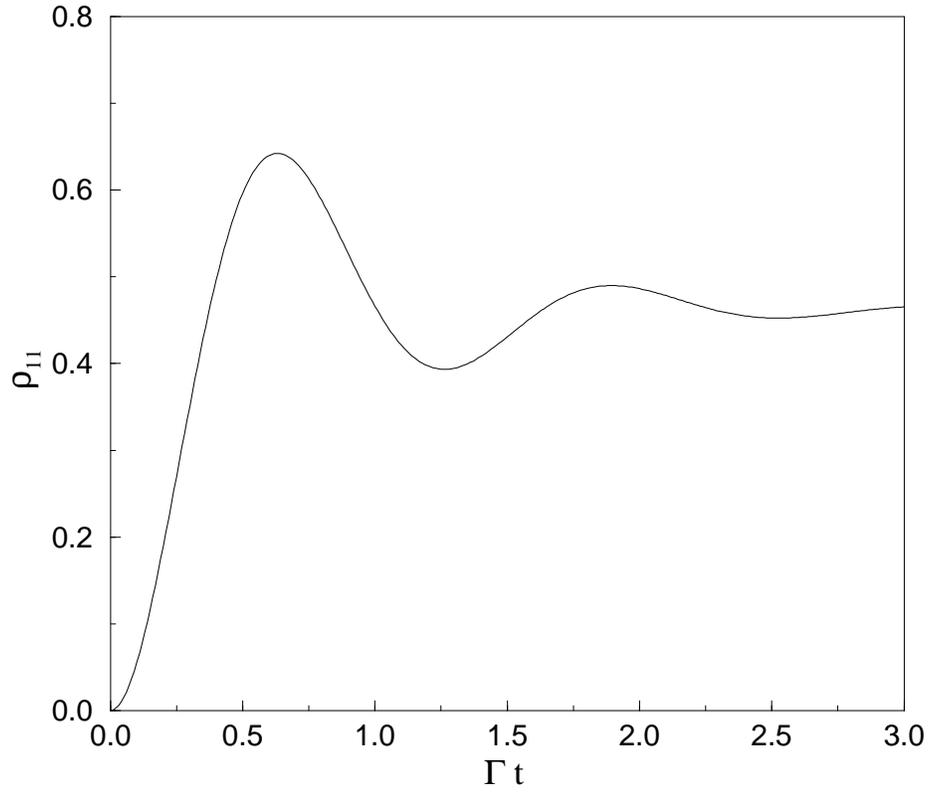}}}
\caption{\label{fig5.1} The upper state population $\rho_{11}$ 
of an ensemble of two-level systems which is driven by a laser with 
Rabi frequency $\Omega=5\,\Gamma$ and 
vanishing detuning $\Delta=0$. The system exhibits some oscillations
and then approaches a nonzero steady state value.}
\end{figure}

\begin{figure}[hbt]
\epsfxsize14.cm
\centerline{{\epsffile{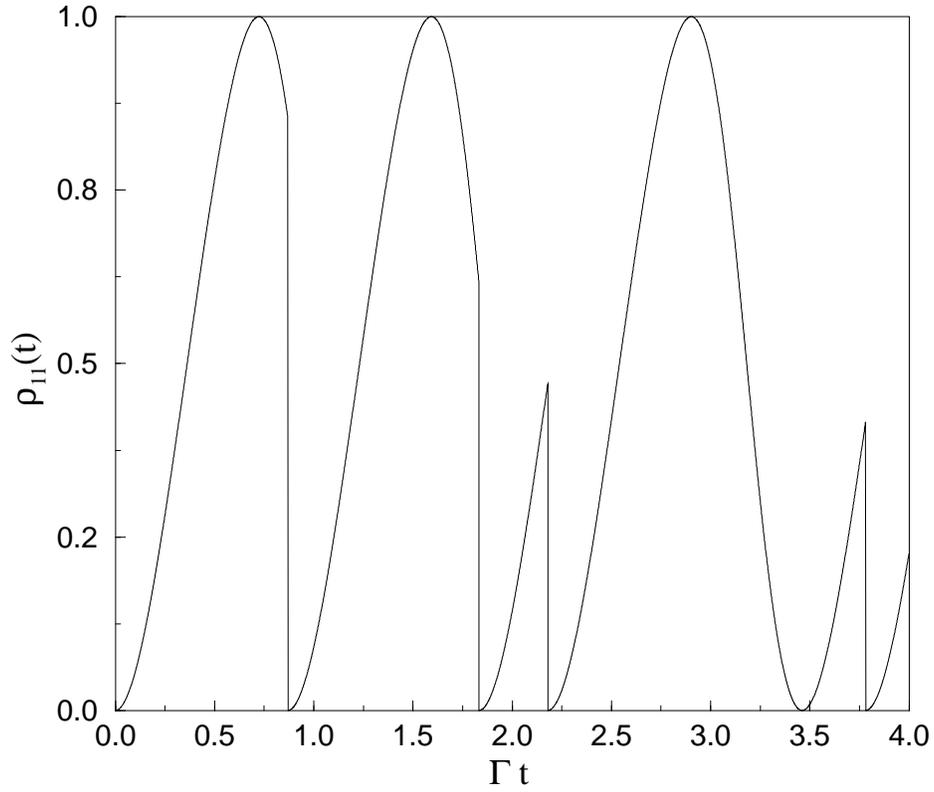}}}
\caption{\label{fig5.2} The time evolution of the upper state population
$\rho_{11}$ of a single driven two-level system. As in 
Fig. \protect\ref{fig5.1} the two-level system is driven by a laser 
with Rabi frequency $\Omega=5\,\Gamma$ and vanishing detuning $\Delta=0$.
The system starts a Rabi oscillation which is then interrupted by a quantum
jump (detection of a photon). After the jump the system is reset in the ground
state and a new Rabi oscillation starts. (After \protect\cite{Garraway5})}
\end{figure}

\begin{figure}[hbt]
\epsfxsize14.cm
\centerline{{\epsffile{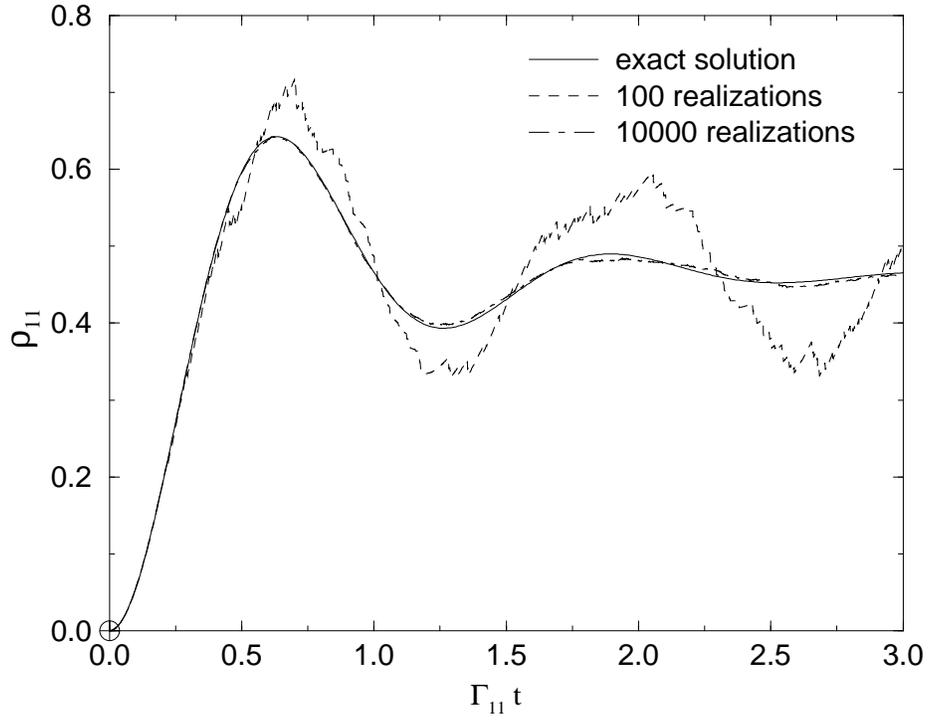}}}
\caption{\label{fig5.3} 
The ensemble result of Fig. \ref{fig5.1} is compared to the average over
$N=100$ and $N=1000$ realizations generated using the quantum jump approach.
The fluctuations of the averages become smaller with increasing $N$ and 
the ensemble average gets approximated more and more
closely. The Rabifrequency 
is $\Omega=5\,\Gamma$ and the detuning is $\Delta=0$.}
\end{figure}

Assuming standard broadband photodetection, the quantum jump approach
gives for the conditional time evolution between detections 
\begin{equation}
	|\psi(t)\rangle = e^{-i H_{eff} t/\hbar} |\psi(0)\rangle \;\; ,
	\label{502}
\end{equation}
with 
\begin{equation}
	H_{eff} = H_{A} - i\hbar\,\Gamma
	\label{503}
\end{equation}
and for the normalized state after the detection of a photon 
\begin{equation}
	|\psi(t_+)\rangle = \frac{\sigma_{01}|\psi(t)\rangle}{||.||} 
	= |0\rangle \;\; .
	\label{504}
\end{equation}
The probability for not having a jump in the time interval $[0,t]$ 
if the initial state is the ground state, $|\psi(0)\rangle = |1\rangle$
is given by
\begin{eqnarray}
	P_0(t) &=& \langle \psi_{eff}(t)|\psi_{eff}(t)\rangle 
	\nonumber\\[.25cm]
	&=& \frac{\lambda_2}{\lambda_2-\lambda_1} e^{\lambda_{1} t}
	  - \frac{\lambda_1}{\lambda_2-\lambda_1} e^{\lambda_{2} t}
	\;\; , \label{505}
\end{eqnarray}
with
\begin{equation}
	\lambda_{1/2} = \frac{-\Gamma + i\Delta}{2} \pm
	\frac{\sqrt{\Gamma^2 - \Delta^2 - \Omega^2 - 2i\Delta\Gamma}}
	{2} \;\; . \label{506}
\end{equation}
Using the parameters $\Omega=5\,\Gamma$ and $\Delta=0$ the upper state
population $\rho_{11}$ of the {\em ensemble} evolves as shown in Fig.
\ref{fig5.1} where the initial state is the ground state.
The population rises from zero, undergoing some Rabi oscillations and
then tends towards a steady state. So much to the ensemble 
average. Now let us look at individual realizations of the time 
evolution using the same parameters and initial conditions. The result
for {\em one} possible realization is shown in Fig. \ref{fig5.2}. The 
picture is
strikingly different from Fig. \ref{fig5.1} in that the time evolution is not
smooth but exhibits jumps and it {\em does not} tend towards a steady state.
We rather observe that initially the system starts to perform a Rabi 
oscillation. As the population in the upper level grows in time so does
the probability for an emission of a photon. This oscillation is then
terminated by the emission of a photon which brings the atom back to its
ground state. Then the whole process starts again. Figs. \ref{fig5.1} and
\ref{fig5.2} show little similarity; however, averaging over many individual
realizations shown in Fig. \ref{fig5.2} lead to a closer and closer
approximation of the ensemble average. To illustrate this we have plotted 
the ensemble result together with the average over $N=100$ and $10000$
individual realizations in Fig. \ref{fig5.3}.
It is found that the root mean square deviation of the simulated 
average from the exact ensemble result are of the order of $1/\sqrt{N}$.
We have examined the photon statistics of a two level system by 
detecting {\em} all the photons emitted by the atom with high time resolution
as we assumed that the counter has unit efficiency and covers the whole 
solid angle. Assuming this, one would be able to measure the next photon
probability density $I_1(t)$. However, in a real experiment the counter
efficiency is much less than unity due to imperfect quantum efficiency 
and because only a finite solid angle is covered
by the counter. Therefore the question is which function we actually 
measure when we perform a real experiment in which we determine the 
detection time of the 'next' photon in our imperfect counter.

To answer this question we would like to employ the quantum jump approach
assuming a perfect photon counter which, however, does not cover all 
space. If it covers a solid angle $\Omega_B$ then this is equivalent to a 
$4\pi$ counter of efficiency $\beta=\Omega_B/4\pi$. The equation
of motions for this setup up have been derived in section IV in connection
with the simulation of spectral information (see Eqs.
(\ref{derivation1})-(\ref{derivation7})). The efficiency $\beta$ of the 
detection process is the fraction of the emitted
photons that are actually detected. We obtain the conditional equation of
motion for the density operator under the assumption that no photon has been 
detected in the counter,
\begin{equation}
	\dot{\rho} = -\frac{i}{\hbar} \left\{ H_{eff} \rho - \rho H_{eff}
	\right\} + 2\Gamma \left(1-\beta \right) \sigma_{01}\rho\sigma_{10}
	\; . \label{508}
\end{equation}
$H_{eff}$ is given by Eq. (\ref{503}) and one should note that now the 
conditional time evolution does {\em not} map pure states onto pure 
states. This  is of course due to the incomplete information gained 
from the imperfect counter. Therefore we have to average over all possible
events that the counter could not detect and this leads to a mixture. One
observes two familiar limits: for zero counter efficiency we recover the
ordinary optical Bloch equations, whereas for unit efficiency $\beta=1$ we 
find the effective time evolution inferred from the assumption that no 
photon has been found in the whole solid angle given a perfect counter. 

\begin{figure}[hbt]
\epsfxsize14.cm
\centerline{{\epsffile{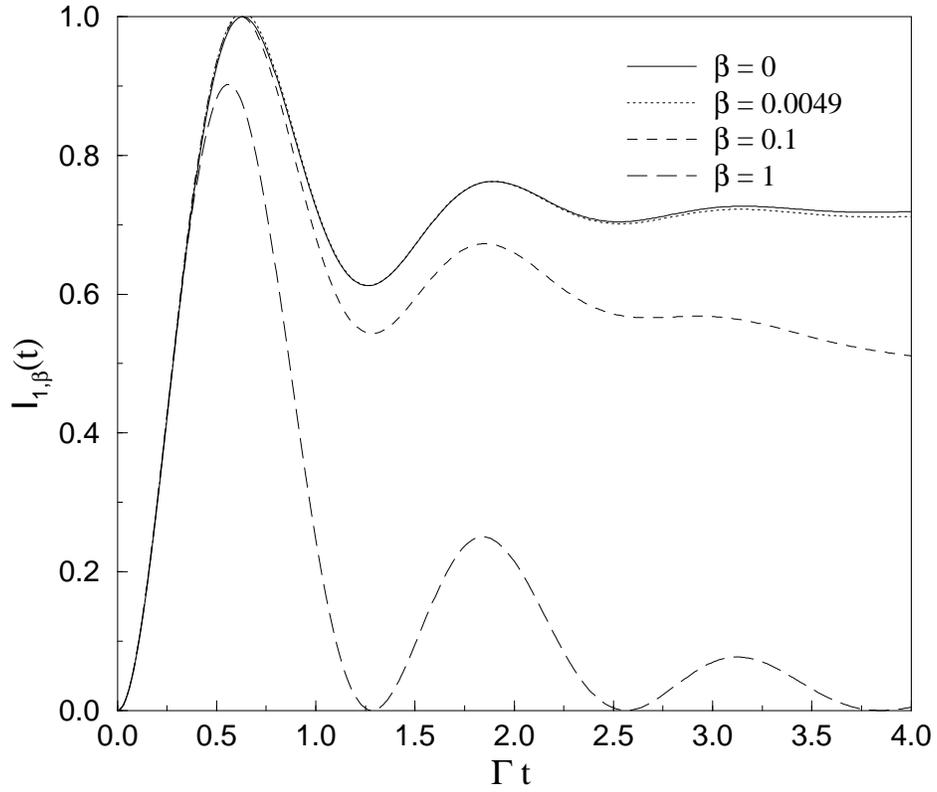}}}
\caption{\label{fig5.4} The next photon rate for different counter
efficiencies $\beta=1,\beta=0.1$ and $\beta=.0049$ and the any photon 
probability which is obtained in the limit $\beta=0$. The Rabifrequency 
is $\Omega=5\,\Gamma$ and the detuning is $\Delta=0$. We have normalized 
the plotted functions to have a maximum value of unity.}
\end{figure}

Having found Eq. (\ref{508}) we can now calculate the probability that the  
detector finds no photon in the time interval $[0,t]$
\begin{equation}
	P_0(t) = tr\{\rho_{0}(t)\} \;\; ,
	\label{509}
\end{equation}
which reduces to Eqs. (\ref{13}) or Eq. (\ref{22a}) in the limit
$\beta\rightarrow 1$. The next count rate is then the negative time 
derivative of Eq. (\ref{509})
\begin{equation}
	I_{1,\beta}(t) = 2\,\Gamma\beta\rho_{11}(t)\;\; ,
	\label{510}
\end{equation}
where $\rho_{22}$ is obtained by solving Eq. (\ref{508}). Instead of solving 
this equation analytically we plot the solution for several different
counter efficiencies $\beta$. We assume $\Gamma=1$ and $\Omega=5\,\Gamma$.
In Fig. \ref{fig5.4} we plot the next detection probability for counter 
efficiency $\beta=1,\beta=0.1$ and $\beta=0.0049$ where the latter value 
of $\beta$ is computed from Mandels antibunching experiments  
\cite{Kimble0,Dagenais1}. We observe that for decreasing counter efficiency the 
function Eq. (\ref{510}) approximates the "any photon" probability of the 
whole ensemble, which becomes more and more closely proportional to the
intensity correlation function $g^{(2)}(t)$. So far we have we have shown 
how to calculate the next detection rate for different counter efficiencies
from the quantum jump approach. We have plotted the results 
numerically. However it is also possible to derive an analytical expression
in a somewhat different way. This was done in Kim {\em et al} (1987)
and was in fact one earlier approach to the problem. As described earlier 
one can utilise the fact that there is
a very simple relation between the "any photon" rate $I_{\beta}(t)$ of 
the complete ensemble and the next detection rate $I_{1,\beta}(t)$ for 
a counter with efficiency $\beta$. This relation can be found using 
\begin{equation}
	\rho_{11}(t) = \rho_{11}^{(\beta)}(t) + \int_{0}^{t} dt'
	\rho_{11}(t-t')\; 2\,\Gamma\beta \rho_{11}^{(\beta)}(t')
	\; ,\label{511}
\end{equation}
and 
\begin{eqnarray}
	I_{1,\beta}(t) &=& 2\,\Gamma\beta\rho_{11}^{(\beta)}(t) \;\; ,
	\label{512}\\
	I_{\beta}(t) &=& 2\, \Gamma\beta\rho_{11}(t)
	\; ,\label{513}
\end{eqnarray}
where $\rho_{11}^{(\beta)}(t)$ is calculated from Eq. (\ref{508}).
Inserting this and taking the Laplace transform this yields 
\begin{equation}
	\hat{I}(z) = \frac{1}{\beta}\;
	\frac{\hat{I}_{1,\beta}(z)}{1-\hat{I}_{1,\beta}(z)}
	\; .\label{514}
\end{equation}
The intensity correlation function $g^{(2)}(t)$ is related to the 
any photon rate $I_{\beta}(t)$ for an imperfect counting process
by
\begin{equation}
	2\,\Gamma\rho_{11}(\infty) \beta g^{(2)}(t) = I_{\beta}(t)
	\; .\label{515}
\end{equation}
Therefore we obtain 
\begin{equation}
	\hat{I}_{1,\beta}(z) = \frac{2\Gamma\rho_{11}(\infty)\beta 
	\hat{g}^{(2)}(z)}
	{2\Gamma\rho_{11}(\infty)\beta \hat{g}^{(2)}(z) +1}
	\; . \label{516}
\end{equation}
where $\hat{g}^{(2)}(z)$ is the Laplace transform of the intensity 
correlation function $g^{(2)}(t)$.

One should note that the intensity correlation function does not depend 
on the efficiency $\beta$ as it is normalized with respect to the 
stationary detection rate. From expression Eq. (\ref{516}) one observes that, 
up to a normalization factor, the next photon probability tends towards
the intensity correlation function. Inserting the well known expressions
for the stationary state of the TLS and the intensity correlation function
for zero detuning
\begin{equation}
	\rho_{22}(\infty) = \frac{\Omega^2}{2\,\Omega^2 + 4\,\Gamma^2}
	\label{517}
\end{equation}
and
\begin{equation}
	\hat{g}^{(2)}(z) = \frac{\Omega^2+2\,\Gamma^2}{z((z+3\,\Gamma/2)^2 
	+\Omega^2 - \Gamma^2/4)}
	\label{518}
\end{equation}
into Eq. (\ref{516}) we obtain
\begin{equation}
	\hat{I}_{1,\beta}(z) = \frac{\beta\Omega^2\Gamma}{(z+\Gamma)
	(z(z+2\,\Gamma) + \Omega^2(z+\beta\Gamma))}
	\; .\label{519}
\end{equation}
This Laplace transform can be inverted easily. Plotting this function
exactly reproduces Fig. \ref{fig5.4} if one normalises the maximum of 
$I_{1,\beta}(t)$ to unity.
\subsection{Intermittent fluorescence}
After this discussion of the photon statistics of the resonance 
fluorescence of a two-level system we now proceed to illustrate the
formalism for a more complex system, namely, a three-level system
in V-configuration as shown in Fig. \ref{fig2.1}. We assume that the 
$0\leftrightarrow 1$ transition is strongly allowed while the 
$0\leftrightarrow 2$
transition is metastable. In a typical experiment one would have a 
lifetime of the order of seconds for the metastable level $2$, while the
unstable level $1$ has a lifetime of several nanoseconds. The system
is irradiated by two lasers, one on each transition. The Rabi frequency
$\Omega_1$ of the laser driving the $0\leftrightarrow 1$ transition
is assumed to be much larger than the Rabi frequency $\Omega_2$ on 
the $0\leftrightarrow 2$ transition. If one observes the intensity 
of resonance fluorescence on the strong transition under these conditions
one typically obtains a result as shown in Fig. \ref{fig2.2} or in a more
schematical representation in Fig. \ref{fig2.4}. Long
periods of brightness with many photon counts (bright periods) are
interrupted by prolonged periods  with no photo--detections (dark periods).
As we discussed previously a simplified treatment of this situation using
rate equations was given by Cook and Kimble \cite{Cook2,Kimble1}. However, as
the described situation involves lasers, a more detailed treatment 
is required as coherences can play a crucial role in the time 
evolution of the system. Such treatments were initially undertaken
using Bloch equations  but this is not the most
natural description of the problem. Such a natural description of 
the problem was provided by the quantum jump approach. In the following 
we will use the quantum jump approach to calculate the photon statistics 
of the V-system and we will use it to gain interesting and sometimes surprising
insights into the single system dynamics. It will turn out that the 
rate equation treatment of Cook and Kimble, while giving a  
qualitative picture, is insufficient to explain many interesting 
properties of the system. A similar analysis can be carried out for other 
systems such as a $\Lambda$ configuration \cite{Agarwal3,Hegerfeldt6,Plenio1},
and in a system where both upper levels
couple strongly to the ground level but are close together 
\cite{Hegerfeldt2,Hegerfeldt4,Hegerfeldt5,Koehler1}.

In the following analysis we will always assume that both bright and
dark periods are much longer than the lifetime of the unstable 
level $1$. This condition is necessary from a physical point of view as 
otherwise 
one would not be able to distinguish between a dark period
and the time interval between two successive emissions in a bright period.
Also a bright period consisting of approximately one photon has
little meaning. With the detuning $\Delta_i$ and Rabi frequencies $\Omega_i$
of the lasers on the $0\leftrightarrow i$ transition  we obtain the condition
\cite{Hegerfeldt7}
\begin{equation}
        \Omega_2^2 \ll \frac{1}{4}\frac{16\,\Delta_2^2\Gamma_{11}^2+
        (\Omega_1^2+4\Delta_2(\Delta_1-\Delta_2))^2}
        {\Gamma_{11}^2+(\Delta_1-\Delta_2)^2}\; .
        \label{520}
\end{equation}  
where $2\Gamma_{11}$ equals the Einstein coefficient of level $1$. This
condition assumes that the Einstein coefficient of level $2$ is negligible.
Again this can be cast into an analytical form. We require \cite{Hegerfeldt7}
\begin{equation}
   \Gamma_{22} \ll \frac{\Omega_1^2\Omega_2^2\Gamma_{11}}
   {16\,\Delta_2^2\Gamma_{11}^2+
   \{\Omega_1^2+4\,\Delta_2(\Delta_1-\Delta_2)\}^2}
   \label{522}
\end{equation}
which expresses the fact that spontaneous emissions from level $2$
are much less frequent than stimulated transitions. This also implies
that there are sufficiently many long dark periods.

\begin{figure}[hbt]
\epsfxsize14.cm
\centerline{{\epsffile{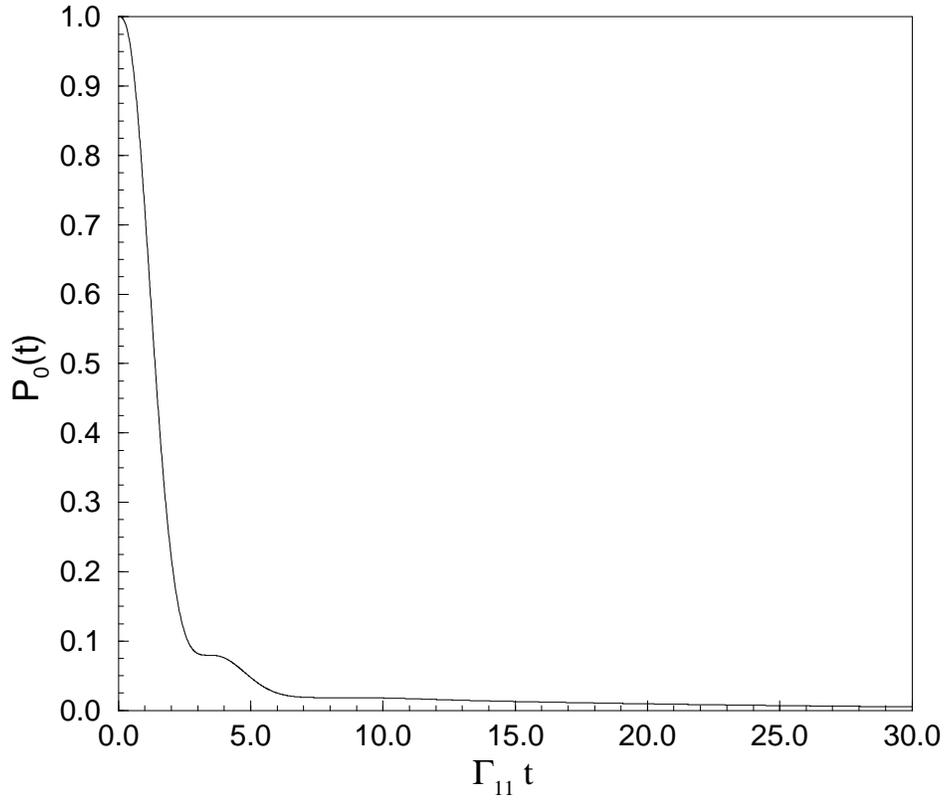}}}
\caption{\label{fig5.13} The delay function $P_0(t)$ of the $V$-system
describing the probability 
that after an emission at $t=0$ no other emission has taken place until 
$t$. Parameters are $\Omega_{1}=2\,\Gamma_{11}$,$\Omega_{2}=0.35\,\Gamma_{11},
\Delta_1=\Delta_2=0$ and $\Gamma_{22}=0$. One observes a slowly decaying 
tail of $P_0(t)$ indicating the possibility of dark periods.}
\end{figure}

To be able to calculate the mean lengths of bright and dark periods
we need to define precisely what we mean by bright and dark periods.
To distinguish between bright and dark periods we introduce a time $T_0$.
If our perfect photo detector fails to detect a photon in a time interval
$[0,T_D]$ where it has found a photon at time $t=0$ we speak of a dark 
period of length $T_D$. If, however, in a time interval $[0,T_L]$ the 
time between successive photon detections is always less than $T_0$ we 
have a bright period of length $T_L$. Using these definitions and the 
{\em next} photon probability density $I_1(t)$ we obtain for the mean
length of a dark period
\begin{equation}
   T_D(T_0) = \frac{\int_{T_0}^{\infty}\!\!dt'\,t'I_1(t')}
                      {\int_{T_0}^{\infty}\!\!dt'\,I_1(t')}
            = T_0 + \frac{\int_{T_0}^{\infty}\!\! dt'\, P_0(t')}
                         {P_0(T_0)}\;\; ,
   \label{523}
\end{equation}
while the mean length of a bright period is
\begin{equation}
   T_L(T_0) = \frac{1}{P_0(T_0)}\, \frac{\int_0^{T_0}\!\!dt'\,t'I_1(t')}
                      {\int_0^{T_0}\!\!dt'\,I_1(t')}\;\; .
   \label{524}
\end{equation}
These expressions appear to be quite complicated but they can be simplified 
when
we use the assumption that bright and dark periods are long compared to the
lifetime of level $1$. In that case it is obvious that the time evolution
of the system has two widely different time scales, one determining the
emission rate during bright periods, the other giving the rate at which
dark periods occur. This is reflected in the fact that the probability to
find no photon in the interval $[0,t]$ has a slowly decaying tail as we can
observe from Fig. \ref{fig5.13} where we plot $P_0(t)$ for the parameters
$\Omega_{1}=2\,\Gamma_{11}$,$\Omega_{2}=0.35\,\Gamma_{11},
\Delta_1=\Delta_2=0$ and $\Gamma_{22}=0$. 
If we chose $T_0$ such that it is much larger than
the mean time between photon detections in a bright period while still 
being much shorter than the mean length of a dark period we can reliably
distinguish between bright and dark periods. The choice of $T_0$ implies
that
\begin{equation}
	P_0(T_0) = p\,e^{-2\,T_0\, Im\,\lambda_1} \cong p \ll 1 \;\; ,
	\label{525}
\end{equation}
where $\lambda_1$ is the smallest eigenvalue of the effective Hamiltonian 
of the system 
\begin{equation}
   H_{eff} = \left( \begin{array}{ccc}
   0                 & -\frac{\Omega_1}{2}          & -\frac{\Omega_2}{2}\\
   -\frac{\Omega_1}{2} & i\,\Gamma_{11} + \Delta_1 & 0                \\
   -\frac{\Omega_2}{2} &          0                 & \Delta_2        
   \end{array}
   \right)\;\; .
   \label{526}
\end{equation}
Simplifying Eqs. (\ref{523}) -- (\ref{524}) we then obtain
\begin{equation}
	T_D(T_0) = \frac{1}{-2\,Im\lambda_1}
	\label{527}
\end{equation}
and 
\begin{equation}
	T_L(T_0) = \frac{\tau_L}{p} \;\; ,
	\label{528}
\end{equation}
where $\tau_L$ is the mean time between successive photo--detections in a 
bright period
\begin{equation}
   \tau_L(T_0) = \int_0^{T_0} dt'\, t' I_1(t')/\int_0^{T_0} dt'\, I_1(t')
	\;\; .
   \label{529}
\end{equation}
The mean time interval between successive emission in a bright period can
be found from Eq. (\ref{529}) or more easily from
\begin{equation}
	\tau_L = \frac{1}{2\Gamma_{11}\rho^{TLS}_{11}} \;\; ,
	\label{530}
\end{equation}
where $\rho^{TLS}_{11}$ is the population of the upper level $1$ of a TLS 
driven by a laser of Rabi frequency $\Omega_1=2\,\Gamma_{11}$ and $\Delta_1=0$.
Now it is quite easy to obtain analytical expressions for $T_D$ and $T_L$.
We find 
\begin{eqnarray}
  	 T_D &=&
  	 \frac{16\Delta_2^2\Gamma_{11}^2+
 	 (\Omega_1^2-4\,\Delta_2(\Delta_2-\Delta_1))^2}
 	 {2\Omega_1^2\Omega_2^2\Gamma_{11}} \;\; , \label{530a}\\
 	 T_L &=& \frac{2\,\Delta_1^2+2\,\Gamma_{11}^2+\Omega_1^2}
  	 {2\Gamma_{11}^2+2(\Delta_1-\Delta_2)^2} T_D\, .
	\label{531}
\end{eqnarray}
We see that under the conditions Eqs. (\ref{520}-\ref{522}) both bright
and dark periods are much longer than the lifetime of the unstable
$0\leftrightarrow 1$ transition. 

\begin{figure}[hbt]
\epsfxsize14.cm
\centerline{{\epsffile{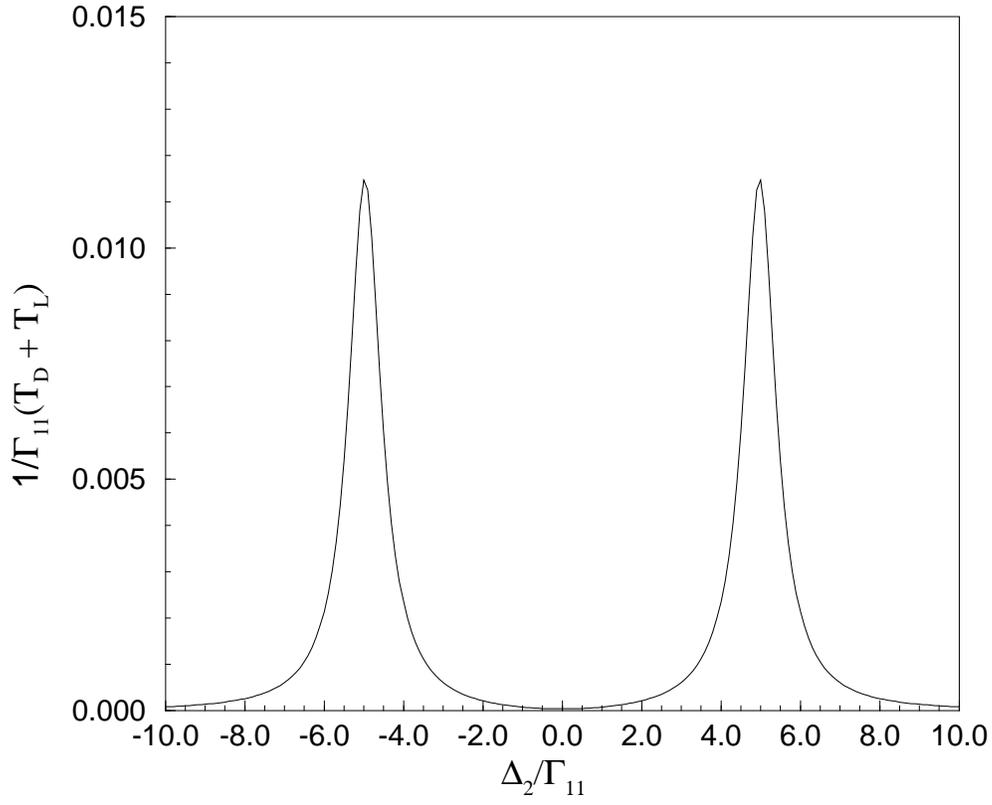}}}
\caption{\label{fig5.6} $\frac{1}{T_D+T_L}$, representing the 
average rate at which quantum jumps will be observed in the 
three--level V--system, 
as a function of the detuning $\Delta_2$ on the weak transition. The 
parameters are $\Omega_1=10\,\Gamma_{11},\Omega_{2}=0.3\,\Gamma_{11},
\Delta_1=0$ and $\Gamma_{22}=0$. One observes that the maximum
quantum jump rate is achieved for detunings $\Delta_2=\pm\Omega_1/2$
illustrating the Autler-Townes splitting of the ground state.}
\end{figure}

\begin{figure}[hbt]
\epsfxsize14.cm
\centerline{{\epsffile{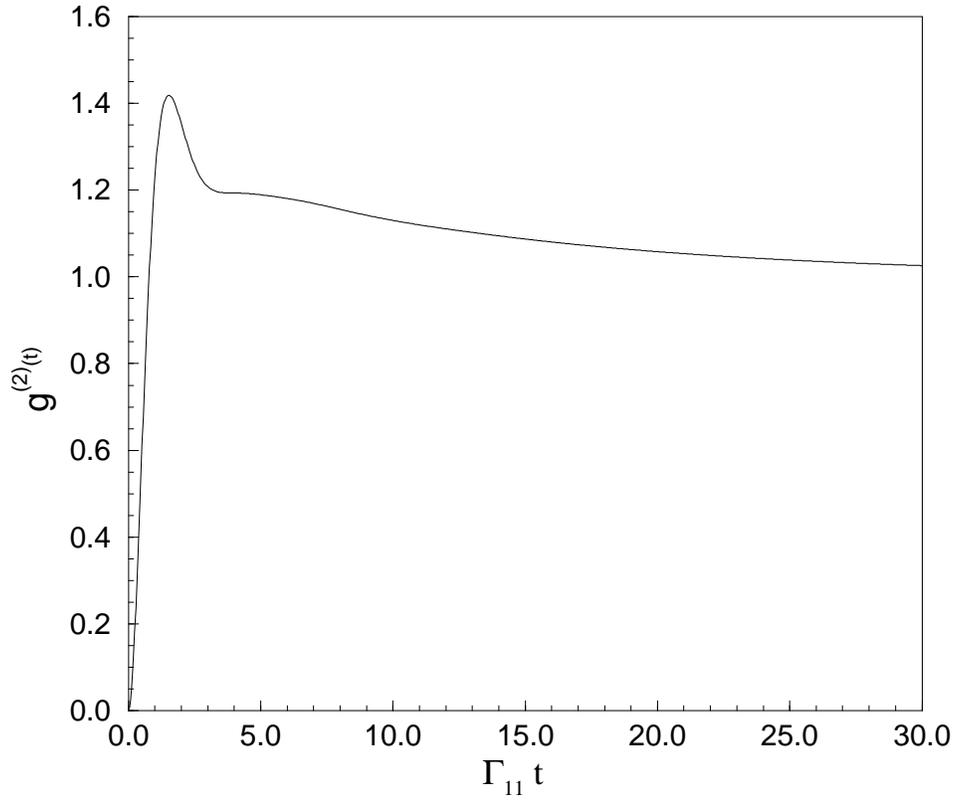}}}
\caption{\label{fig5.7} The intensity correlation function $g^{(2)}(\tau)$ of
a $V$ -- system for the same parameters as in Fig. \ref{fig5.13}. One clearly
observes that $g^{(2)}(t)$ first falls off quickly to a value of around
$1.2$ for times around $\tau\approx 5\Gamma_{11}^{-1}$ and then starts to 
fall off slowly towards the stationary value of $1$.}
\end{figure}

Investigating the average rate $\frac{1}{T_D+T_L}$ at which we observe 
quantum jumps, i.e., the onset of dark periods, we see that it depends on the 
detuning of the laser on the $0\leftrightarrow 2$ transition. If we assume
that the strong laser ($\Omega_1\gg\Gamma_1$) is resonant ($\Delta_1=0$) 
we observe that the
ratio $\frac{1}{T_D + T_L}$ becomes minimal when $\Delta_2=0$ and 
maximal for the Autler--Townes or Stark split detunings
$\Delta_2=\pm\frac{\Omega_1}{2}$. 
\begin{equation}
	\frac{1}{T_D + T_L} = \frac{4\Omega_1^2\Omega_2^2\Gamma_{11}}{
	16\Delta_2^2\Gamma_{11}^2 + (\Omega_1^2 - 4\Delta_2^2)^2}\,
	\frac{\Gamma_{11}^2 + \Delta_2^2}{4\Gamma_{11}^2 + 2\Delta_2^2
	+\Omega_1^2}\;\; .
	\label{531a}
\end{equation}
This dependence
on the detuning $\Delta_2$ reflects the fact that due to the strong
driving of the $0\leftrightarrow 1$ transition the lower level exhibits
Autler-Townes splitting. The two effective levels are shifted by
$\pm\Omega_1/2$ with respect to the original level $0$. To obtain long dark
periods one needs to bring the weak laser on the $0\leftrightarrow 2$ 
transition into resonance with one of the dressed states of the 
$0\leftrightarrow 1$
transition. The resulting frequency dependence is 
shown in Fig. \ref{fig5.6}. Before we proceed with the investigation of 
the single system behaviour, we show again, now quantitatively, 
that the existence of bright and dark periods in the
resonance fluorescence has a visible effect on the ensemble quantities, 
too. To see this, we plot the intensity correlation function $g^{(2)}(\tau)$ 
in Fig. \ref{fig5.7}. We observe that there is a slowly decaying contribution
for times $\tau\gg\Gamma_{22}^{-1}$ quite similar to the behaviour of the 
{\em next} photon probability density $I_1(t)$ shown in Fig. \ref{fig5.13}. 
One can show that in this regime we can approximate $g^{(2)}(\tau)$ by 
\begin{equation}
	g^{(2)}(\tau) \cong 1 + \frac{T_D}{T_L}exp\left\{\left(\frac{1}{T_D}
	+ \frac{1}{T_L}\right)\;\tau\right\} \;\; ,
	\label{532}
\end{equation}
where $T_D$ and $T_L$ are given by Eqs. (\ref{530a})-(\ref{531}). Therefore 
from the measurement of the $g^{(2)}$-function we can infer the existence of 
long 
bright and dark periods and one can determine their mean lengths.
The fact that this information is hidden in the intensity correlation
function is not surprising as it is proportional to the {\em a} photon rate
$I(\tau)$ which is related to the {\em next} photon rate $I_1(\tau)$ via
\begin{equation}
	I(\tau) = I_1(\tau) + \int_0^{\tau} d\tau' I(\tau-\tau')I_1(\tau')
	\;\; .
	\label{533}
\end{equation}
Later we will also show that the existence of bright and dark periods also 
leaves 
its fingerprints in the spectrum of resonance fluorescence as well
as in the absorption spectrum.

\begin{figure}[hbt]
\epsfxsize14.cm
\centerline{{\epsffile{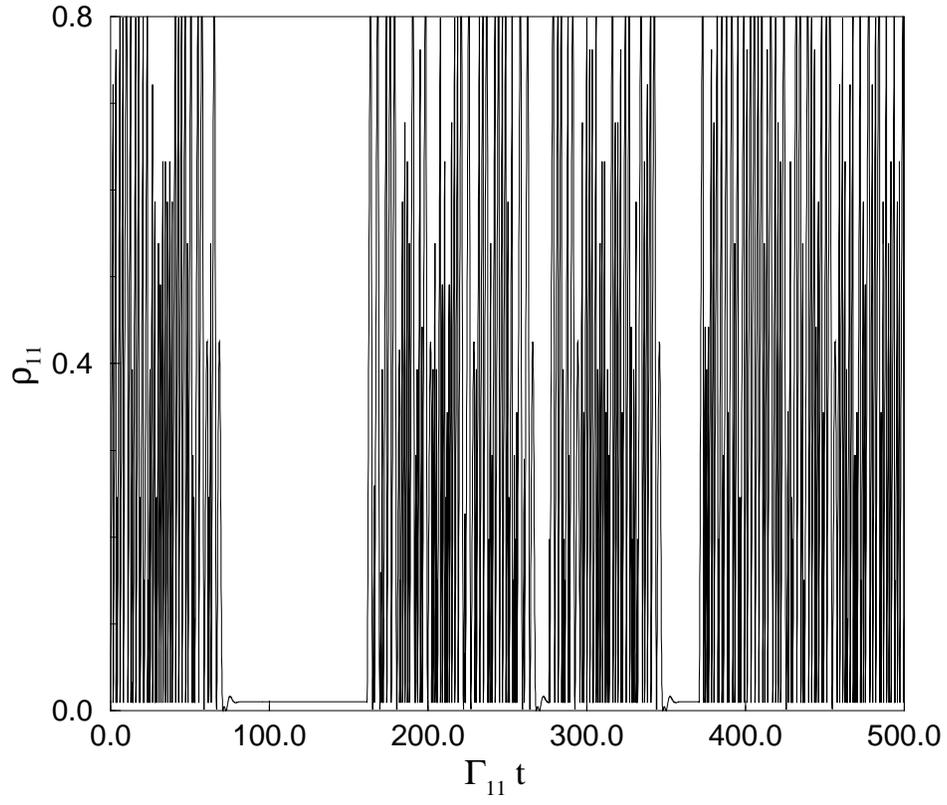}}}
\caption{\label{fig5.15}The time evolution of the population $\rho_{11}$
of the rapidly decaying level in the $V$ system. Periods where the time
evolution exhibits rapid Rabi oscillations interrupted by quantum jumps
can suddenly stop and lead to periods of no Rabi oscillations and no jumps.
The parameters are $\Omega_1=2\,\Gamma_{11},\Omega_2=0.2\,\Gamma_{11},
\Delta_1=0,\Delta_2=0$ and $\Gamma_{22}=0$. }
\end{figure}

\begin{figure}[hbt]
\epsfxsize14.cm
\centerline{{\epsffile{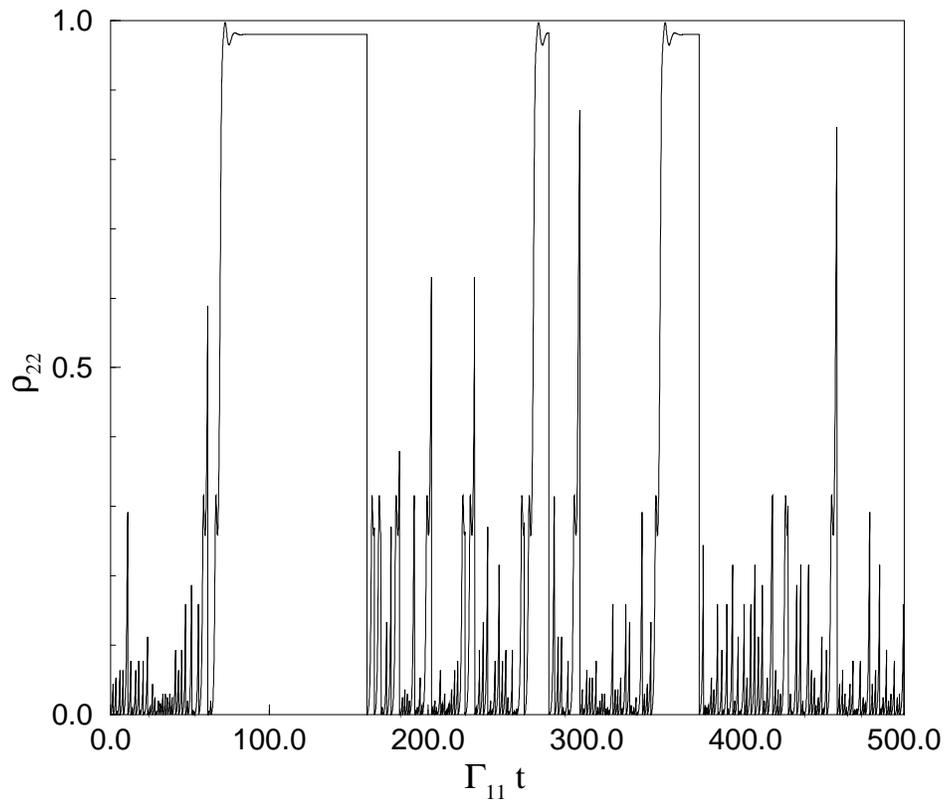}}}
\caption{\label{fig5.16} The same parameters as in Fig. \protect\ref{fig5.15}
but now for the population $\rho_{22}$ of the metastable state $2$. One clearly
observes that during a dark period the population evolves smoothly into the
metastable state $2$.}
\end{figure}

\begin{figure}[hbt]
\epsfxsize14.cm
\centerline{{\epsffile{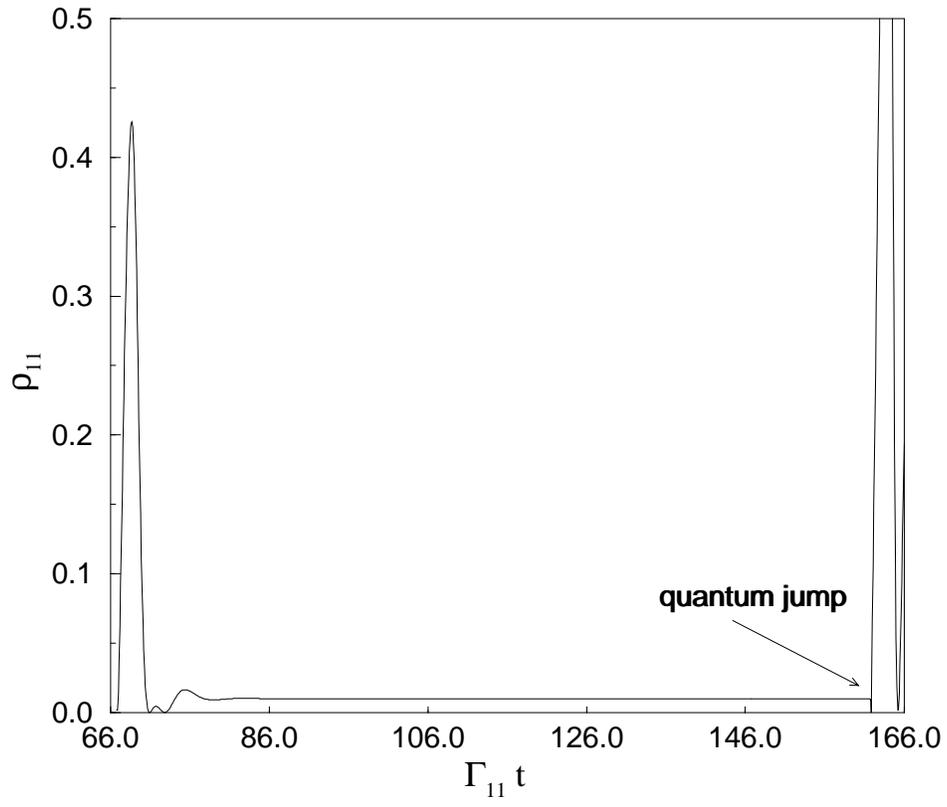}}}
\caption{\label{fig5.17} For the same parameters as in Fig.
\protect\ref{fig5.15}. The population of the unstable state $1$ in a dark
period. The population does not jump out of level $1$ but evolves continuously.
However, at the end of the dark period a jump occurs which is due to an 
emission for level $1$ because we assumed the shelving level $2$ to be stable.}
\end{figure}

After this analytical discussion of the photon statistics of the V-system 
we will
now investigate how the system time evolution for a single realization 
will typically look like. For the parameters $\Omega_1=2\,\Gamma_{11},
\Omega_2=0.2\,\Gamma_{11},\Delta_1=0,\Delta_2=0$ and $\Gamma_{22}=0$ 
we have plotted 
both the time evolution of the population $\rho_{11}$ of the unstable
level $1$ and population $\rho_{22}$ of the metastable level $2$. In 
Fig. \ref{fig5.15} one observes that in some regions a rapid change of 
$\rho_{11}$
accompanied by many detections takes place. However, there are also 
regions where no photon is found and most population is found in level
$2$ while there is still a remnant in level $1$. One should note that, as
seen in Fig. \ref{fig5.16} the population in level $2$ grows 
{\em continuously} and {\em does not}
jump from level $0$ into level $2$ as one might expect from a rate 
equation picture. Nevertheless there is a jump, but not at the beginning
of the dark period, but at its {\bf end} \cite{Wilser1,Garraway4} as we 
can see in Fig. \ref{fig5.17} where we 
have magnified the time evolution of $\rho_{11}$ in a dark period. The
jump occurs after a long time and marks the end of the dark period as after
the jump the population is in the ground state which is strongly coupled
to state $1$. It should be noted that in the simulation shown in
Figs. \ref{fig5.15}-\ref{fig5.17} the decay rate of the metastable level 
was assumed to be 
$\Gamma_{22}=0$ which implies that the dark period {\em always} ends
with an emission from level $1$. The continuous change of population 
towards the metastable state in a dark period nicely clarifies
the importance of the failure to detect photons (null measurements)
for the time evolution of the wavefunction. The failure to detect 
a photon provides us information about the system state which is 
described by the wave function. In this case the no-detection event tells
us that we are more likely to find the system in the metastable state.
Accordingly the wavefunction quickly (within several lifetimes of the
unstable level) tends towards the metastable state if we have not found
a photon for this time. The evolution is continuous as we assume 
(in the limit of the Markov approximation) continuous measurements
of the radiation field. Therefore the failure to detect a photon in 
an infinitesimal time interval $[t,t+\delta t]$ leads to an 
infinitesimal change proportional to $\delta t$. On the contrary the
detection of a photon leads to a discontinuous change of the
wave function because one quantum of energy has leaked out of the system 
into the environment. Therefore the system has to change discontinuously
and jumps back into its ground state. Again this change of the wavefunction
is due to our increased knowledge about the system.
\subsection{From quantum jumps to quantum state diffusion}
\begin{figure}[hbt]
\epsfxsize14.cm
\centerline{{\epsffile{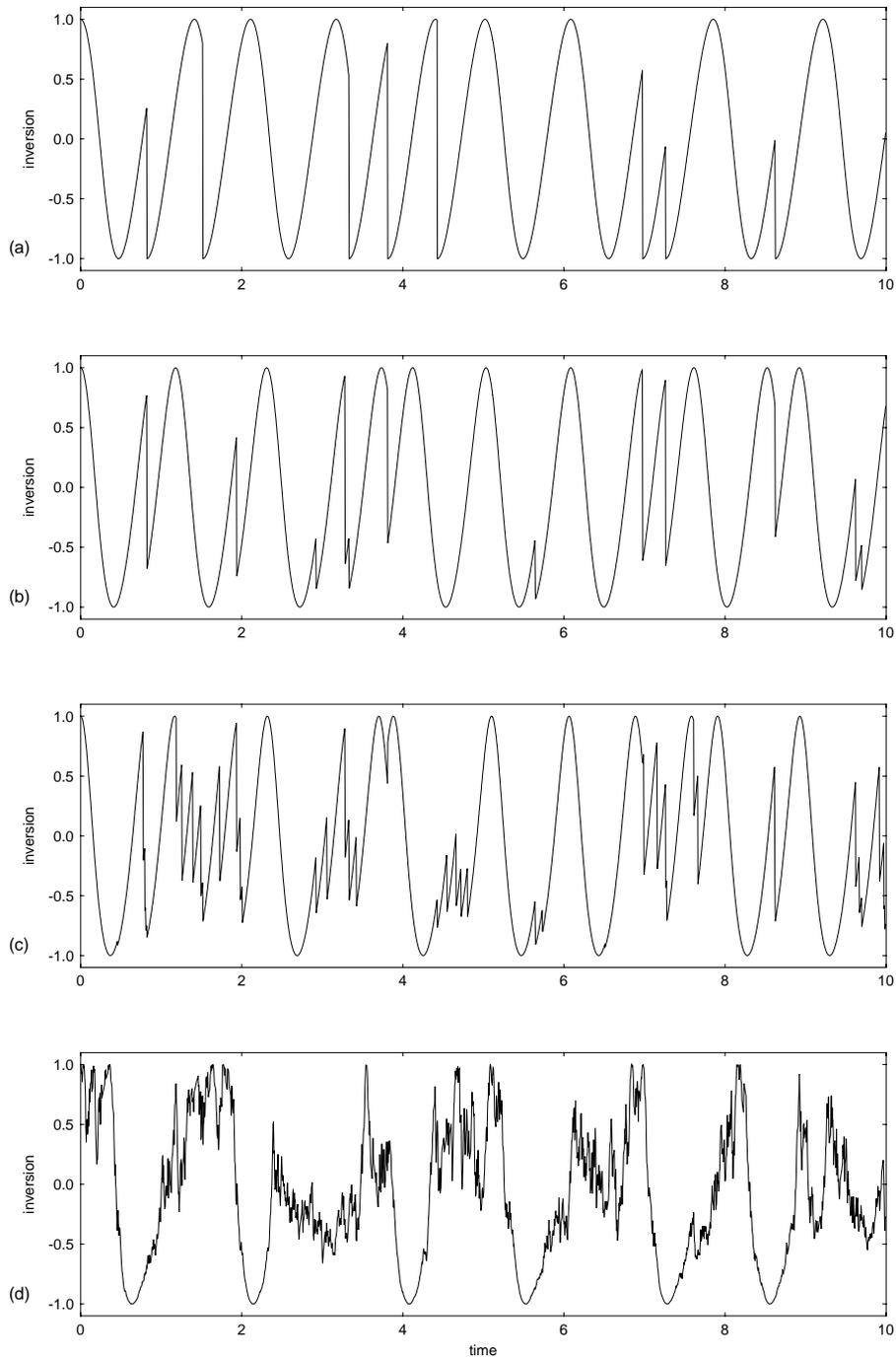}}}
\caption{\label{Granzow} Single realizations of a driven two-level system 
whose resonance fluorescence is observed in homodyne detection. 
The Rabi frequency is $\Omega=4\,\Gamma_{11}$ and the amplitude of the 
local oscillator is a) $\alpha=0$, b) $\alpha=0.5$, c) $\alpha=1$ and
$\alpha=10$. With increasing $\alpha$ jumps become more frequent and smaller
in amplitude. From \protect\cite{Granzow96}.}
\end{figure}

So far we have only illustrated the quantum jump approach to the photon
statistics of a single ion. In Section IV we have seen that other pictures
are possible, namely, the quantum state diffusion picture. Instead of 
presenting a large number of examples for the quantum state diffusion
model we rather want to illustrate how the transition from the quantum 
jump picture to the quantum state diffusion picture takes place. In Section IV
we have seen how to derive the quantum state diffusion equations from a quantum 
jump description of a balanced heterodyne detection experiment. In this 
derivation we had to assume the limit in which the photon number in the
local oscillator tends to infinity. In the following we will illustrate 
this limit by choosing a number of finite values for the photon number of
the local oscillator. To keep the following short analysis as simple as
possible we in fact do not consider the case of balanced heterodyne detection
but of homodyne detection \cite{Vogel1}. Instead of a cavity we will 
investigate the time evolution of a laser driven two-level system with upper
level $1$ and lower level $0$. 
We follow the presentation give in \cite{Granzow96}. For the simulations 
we need to know two quantities. The Lindblad operator for the homodyne 
detection scheme is given by
\begin{equation}
	L = |0\rangle\langle 1| + \alpha {\bf 1}
	\label{1500}
\end{equation}
where $\alpha$ is the field amplitude of the local oscillator. As the 
Lindblad operator is changed and depends on $\alpha$ so does the 
effective Hamilton operator between photon detections. We have
\begin{equation}
	H_{eff} = -\hbar \Delta |1\rangle\langle 1| - 
	\frac{\hbar\Omega}{2}(|1\rangle\langle 0| + |0\rangle\langle 1|)
	- i\hbar \Gamma_{11} ( L^{\dagger} L + 2\alpha^{*}L +|\alpha|^2)
	\label{1501}
\end{equation}
where $2\Gamma_{11}$ equals the Einstein coefficient of the upper level $1$,
$\Omega$ is the Rabi frequency on the $0\leftrightarrow 1$ transition
and $\Delta$ the detuning. The probability for  a jump in the time interval
$\delta t$ is given by
\begin{equation}
	p_c = 2\Gamma_{11} \left[ \langle L^{\dagger} L + 
	\alpha^{*} \langle L \rangle + \alpha \langle L^{\dagger} \rangle
	+|\alpha|^2 \right] \delta t
	\label{1502}
\end{equation}
These expressions can be derived analogously to the procedure 
that we applied in the description of the balanced heterodyne detection.
In Fig. \ref{Granzow} we have plotted single realizations of a two-level 
system with Rabi frequency $\Omega=4\Gamma$ and local oscillators amplitudes
a) $\alpha=0$, b) $\alpha=0.5$, c) $\alpha=1$ and d) $\alpha=10$.
One clearly observes that with increasing $\alpha$ the number of 
jumps increase while their amplitudes decrease. For $\alpha=10$ we already 
see a behaviour very close to that one would obtain in the limiting case
$\alpha=\infty$. 
\subsection{A decaying cavity}
\begin{figure}[hbt]
\epsfxsize14.cm
\centerline{{\epsffile{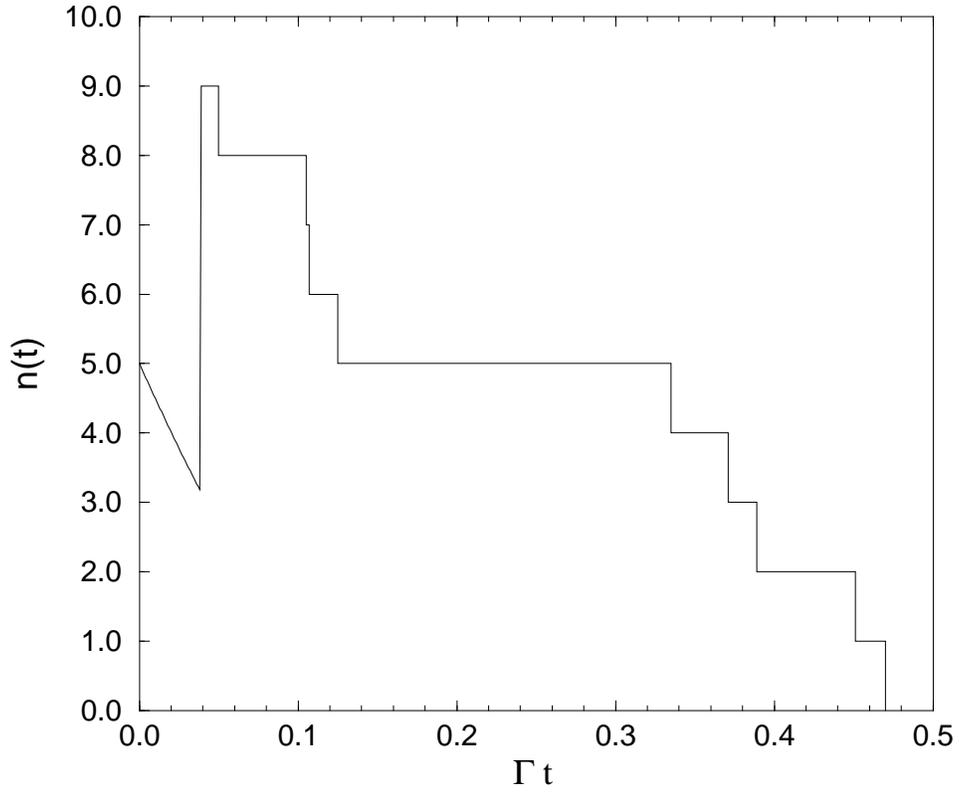}}}
\caption{\label{fig5.18} A cavity with decay rate $2\Gamma$ is prepared in
an initial
state $|\psi\rangle = (|0\rangle + |10\rangle)/2$. We plot the expected photon 
number of the state of the cavity as a function of time. Before we observe the
first photon outside the cavity, the expected photon number decreases. The 
first jump increases the expected photon number because we now know that 
the state has to be $|9\rangle$. Subsequently each photo detection decreases 
the expected photon number by one.}
\end{figure}

So far we have illustrated the quantum jump approach in the context of 
single ion resonance fluorescence. Now we would like to discuss a
different kind of problem, namely, cavity QED \cite{Haroche1}. 
That means we are interested
in the field states inside a cavity as well as the time evolution of
an atom interacting with such a cavity 
\cite{Grochmalicki2,Imamoglu1}.
Again these problems can be 
formulated in the framework of the quantum jump theory because both
the losses of the cavity to the outside world as well as spontaneous 
emission of an atom give rise to a master equation of Lindblad form
which can be simulated using wavefunctions. A derivation of the 
simulation equations is possible using the physical picture of
photon counters surrounding the system. For example, a broad band
photon counter outside the cavity will detect photon losses of the cavity. 
As an initial example to illustrate the physical insight that we gain from
the quantum jump approach, we would like to investigate the time evolution
of a field state in a lossy cavity. As an initial state we chose the
state
\begin{equation}
	|\psi\rangle = \frac{(|0\rangle + |10\rangle)}{\sqrt{2}}
	\; .
	\label{534}
\end{equation}
In an appropriate interaction picture, the time evolution under the condition 
that no photon has been found by the photon counter outside the cavity is 
given by the effective Hamiltonian
\begin{equation}
	H_{eff}	= -i\Gamma a^{\dagger} a
	\label{535}
\end{equation}
while the normalized state after the detection of a photon is given by
\begin{equation}
	|\psi_R\rangle = \frac{a|\psi\rangle}{\sqrt{\langle \psi| a^{\dagger}
	a|\psi \rangle}}\;\; .
	\label{536}
\end{equation} 
In Fig. \ref{fig5.18} the mean photon number for a single realisation 
is shown. Initially no jump takes place and the effect of this
failure to detect a photon outside the cavity is that the mean 
photon number inside the cavity decreases; it becomes more and more 
unlikely to find a photon inside the cavity because if there was a
photon in the cavity then it would leave the cavity, leading to a photon count.
However, in the simulation shown in Fig. \ref{fig5.18} we finally observe a
photon outside the cavity and at that moment we know that there 
have been photons in the cavity. Calculating the state after the
detection, given that we started with the superposition Eq. (\ref{534}) 
we find that the post--detection state is a Fock state containing $9$ photons.
This implies that the mean photon number {\em after} the photon detection 
is actually higher than before the detection. After this first
detection, the cavity continues to decay and now each detection of a photon 
outside the cavity decreases the mean number of photons inside the 
cavity, while now the number of photons remains invariant under the 
conditional time evolution from the relevant Fock states. 

In the previous example we saw that the exponential decay of the field
mode in the ensemble average is the result of the superposition of many
single realizations in which the cavity excitation changes discontinuously
at random times. Coherence between component parts of a superposition 
state changes in amplitude and is eventually destroyed. However, this is 
not the only
way in which coherence between superposition states is destroyed.
In the following example we show that the decay of coherence of a 
Schr{\"o}dinger cat of the form
\begin{equation}
	|\psi\rangle = (|\alpha\rangle + |-\alpha\rangle\rangle)/||.||
	\label{537}
\end{equation}
is due to a randomisation of the relative phase between the two coherent 
states while the modulus of the relative phase remains invariant under the
no--jump evolution
\cite{Garraway2}. In each individual realization the cat state remains a 
cat state. Although
the amplitude of the two coherent states decays the relative phase
between the two coherent states just changes its sign. In fact if the 
state before the detection is given by Eq. (\ref{537}) then after the
detection of a photon we have the state
\begin{eqnarray}
	|\psi_+\rangle &=& a|\psi\rangle/||.||
	\nonumber\\	
	               &=& (|\alpha\rangle - |-\alpha\rangle\rangle)/||.||
	\;\; .\label{538}
\end{eqnarray}
The conditional time evolution when no photon has been found is given by
\begin{equation}
	|\psi(t)\rangle = (|\alpha e^{-\Gamma(t-t')} \rangle -
			   |-\alpha e^{-\Gamma(t-t')}\rangle\rangle)/||.||
	\label{539}
\end{equation}
so that the amplitudes of the coherent states decay exponentially while
their relative phase remains unaffected. Averaging over many realizations,
however, leads to a decaying relative phase as random phases tend to
cancel out. It is interesting to note that the rate of decay for the 
relative phase of the Schr{\"o}dinger cat is given by $\Gamma|\alpha|^2$
while the amplitude decay rate is given by $\Gamma$. Therefore in the 
ensemble average the cat decoheres before the amplitudes of its 
constituents are significantly affected \cite{Garraway3}. 

\begin{figure}[hbt]
\epsfxsize14.cm
\centerline{{\epsffile{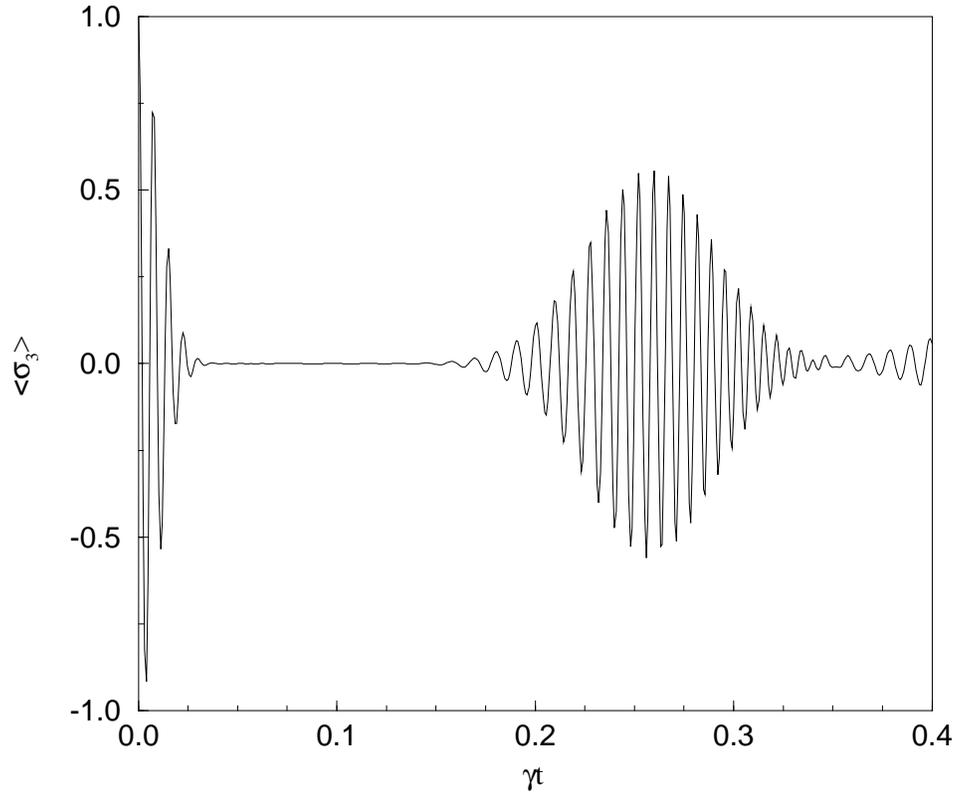}}}
\caption{\label{fig5.19a} Revivals in the inversion $\langle \sigma_3
\rangle= \langle |1\rangle\langle 1| - |0\rangle\langle 0| \rangle/2$ 
of a two-level atom in a cavity with an initial field prepared
in a coherent state $|\alpha \rangle$ with $\alpha =4$. The parameters 
of the simulation were $\, \Delta =0,\, g / \gamma = 100 $, where $g$ is
the atom--field coupling constant. $\gamma$ is used to scale time while
in the next figure it is the decay constant of the cavity.  }
\end{figure}

\begin{figure}[hbt]
\epsfxsize14.cm
\centerline{{\epsffile{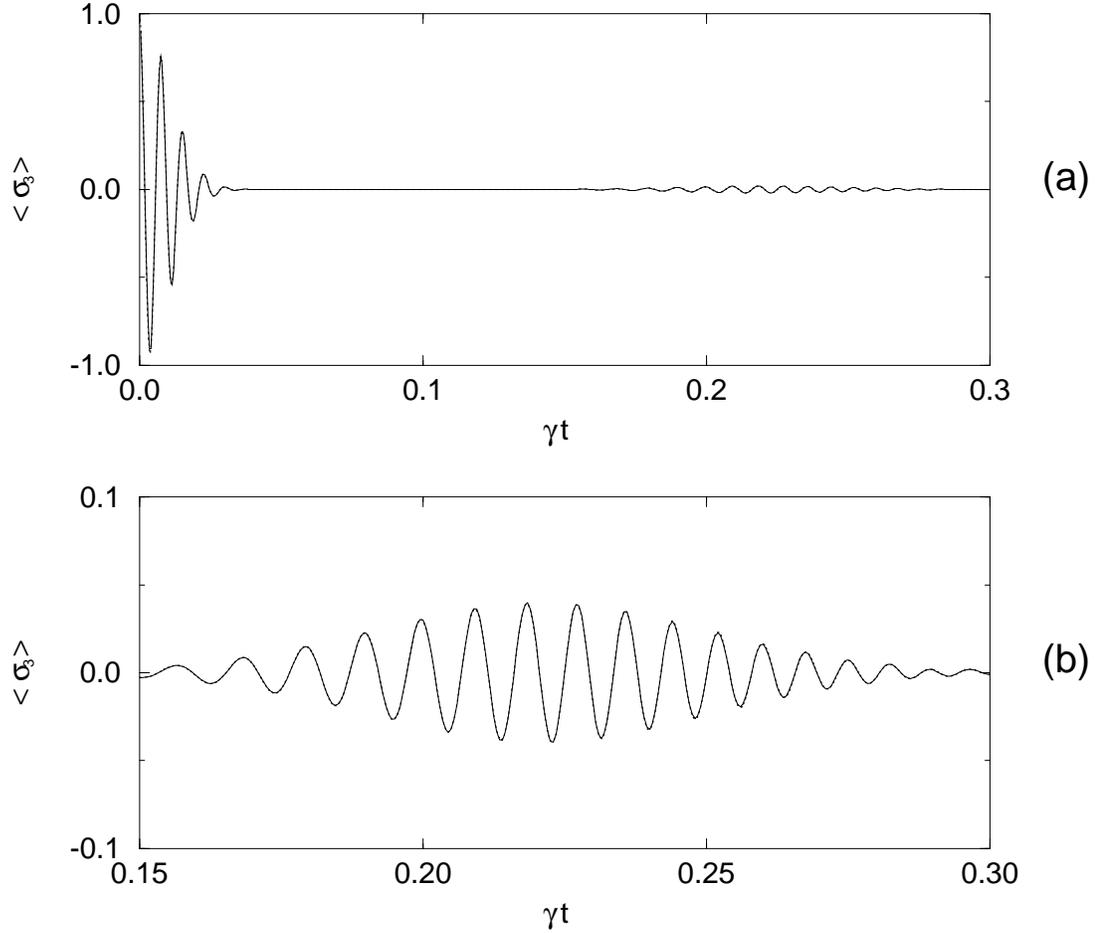}}}
\vspace*{1.cm}
\caption{\label{fig5.19} Revivals in the inversion $\langle \sigma_3
\rangle = \langle |1\rangle\langle 1| - |0\rangle\langle 0| \rangle/2$ 
of a two-level atom in a cavity with an initial field prepared
in a coherent state $|\alpha \rangle$ with $\alpha =4$. Even a modest
decay rate of the cavity leads to a rapid destruction of the revivals.
The lower figure (b) shows the revival. It was obtained by a quantum
jump simulation using $320000$ runs and is indistinguishable from the
numerical integration of the master equation. The parameters of the
simulation were 
$(\, \Delta =0,\, g / \gamma = 100,\, \gamma \delta t = 5 
\times 10^{-4}\,.)$. }
\end{figure}

\begin{figure}[hbt]
\epsfxsize14.cm
\centerline{{\epsffile{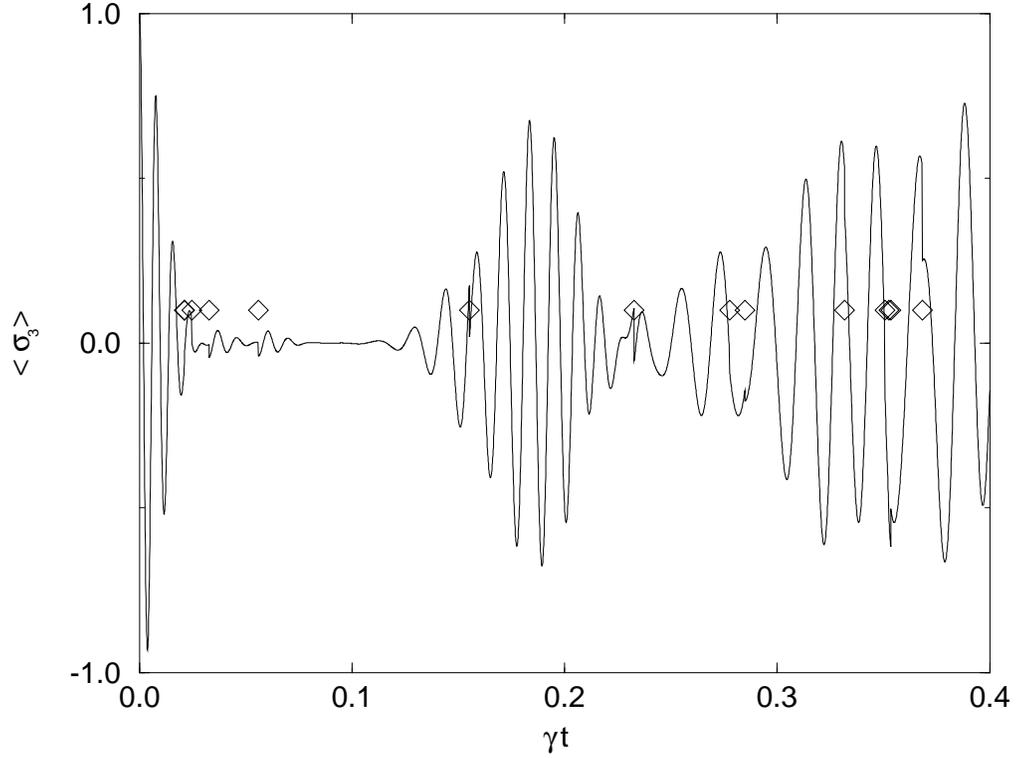}}}
\caption{\label{fig5.20} For the same parameters as in 
Fig. \protect\ref{fig5.19} we plot a single realization of the time
evolution. The diamonds mark the instants where a photon was detected
outside the cavity. Note that the revivals in a single realization 
survive. The quantum jumps lead to a phase jump of the time
evolution of the inversion of the atom which leads to a destruction
of the revivals in the ensemble average. From \protect\cite{Garraway4}}
\end{figure}

Another example that shows that the loss of coherence can be due to phase
shifts at random times is the phenomenon of revivals in the Jaynes-Cummings 
model (Shore and Knight (1993) and references therein). In this phenomenon 
a two-level atom is initially in its excited state $1$ while the cavity 
field is in a coherent state with amplitude $\alpha$,
\begin{equation}
	|\psi\rangle = |1\rangle\langle 1| \otimes 
		       |\alpha\rangle\langle\alpha | \;\; .
	\label{540}
\end{equation}
This initial product state becomes entangled (Ekert and Knight (1995) and
references therein)
and the average excitation of 
the atom rapidly tends towards $1/2$. The reason for this is the fact that
the frequency of Rabi oscillations depends on the number of photons in
the field mode \cite{Narozhny1}. The different Rabi 
frequencies quickly decohere so that 
one observes an average excitation of $1/2$ of the atom. However, as the
frequencies are discrete they partially rephase after a longer time and
the excitation of the atom rises again. This revival is shown , for example,
in Fig. \ref{fig5.19a}. Revivals of this kind have been observed 
experimentally by \cite{Meekhof1,Brune1} and of the micromaser kind by 
Rempe {\em et al} (1987). If the cavity is damped, however, these revivals 
are
much weaker and will vanish for strong damping as shown in Fig. \ref{fig5.19}
where we have chosen the same parameters as in Fig. \ref{fig5.19a} but with
an additional cavity damping constant of $\gamma=0.01 g$, where $g$ is the 
atom--field coupling constant. 
We assumed that the atom does not decay spontaneously. No substantial 
revivals can be observed \cite{Barnett1}. 
What will we see for an individual realization? If we assume that the
atom does not decay spontaneously we obtain the effective Hamiltonian 
\begin{equation}
	H_{eff} = H_{sys} - i\hbar\gamma a^{\dagger} a
	\label{541}
\end{equation}
that generates the conditional time evolution of the atom cavity system if
no photons are detected outside the cavity. $H_{sys}$ generates the free 
evolution of an undamped atom cavity field system while the second term
on the right hand side describes the damping of the cavity at a rate 
$2\gamma$. The state after the detection of a cavity photon which is
\begin{equation}
	|\psi_+\rangle = \frac{a|\psi\rangle}{||.||}\;\; .
	\label{542}
\end{equation} 
A single realization for the parameters of Fig. \ref{fig5.19} is shown as 
in Fig. \ref{fig5.20}. Now the revivals persist,
however, each jump that occurs at a random time introduces a phase shift.
An average taken over many individual realizations then leads to
a quick decay of the coherence and revivals are not observed any more.
A similar analysis can be performed for a lossless cavity with an atom that
may spontaneously decay \cite{Burt1}. 

In these examples we only considered undriven atoms in a cavity. 
Driven two-state systems inside a cavity can of course also be investigated
using the quantum jump approach \cite{Alsing91,Tian1} but we do not 
discuss this in detail here. 
\subsection{Other applications of the quantum jump approach}
So far we have illustrated the quantum jump approach for a number of
examples in single ion physics and cavity QED. The quantum jump approach
can and has been applied to a large number of problems which we cannot 
discuss in detail here. Amongst these examples are the discussion of 
correlations between quantum jumps in two stored ions. 
Such correlations were observed in an experiment \cite{Sauter3} 
in which the intermittent fluorescence of two $Ba^{+}$ ions in an ion trap 
were investigated and correlations were found between the quantum jumps in 
the two ions exceeding those expected from independent jumps. However, 
Itano {\em et al} (1988) could not confirm these correlations in their
experiment \cite{Itano0}. Theoretical
investigations 
\cite{Agarwal2,Hendriks1,Lawande2,Lawande4,Lewenstein1,Lewenstein2} 
predicted that correlations should be small if the ions are separated by 
many optical wavelengths. If the ions are closer than an optical wavelength
together, correlations might be observable. For such small separations even 
two two-level ions could exhibit quantum jumps \cite{Yamada1,Kim4,Lawande5}.
 
Another example of the application of the Quantum jump approach is the
investigation of the Quantum Zeno effect \cite{Misra1}. The Quantum Zeno
effect was measured in an experiment by Itano {\em et al} (1990)
which was originally proposed by Cook (1988). The experiment used a 
two-level system. The ground state population was measured via coupling 
the lower level strongly to a rapidly decaying third level. Observation 
of resonance fluorescence then indicates that the system was in its 
ground state. This experiment was investigated theoretically independently
by Power and Knight \cite{Power1,Power2,Power3}, by Beige and Hegerfeldt
\cite{Beige1,Beige2} and by Mahler and coworkers \cite{Mahler1}
using the quantum jump approach and, e.g., by
Frerichs and Schenzle (1991) using Bloch equations. The investigations
using the quantum jump approach helped to understand the experiment from
the single particle point of view and will be important in the analysis
of future Quantum Zeno experiments \cite{Plenio3} using single ions instead 
of around $5000$ in the experiment by Itano {\em et al} (1990).

Cohen-Tannoudji and coworkers have applied the quantum jump approach to
the problem of lasing without inversion 
\cite{Cohen-Tannoudji4,Cohen-Tannoudji5,Cohen-Tannoudji3} to understand the
processes involved from the point of view of single systems.

Apart from these more analytical applications of the quantum jump approach 
also its numerical usefulness has been demonstrated , for example, in 
numerical simulations of laser cooling experiments in two or three 
dimensions (see, for example, \cite{Castin1,Marte1,Marte2}). It is the fact 
that
the quantum jump approach allows the description of the system using 
a wave function instead of the density operator that made these
investigations possible. Both the gain in computational speed and the saving
in memory space is considerable as in a quantum jump simulation only
$N$ differential equations have to be propagated instead of $N^2$ in the
density operator simulation. 

The quantum state diffusion model, apart from its importance in 
fundamental issues such as the measurement process or intrinsic 
decoherence, is now widely used to investigate the transition between
classical and quantum behaviour and in the field of quantum chaos 
(see, for example, \cite{Rigo96,Spiller1,Brun3}). As we have already mentioned 
in Section IV the quantum state diffusion model also exhibits interesting
localisation properties in both position and phase space 
\cite{Gisin6,Gisin7,Herkommer96,Percival1} which can be
used to implement very fast simulation algorithms \cite{Schack0,Schack1}.
Similar localisation properties exist also for variants of the quantum jump
approach \cite{Holland2}.

A more exotic application of the quantum jump approach, or more precisely
of the experiments in which quantum jumps were observed, is the fact
that these experiments can provide a ''perfect'' random number generator.
This is because these experiments allow the observation of single 
quantum jumps which occur at absolutely random times due to the fundamental
indeterminacy of quantum mechanics \cite{Erber1,Erber2}. Whether this idea
is useful is, however, doubtful although in principle there could be 
applications, e.g., in cryptography.
\subsection{The spectrum of resonance fluorescence and single system dynamics}
\begin{figure}[hbt]
\epsfxsize14.cm
\centerline{\epsffile{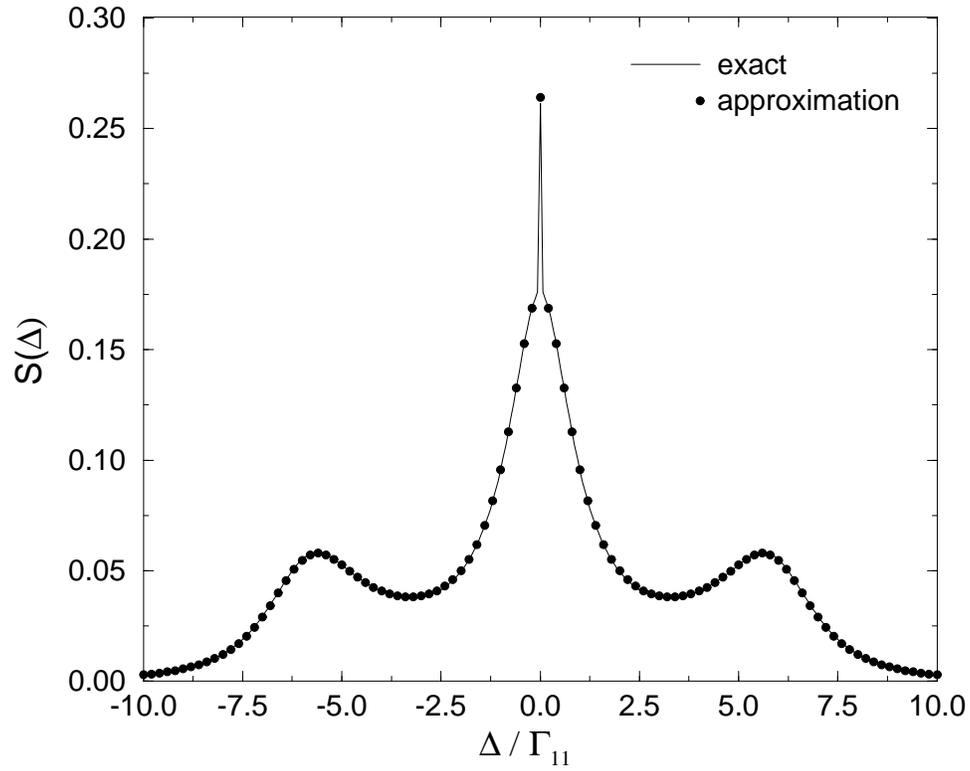}}
\caption{\label{figspec1} The spectrum of resonance fluorescence on the
$0\leftrightarrow 1$ transition. The parameters are
$\Omega_1=6\,\Gamma_{11},\Omega_2=0.4\,\Gamma_{11}$ and 
$\Delta_1=\Delta_2=0.$ One clearly observes the Mollow triplet and the
additional narrow peak in the spectrum. From \protect\cite{Hegerfeldt7}.}
\end{figure}

\begin{figure}[hbt]
\epsfxsize14.cm
\centerline{\epsffile{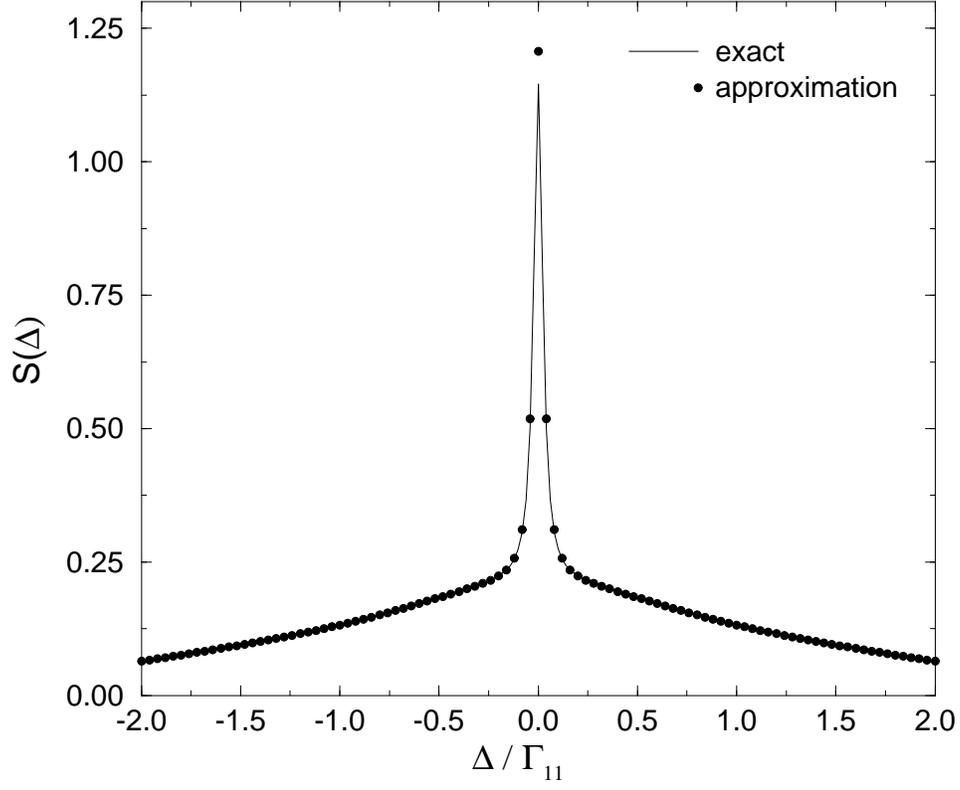}}
\caption{\label{figspec2} The spectrum of resonance fluorescence on the
$0\leftrightarrow 1$ transition. The parameters are
$\Omega_1=2\,\Gamma_{11},\Omega_2=0.2\,\Gamma_{11}$ and $\Delta_1=\Delta_2=0.$
Due to the weaker driving as compared to \protect\ref{figspec1} the sidebands
in the Mollow triplet are no longer resolved. The additional narrow peak 
is again observable and has a higher relative weight. From 
\protect\cite{Hegerfeldt7}.}.
\end{figure}

\begin{figure}[hbt]
\epsfxsize14.cm
\centerline{\epsffile{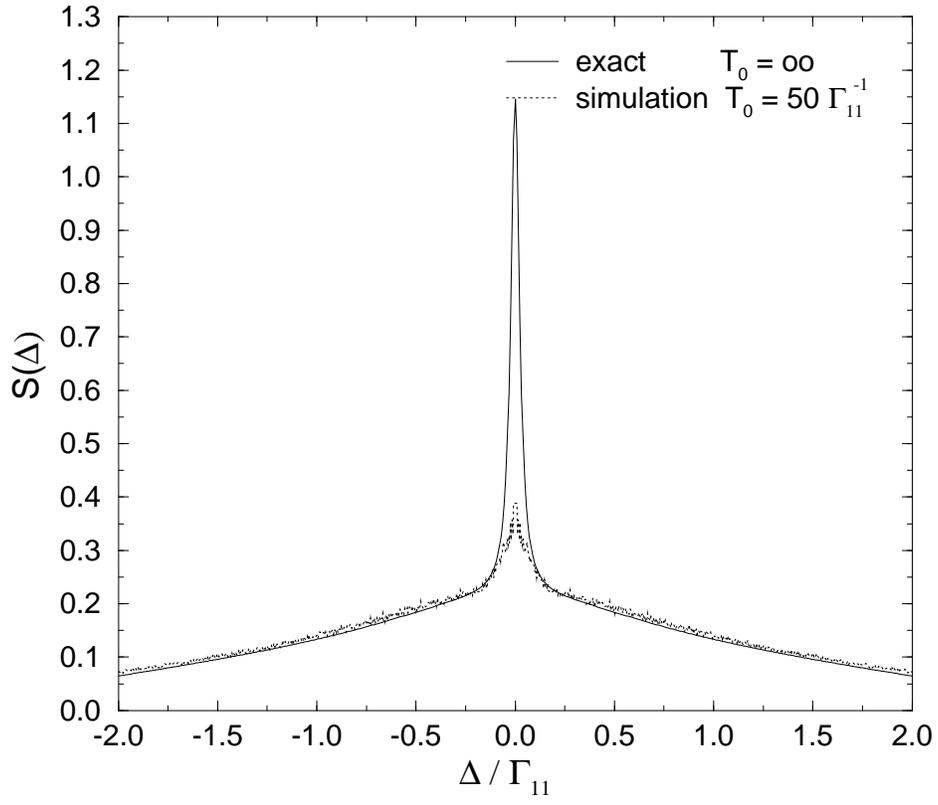}}
\caption{\label{simspec2} Simulation of the conditional spectrum of 
resonance fluorescence
for $T_0=\infty$ and $T_0=50\,\Gamma_{11}^{-1}$. One observes that only 
the narrow central peak is affected by the conditioning of the photon
statistics while the wings of the spectrum are essentially independent
of the choice of $T_0$.}.
\end{figure}

So far our examples of the quantum jump approach were limited to the 
investigation 
of the photon statistics of the radiation emitted by a 
system and of the internal dynamics of a system conditioned on the 
measurement record observing the radiation emitted by the system.
As a further application of the quantum jump approach we would now like to 
consider the spectrum of resonance fluorescence of a three-level atom in
the V-configuration as shown in Fig. \ref{fig2.1}. This system, whose 
photon statistics we already 
discussed in the context of bright and dark periods exhibits interesting 
features in the spectrum of resonance fluorescence on the strong 
$0\leftrightarrow 1$ transition. In the following we will discuss briefly the 
ensemble behaviour of the spectrum and then show how we can understand this
behaviour from the point of view of the single system dynamics. We will
closely follow the analysis in Hegerfeldt and Plenio (1995b) and Plenio 
(1994,1996).
We consider the system shown in Fig. \ref{fig2.1} and assume that the 
Rabi frequencies $\Omega_i$ of the lasers and the decay constants 
$\Gamma_{ii}$ of the two upper levels satisfy the following conditions.
\begin{itemize}
\item
The Rabi frequency of laser driving the $0\leftrightarrow 2$ transition is
weak, that is
\begin{equation}
	\Omega_2^2 \ll \frac{1}{4}\frac{16\,\Delta_2^2\Gamma_{11}^2+
	(\Omega_1^2+4\Delta_2(\Delta_1-\Delta_2))^2}
	{\Gamma_{11}^2+(\Delta_1-\Delta_2)^2}
	\label{cond1}
\end{equation}
which simplifies for $\Omega_1\gg\Delta_1,\Delta_2$ to
\begin{equation}
   \Omega_2^2 \ll \left(\frac{\Omega_1^2}{2\,\Gamma_{11}}\right)^2\; .
   \label{cond1a}
\end{equation}
\item
Spontaneous emission on the $2\leftrightarrow 0$ transition should be
negligible, that is
\begin{equation}
   \Gamma_{22} \ll \frac{\Omega_1^2\Omega_2^2\Gamma_{11}}
   {16\,\Delta_2^2\Gamma_{11}^2+
   \{\Omega_1^2+4\,\Delta_2(\Delta_1-\Delta_2)\}^2}\; .
   \label{cond2}
\end{equation}
\end{itemize}
These conditions have the following interpretation. If Eq. (\ref{cond1}) is 
satisfied the V system exhibits long bright and dark periods as discussed
in Eqs. (\ref{520}) -(\ref{531}). If Eq. (\ref{cond2}) is satisfied stimulated
transitions from level $2$ to level $0$ are much more frequent than 
spontaneous emissions on the same transition. Under the assumption of Eqs.
(\ref{cond1})-(\ref{cond2}) the spectrum of resonance fluorescence takes on 
the following approximate analytical form
\begin{eqnarray}
   S(\Delta) &=& \frac{1}{\pi\rho_{11}^{ss}}
   Re\int_0^{\infty}\!\!\! d\tau\,
   \langle\sigma_{10}(\tau)\sigma(0)\rangle_{ss} e^{-i\Delta \tau}
   \nonumber\\[.2cm]
   &=& S_{coh}(\Delta) + S_{Mollow}(\Delta) + S_{peak}(\Delta)
   \label{spec1}
\end{eqnarray}
with
\jot.35cm 
\begin{eqnarray}
   B &=& 6\,\Gamma_{11}^2-2\,\Omega_1^2-2\,\Delta_1^2 \;\; ,
   \label{spec2}\\
   C &=& \Omega_1^4
   +2\,\Omega_1^2\Gamma_{11}^2+9\,\Gamma_{11}^4+\Delta_1^4+
   2\,\Delta_1^2
   \Omega_1^2-6\,\Gamma_{11}^2\Delta_1^2 \;\; ,
   \label{spec3}\\
   D &=& \Gamma_{11}^2(\Omega_1^2+2\,\Delta_1^2+2\,\Gamma_{11}^2)^2 \;\; ,
   \label{spec4}\\[.1cm]
   A_{p} &=&
   2 \frac{(\Delta_1^2+\Gamma_{11}^2)\,
   \left[(\Delta_1-\Delta_{2})^2+\Gamma_{11}^2\right]\,
   \left[(\Omega_1^2-4\Delta_2^2+4\Delta_1\Delta_2)^2 + 
   16\Delta_2^2\Gamma_{11}^2\right]}
   {\Omega_1^2\Omega_2^2\Gamma_{11}
   \left[\Omega_1^2+2\,\Delta_2^2+4\,\Delta_1^2-4\,\Delta_1\Delta_2
   +4\,\Gamma_{11}^2\right]^2} \;\; ,
   \label{spec5}\hspace*{.5cm}\\[.1cm]
   \Gamma_{p}  &=&
   \frac{2\,\Omega_1^2\Omega_2^2 \Gamma_{11}
   \left( 2\Delta_2^2 +4\Delta_1^2-4\Delta_1\Delta_2+
   \Omega_1^2+4\Gamma_{11}^2 \right)}
   {\left[
   (\Omega_1^2-4\,\Delta_2^2+4\Delta_1\Delta_2)^2 + 
   16\,\Delta_2^2\Gamma_{11}^2\right]
   (\Omega_1^2 +2\Delta_1^2+ 2\,\Gamma_{11}^2)}
   \label{spec6}
\end{eqnarray}
and
\begin{eqnarray}
   \pi\, S_{coh}(\Delta) \!\!&=&\! \pi\, \frac{2(\Delta_1^2+
   \Gamma_{11}^2)}
   {4\,\Gamma_{11}^2+\Omega_1^2+2\,\Delta_2^2+4\,\Delta_1^2-
   4\Delta_1\Delta_2}
   \,\delta(\Delta) \;\; ,
   \label{spec7}\\[.1cm]
   \pi\, S_{Mollow}(\Delta) \!\!&=&\!
   \frac{\Gamma_{11}\Omega_1^2(\Omega_1^2+2\,\Delta^2+8\,\Gamma_{11}^2)}
   {2\left( \Delta^6 + B \Delta^4 + C \Delta^2 + D \right)} \;\; ,
   \hspace*{1.25cm}\label{spec8}\\[.1cm]
   \pi\, S_{peak}(\Delta) \!\!&=&\!
   \frac{A_{p}\Gamma_{p}^2}{\Delta^2+\Gamma_{p}^2}\: .
   \label{spec9}
\end{eqnarray}
The contributions Eqs. (\ref{spec7}) and (\ref{spec8}) just represent the 
well known Mollow triplet and the Rayleigh peak
\cite{Mollow1,Mollow2,Mollow3,Mollow4,Mollow5}. These contributions 
are expected and are well--understood. The third contribution however, 
a narrow Lorentzian,
is a new feature in the spectrum of resonance fluorescence. In Figs. 
\ref{figspec1} and \ref{figspec2} we see the spectra for the cases of
strong driving of the $0\leftrightarrow 1$ transition and for a medium
strong laser on the same transition. One clearly observes the narrow peak 
, which should not be confused with the Rayleigh peak, in both spectra.
In the following
we will focus our attention to this new feature and use it as an example
for the application of the quantum jump approach for spectral information 
and to illustrate how ensemble properties can be understood better from a
single system point of view.
One might, for example, be interested in the following question. What happens
if we observe the spectrum of resonance fluorescence in a bright period 
exclusively? Experimentally this may be measured by triggering the 
spectrometer with the broadband counter. The spectrometer will be opened if 
we detect photons in the broadband counter at a sufficiently high rate and 
it will be closed when we fail to find photons for a certain time $T_0$. 
Using this time constant $T_0$ we have a means either to observe the spectrum
in a bright period ($T_0\approx 10/\Gamma_{11}$) or with no restriction on
the emission times ($T_0=\infty$), i.e., the ensemble spectrum. Changing 
$T_0$ we can continuously switch between the two regimes. The spectrum of
resonance fluorescence can be simulated using the quantum jump approach 
\cite{Hegerfeldt8}. In Fig. \ref{simspec2} the effect of the
change in $T_0$ is clearly visible. For small $T_0$ the
amplitude of the narrow central peak becomes small. However only the central
part of the spectrum is strongly affected while the wings of the spectrum 
are more or less independent of $T_0$. This is an example of conditional 
fluorescence spectra and it shows that spectra can be dependent on the
conditions imposed on the photon statistics. Here we have shown the effect
for the spectrum of resonance fluorescence but similar results may be 
obtained by investigating the absorption spectrum of a weak probe beam on the
strongly driven $0\leftrightarrow 1$ transition \cite{Plenio2}. Again a 
narrow peak is observed which vanishes if we try to measure the same feature 
in a bright period of the system. 
After we have investigated the spectrum of resonance fluorescence of the 
V system and some of its properties, especially the occurrence of the 
narrow peak, it is now instructive to see how we can understand this feature
from the photon statistics of the single atomic system. The photon statistics
are provided by the quantum jump approach. Explicit results have been given
in this section in Eqs. (\ref{520}) -(\ref{531}). 
 
Now that we know that the V configuration exhibit light
and dark periods for those parameters for which the narrow peak in
the spectrum appears we proceed with a somewhat simplified model of
resonance fluorescence of the V-system. We assume that the lengths of 
light and dark periods obey exactly Poissonian distributions, e.g., the 
probability density that a light period has the length $t$ is
\begin{equation}
	I_L(t) = \frac{1}{T_L}\,e^{-t/T_L}
	\label{62}
\end{equation}
and for a dark period to have a length of $t$ is
\begin{equation}
	I_D(t) = \frac{1}{T_D}\,e^{-t/T_D}\; .
	\label{63}
\end{equation}
These probability densities have been derived from a rate equation model
of the time evolution by Cook and Kimble \cite{Cook2}.
Additionally we assume that in a bright period the system behaves
exactly like a two-level system made up of the two levels $0$ and $1$. 
This assumption has to be checked and turns out to be good in the case
of the spectrum of resonance fluorescence treated here. However, it is
not always correct as a careful analysis of the absorption spectrum 
on the strong $0 \leftrightarrow 1$ transition proves and deviations 
from this idealised assumption may lead to significant contributions 
in the absorption spectrum \cite{Plenio2}.

It is well known \cite{Loudon1} that the stationary spectrum of resonance
 fluorescence of such a two-level system is given by
\begin{equation}
	S^{(2)}(\Delta) = \lim_{T\rightarrow\infty}\frac{C}{T}
	\int_0^{T} \!\!dt_1\int_0^{T} \!\!dt_2 \langle E^{(-)}(t_1)
	E^{(+)}(t_2)\rangle e^{-i\Delta(t_1-t_2)}\;\; ,
	\label{64}
\end{equation}
where $E^{(-)}(t)$ and $E^{(+)}(t)$ denote the negative and positive
frequency part of the electric field operator and $C$ is chosen in 
such a way that with 
\begin{equation}
	E^{(+)}(t) \sim \sigma_{01}(t)
	\label{65}
\end{equation}
one obtains
\begin{equation}
	S^{(2)}(\Delta) = \frac{2\Gamma_{11}}{\pi}\mbox{Re}
	\int_0^{\infty}\!d\tau e^{-i\Delta\tau}
	\langle\sigma_{10}(\tau)\sigma_{01}(0)\rangle_{ss}\; .
	\label{65a}
\end{equation}

For the nonnormalized resonance fluorescence spectrum of the
two-level system one then obtains
\begin{equation}
	S^{(2)}(\Delta) = 
	\frac{\Gamma_{11}\Omega_1^2}{\Omega_1^2+2(\Delta_1^2+
	\Gamma_{11}^2)}
	\left\{ \frac{1}{\pi}
	\frac{\Gamma_{11}\Omega_1^2(\Omega_1^2+2\Delta^2+
	8\Gamma_{11}^2)}
	{2(\Delta^6+B\Delta^4+C\Delta^2+D)} +
	\frac{2(\Delta_1^2+\Gamma_{11}^2)}
	{\Omega_1^2+2(\Delta_1^2+\Gamma_{11}^2)}\,\delta(\Delta)     
	\right\}\;\; ,
	\label{66}
\end{equation}
where the constants $B,C$ and $D$ are given in Eqs. (\ref{spec2})-(\ref{spec4}).
As the light emitted by the atom switches on and off due to the
light and dark periods we assume that the electric field radiated by
the 3-level configurations is given by
\begin{equation}
	\hat{E}^{(\pm)}(t) := E^{(\pm)}(t)f(t)\;\; ,
	\label{67}
\end{equation}
where $f(t)$ is a two state jump process with values 0 and 1. The
probability density for the length of a period where $f(t)=0$ is given
by Eq. (\ref{62}) and that for $f(t)=1$ is given by Eq. (\ref{63}). Therefore
we have to substitute 
\begin{equation}
	\hat\sigma_{ij}(t) := \sigma_{ij}(t)f(t)
	\label{67a}
\end{equation}
in Eq. (\ref{65a}) and expect the spectrum
of the three-level configurations to be given by
\begin{equation}
	S^{(3)}(\Delta) = \frac{2\Gamma_{11}}{\pi}
	\mbox{Re} \int_0^{\infty}\!\!d\tau\, e^{-i\Delta\tau}
	\langle\langle\hat\sigma_{10}(\tau)\hat\sigma_{01}(0)\rangle
	\rangle_{ss}\;\; ,
	\label{68}
\end{equation}
where $\langle\langle.\rangle\rangle$ denotes both the quantum
mechanical average as well as the stochastic average over all
realizations of the process $f(t)$. This can be simplified to
\begin{eqnarray}
	S^{(3)}(\Delta) &=& \frac{2\,\Gamma_{11}}{\pi}\mbox{Re}
	\int_0^{\infty}\!\!d\Delta\, e^{-i\Delta\tau} 
	\langle \sigma_{10}(\tau)\sigma_{01}(0)\rangle_{ss}
	\langle f(\tau)f(0)\rangle_{stoch.}
	\nonumber\\
	&=& \int_{-\infty}^{\infty}\!\!d\Delta'\, 
	S^{(2)}(\Delta-\Delta')
	k(\Delta')
	\label{69}
\end{eqnarray}
with 
\begin{eqnarray} 
	k(\Delta) &=& \frac{1}{\pi}
	\int_{0}^{\infty}\! d\tau e^{-i\Delta\tau}
	\langle f(\tau)f(0)\rangle_{stoch.}
	\nonumber\\
	&=&
	\frac{T_L}{T_L+T_D}\left\{\frac{T_L}{T_L+T_D}\delta(\Delta)
	+\frac{T_D}{T_D+T_L}{\cal L}(\Delta)\right\}\;\; ,
	\label{70}
\end{eqnarray}
where
\begin{equation}
	{\cal L}(\Delta) =\frac{1}{\pi}
	\frac{\frac{1}{T_D}+\frac{1}{T_L}}{(\frac{1}{T_D}+
	\frac{1}{T_L})^2
	+\Delta^2}\; .
	\label{70a}
\end{equation}
Because both $T_D$ and $T_L$ (Eqs. (\ref{530})-(\ref{531})) are assumed to 
be much longer than the
mean emission time of a two-level system, which is of the order of
$\Gamma_{11}^{-1}$, we can deduce from this
\begin{eqnarray}
	S^{(3)}(\Delta) &\cong&
	\left(\frac{T_L}{T_D+T_L}\right)^2
	\left\{ 
	\,S^{(2)}(\Delta) +
	\frac{T_D}{T_L}\,S_{inc}^{(2)}(\Delta) +
	\frac{T_D}{T_L}{\cal L}(\Delta)
	\frac{2(\Delta_1^2+\Gamma_{11}^2)}
	{\Omega_1^2+2(\Delta_1^2+\Gamma_{11}^2)}\right\}
	\nonumber\\
	&=& \frac{T_L}{T_D+T_L}\left\{
	\!S_{inc}^{(2)}(\Delta)
	+ \frac{T_L}{T_D+T_L}S_{coh}^{(2)}(\Delta) +
	\frac{T_D}{T_D+T_L}\frac{2(\Delta_1^2+\Gamma_{11}^2)}
	{\Omega_1^2+2(\Delta_1^2+\Gamma_{11}^2)}{\cal L}(\Delta)
	\!\right\}\;\; .
	\label{71}
\end{eqnarray}
This expression has to be compared with the results Eqs. 
(\ref{spec1})-(\ref{spec9}). In fact inserting Eqs. 
(\ref{530})-(\ref{531}) into Eq. (\ref{71}) yields an expression
in very good agreement with the spectra of the V-system.
The width $\Gamma_p$ of the resulting narrow peak is 
\begin{equation}
	\Gamma_p = \frac{1}{T_D} + \frac{1}{T_L}\; .
	\label{72}
\end{equation}
The amplitude $A_p$ of the narrow peak in the normalized spectrum is
then given by
\begin{equation}
	A_p = \frac{2(\Delta_1^2+\Gamma_{11}^2)}
	{\Omega_1^2+2(\Delta_1^2+\Gamma_{11}^2)}
	\,\frac{T_D^2 T_L}{(T_D+T_L)^2}\; .
	\label{73}
\end{equation}
Now the interpretation of the narrow peak is obvious. The stochastic
modulation of the resonance fluorescence due to dark periods leads to
a partial broadening of the Rayleigh peak. The small width of the 
additional peak in the resonance fluorescence spectrum is then
understood  from the fact that the correlation time $\tau_c$ of the
random telegraph process f(t) which simulates the light and dark
periods is very large. In fact it is easy to show that
\begin{equation}
	\tau_c = \left(\frac{1}{T_D}+\frac{1}{T_L}\right)^{-1}\; .
\end{equation}
This results in an extremely narrow distribution in frequency space
with a width $\Gamma_p=\tau_c^{-1}$. It is this structure which is
observable in  the spectrum of  resonance fluorescence.
It is interesting to note that the narrow peak is not easily interpreted
in a dressed-states picture. Indeed, in secular approximation and for 
$\Delta_1=\Delta_2=0$ one obtains a zero weight for the narrow peak. 
Even if we tune the laser on the $0\leftrightarrow 2$ transition
to resonance with one of the dressed states, i.e.. 
$\Delta_2=\pm\Omega_1/2$, the weight of the narrow peak comes out much 
too small. In fact it would then be predicted to be proportional to
$\Omega_2^2$. The narrow peak found here has clearly a different
origin than the line narrowing effects found by others
\cite{Manka1,Narducci1,Narducci2} where the systems do not exhibit bright
and dark periods in their resonance fluorescence but simply a decrease in 
intensity. The 
existence of bright and dark periods in the resonance fluorescence leads
to a narrow peak in the spectrum but it should be noted that the 
converse is not necessarily true. There are situations (eg a  laser driven 
TLS in a squeezed vacuum \cite{Swain1}) in which the spectrum of resonance 
fluorescence
exhibits narrow peaks but where the photon statistics does not show bright
and dark periods. The reason for that can be found in the fact that the
photon statistics is governed by the population decay rates while the
spectrum of resonance fluorescence is derived from the $g^{(1)}(t)$ 
correlation function which is strongly influenced by the decay rates 
of the coherences in the system. In the case of a TLS in squeezed light 
the coherences have a slowly decaying component while the population decay 
rate is still large.
\subsection{Spontaneous emission in quantum computing}
\begin{figure}[hbt]
\epsfxsize14.cm
\centerline{\epsffile{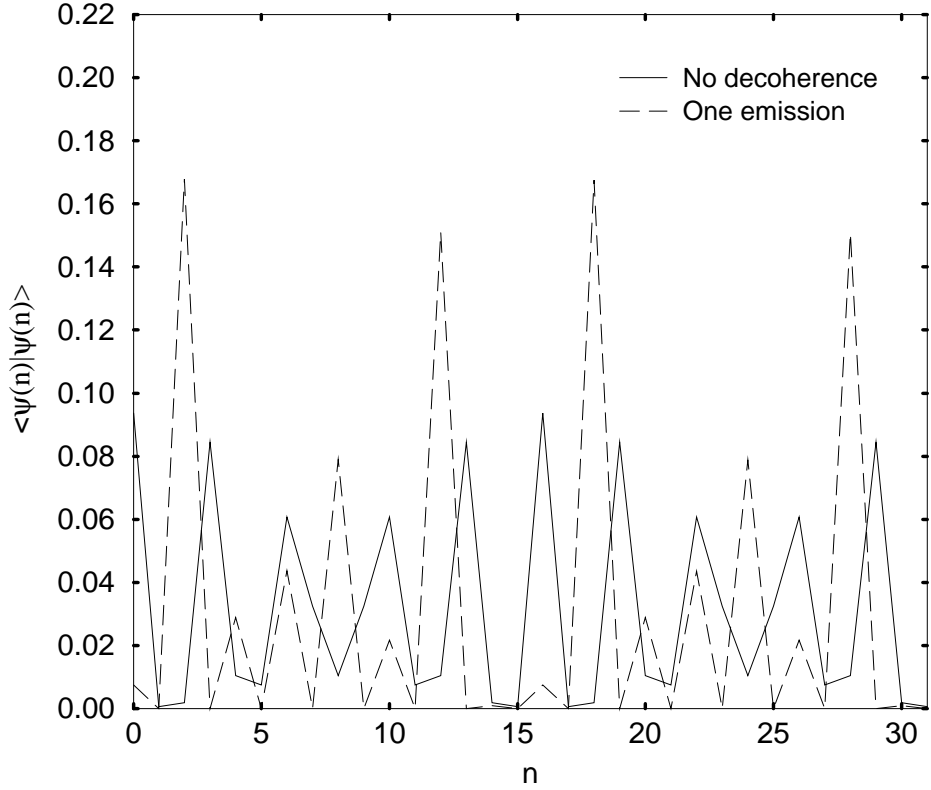}}
\caption{\label{fourier1}
Results of a discrete Fourier transform (DFT) of a function 
$f(n)=\delta_{8,(n mod 10)}$ with $n=0,1,\ldots,31$. The solid line
is the result for a quantum computer with stable qubits and represents
the correct result. The dashed line shows the result of the same 
computation using a quantum computer with unstable qubits, one of which
has suffered a spontaneous emission during the calculation. The results 
clearly differ and show the impact of a single spontaneous emission on a 
quantum computation. For the parameters chosen on average the quantum 
computer will suffer one emission per DFT, ie $\tau_{sp}=T$ in this case.
From \protect\cite{Plenio5}.}
\end{figure}

\begin{figure}[hbt]
\epsfxsize14.cm
\centerline{\epsffile{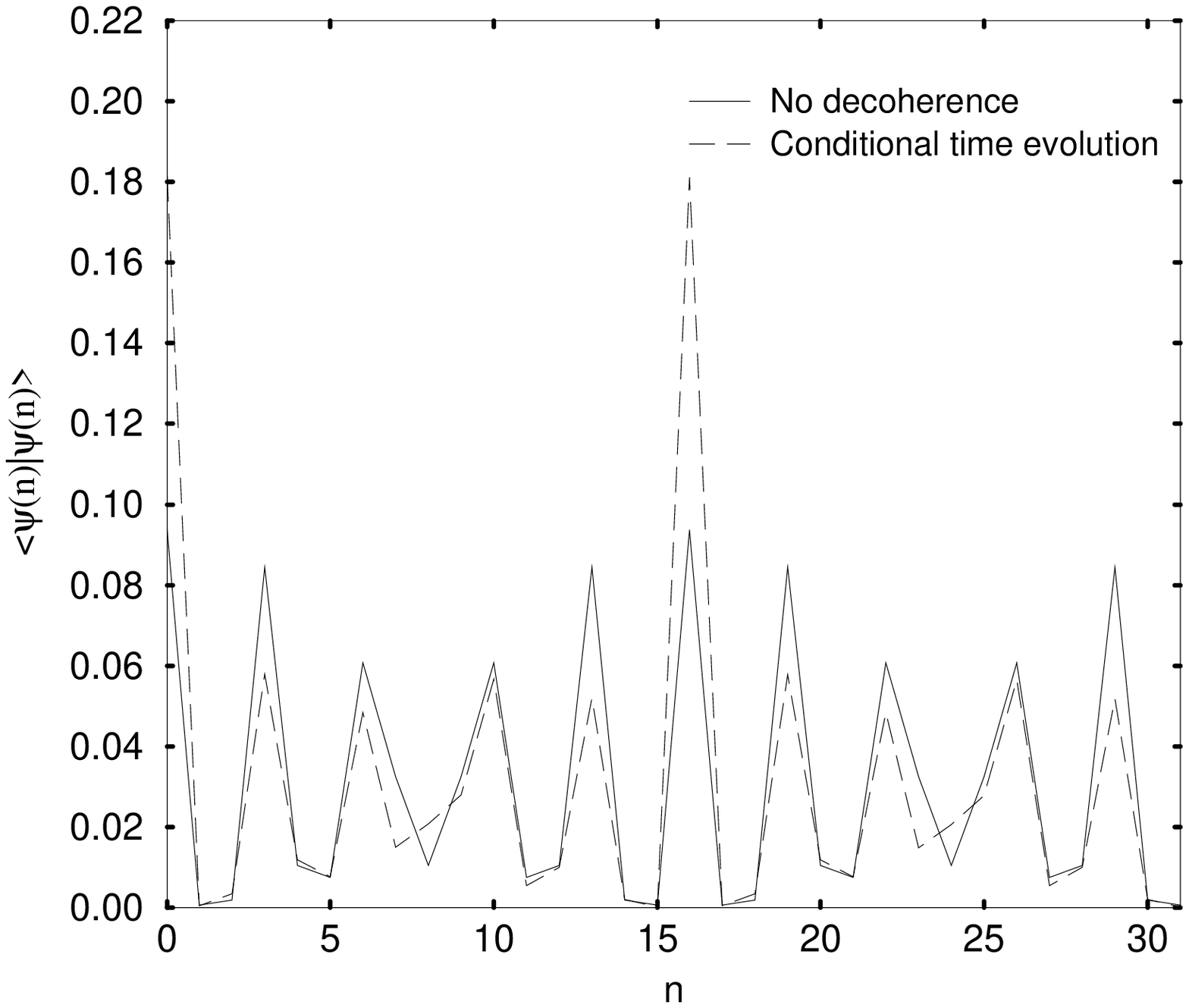}}
\caption{\label{fourier2}
The same quantum computation as in Fig. 5. The solid line again represents 
the result using a quantum computer with stable qubits, while the dashed
line shows the result using a quantum computer with unstable qubits. This
time, however, the unstable quantum computer does not suffer an emission
during the whole calculation. Again the results differ illustrating the 
impact of the conditional time evolution between spontaneous emissions.
From \protect\cite{Plenio5}. }
\end{figure}

As a last application of the quantum jump approach we would like to 
investigate the influence that spontaneous emission has on the function
of a quantum computer. Quantum computing is an idea that has attracted 
enormous interest in the last two years. It was elevated from the obscurity 
of theoretical idealisation to possible practical applications by the 
discovery of an algorithm by Shor (1994) (see also Ekert and Josza (1996)
and references therein) that allows the factorization of 
large numbers in polynomial time on a quantum computer as compared to 
the exponential time required on a classical computer. However, this 
achievement in computing speed is only possible due to the massive use of
the superposition principle in quantum mechanics. The basic idea is that
a qubit (a two level system) can exist in a superposition of the two values
$0$ and $1$. $N$ qubits can then exist in a superposition of $2^N$ values.
These values can then be manipulated by a series of unitary transformations.
A final readout can then provide us with information about global properties 
of the function implemented by the unitary transformation. Such
a global property of the function is, for example, its period which a quantum
computer determines by performing a discrete Fourier transform, something 
which can also be implemented on a quantum computer \cite{Coppersmith1}.
In the course of its time evolution (computation) the quantum computer
evolves into a highly entangled state. 
However, it is known that any entangled state is very 
sensitive to dissipation. Therefore one expects 
that the quantum computer is highly sensitive to spontaneous emission
and other sources of dissipation.
This is in fact the case and currently research in quantum error correction
methods has concentrated on attempts to find methods to correct for these 
errors \cite{Shor2,Steane1,Calderbank1}. The quantum jump approach is 
ideally suited For the investigation of this problem because it is able 
to describe single runs of a quantum computer rather than an ensemble of 
quantum computers as in the Bloch equation description. We will
illustrate the problems caused by spontaneous emission in quantum computers 
by examining the example of the discrete Fourier transform mentioned above.
There are two effects contributing to the decoherence of the quantum computer.
One is the obvious fact that a spontaneous emission will destroy at least
part of the coherence in the quantum computer. The second decohering 
effect, however, originates from the conditional time evolution between
spontaneous emissions \cite{Plenio5}. We have learnt above that this time 
evolution is 
actually different from the unit operation because even the non-detection
of a photon represents a gain in our knowledge about the system. Therefore
the wavefunction of the system, which represents our knowledge
of the system, has to change. This leads to a distortion of the time 
evolution, which will then affect the result of our calculation.
In Figs. \ref{fourier1} and \ref{fourier2} we simulate a quantum computer 
(we do not go into detail concerning its implementation here) that 
calculates the discrete Fourier transform of a function that is evaluated 
at $32$ points \cite{Plenio5}. 
The resulting square modulus of the wavefunction of the quantum computer 
is compared to the exact result obtained from a absolutely stable 
quantum computer. The function on which we perform the DFT is given for
definiteness in this example by 
$f(n)=\delta_{8,(n\,mod  10)}$ for $n=0,1,\ldots,31$. One can implement
the Hamilton operators (in the Lamb-Dicke limit) for all the necessary 
quantum gates in a linear ion trap \cite{Cirac1} to realize this DFT. In
addition to the coherent time evolution, possible spontaneous emissions 
from the upper levels of the ions are taken into account
but all other sources of loss are neglected. 

In Fig. \ref{fourier1} one emission has taken place during the calculation time
of the quantum computer. If we compare the resulting wavefunction with the
correct wavefunction, we observe a marked difference between the two. 
In Fig. \ref{fourier2} we show the wavefunction of an unstable quantum 
computer which has not suffered a spontaneous emission during the 
calculation of the DFT. We clearly see that even when no spontaneous 
emission has taken place, the wavefunction of the quantum computer differs
substantially from the correct result. This difference becomes stronger
and stronger the larger the ratio between the computation time $T$
and the spontaneous lifetime $\tau_{sp}$ of the quantum computer becomes.
Therefore the wavefunction of the quantum computer will be sufficiently
close to the correct result only if the whole computation is finished 
in a time $T$ that is much shorter than the spontaneous lifetime $\tau_{sp}$
of the quantum computer.

The fact that even one spontaneous emission will usually make the result
of the quantum computation completely incorrect can then be used to derive 
stringent upper limits to the numbers that can be factorized on a quantum 
computer \cite{Plenio4,Plenio5}. This again shows that knowledge of the 
single system behaviour gained from the quantum jump approach gives us 
useful new insights into important properties of quantum systems. 
\section{Conclusions}
\label{Conclusions}
Recent work in quantum optics has forced us to re-examine the dynamics
of individual quantum systems in which single realizations (single atoms
or trapped ions, single cavity field modes and so on) are described in
quantum mechanics. In these situations, the dynamics is always dissipative, 
and leaves a record of its history accessible in the wider world of the 
outside environment. If this record is read, so that we acquire specific 
information, then we can associate a specific quantum trajectory to that 
conditional record. In this way we "unravel" the dissipative master equation 
into a family of records. We have reviewed the new technique developed to 
describe this unravelling which go under the name of quantum jumps, Monte 
Carlo wavefunction simulations and so on. We have further demonstrated 
how they can be used to describe entirely non-classical behaviour in a 
wide range of situations in quantum optics. Future applications will surely 
emerge from these powerful approaches.
\section*{Acknowledgements}
\addcontentsline{toc}{section}{Acknowledgements}
We are grateful to J. Dalibard, B.M. Garraway, G.C. Hegerfeldt, M.S. Kim,
R. Loudon, K. M{\o}lmer, D.T. Pegg, I.C. Percival, W.L. Power, R. Reibold, 
D. Sondermann, J. Steinbach, S. Stenholm, R.C. Thompson and 
K. W{\'o}dkiewicz for 
discussions over the years on quantum jumps. This work was supported 
in part by the UK Engineering and Physical Sciences Research Council, 
the European Union and the Alexander von Humboldt Foundation.
%
%
%
%
%
%
%
%
%
%
%
%
%
%
%
%
%
%
%
%
%
%
%
%
%
%
%
%
%
%
%
%
%
%
%
%
\newpage
%
%
%
%
%
%
%
%
%
%
%
%
%
%
%
%
%
%
%
%
%
%
%
%
%
%
%
%
%
%
%
%
%
%
%
%
%
%
%
%
%
%
%
%
%
%
%
%
%
%
%
%
%
%
%
%
%
%
%
%
%
%
%
%
%
%
%
%
%
%
%
%
%
%
%
%
%
%
%
%
%
%
%
%
%
%
%
%
%
%
%
%
%
%
%
%
%
%
%
%
%
%
%
%
%
%
%
%
%
%
%
%
%
%
%
%
%
%
%
%
%
%
%
%
%
%
%
%
%
%
%
%


\begin{references}
\addcontentsline{toc}{section}{References}
%
\harvarditem{Agarwal}{1974}{Agarwal1} 
Agarwal, G. S., 1974, 
{\em Quantum Statistical Theories of Spontaneous Emission and their 
Relation to other Approaches}, Springer Tracts Mod. Physics {\bf 70} 
(Springer, Berlin), p. 1.
%
\harvarditem{Agarwal {\em et al.}}{1988a}{Agarwal2}
 Agarwal, G.S., S.V. Lawande, and R. D'Souza, 1988a, 
IEEE Journal of Quant. Electronics {\bf 24}, 1413.
%
\harvarditem{Agarwal {\em et al.}}{1988b}{Agarwal3}
Agarwal, G.S., S.V. Lawande, and R. D'Souza, 1988b, 
Phys. Rev. A {\bf 37}, 444.
%
\harvarditem{Alsing and Carmichael}{1991}{Alsing91}
Alsin, P., and H.J. Carmichael, Quant. Opt. {\bf 3}, 13.
%
%
%
\harvarditem{Arecchi {\em et al.}}{1986}{Arecchi1}
Arecchi, F.T., A. Schenzle, R.G. De Voe, K. Jungmann, and 
R.G. Brewer, 1986, Phys. Rev. A {\bf 33}, 2124.
%
\harvarditem{Autler and Townes}{1955}{Autler1}
Autler, S.H., and C.H. Townes, 1955, Phys. Rev. {\bf 100}, 703.
%
%
\harvarditem{Barchielli}{1986}{Barchielli1}
Barchielli, A., 1986, Phys. Rev. A {\bf 34}, 1642.
%
\harvarditem{Barchielli}{1993}{Barchielli3}
Barchielli, A., 1993, Quantum Optics {\bf 2}, 423.
%
\harvarditem{Barchielli and Belavkin}{1991}{Barchielli2}
Barchielli, A., and V.P. Belavkin, 1991, J. Phys. A {\bf 24},1495.
%
\harvarditem{Barnett and Knight}{1986}{Barnett1}
Barnett, S.M., and P.L. Knight, 1986, Phys. Rev. A {\bf 33}, 2444
%
\harvarditem{Beige and Hegerfeldt}{1996a}{Beige1}
Beige, A., and G.C. Hegerfeldt, 1996, Phys. Rev. A {\bf 53}, 53.
%
\harvarditem{Beige and Hegerfeldt}{1996b}{Beige2}
Beige, A., and G.C. Hegerfeldt, 1996, Quant. Optics {\bf 8}, 999.
%
%
\harvarditem{Bergquist {\em et al.}}{1986}{Bergquist1}
Bergquist, J.C., R.B. Hulet, W.M. Itano, and 
D.J. Wineland, 1986, Phys. Rev. Lett. {\bf 57}, 1699.
%
\harvarditem{Bergquist {\em et al.}}{1994}{Bergquist2}
Bergquist, J.C., W.M. Itano, and D.J. Wineland, in 
{\em Frontiers in Laser Spectroscopy}, Proceedings of 
the International School of Physics ''Enrico Fermi'',
Course CXX, ed. by T.W. H{\"a}nsch and M. Inguscio (North
Holland, Amsterdam, 1994), pp. 359-376.
%
\harvarditem{Blatt {\em et al.}}{1986}{Blatt1}
Blatt, R., W. Ertmer, P. Zoller, and J.J. Hall, 1986, Phys. Rev. A
{\bf 34}, 3022.
%
\harvarditem{Bouwmeester {\em et al.}}{1994}{Bouwmeester1}
Bouwmeester, D., R.J.C. Spreeuw, G. Nienhuis, and 
J.P. Woerdman, 1994, Phys. Rev. A {\bf 49}, 4170.
%
%
%
%
\harvarditem{Brun}{1996}{Brun1}
Brun, T.A., preprint quant-ph/9606025 (1996)
%
\harvarditem{Brun and Gisin}{1996}{Brun2}
Brun, T.A., and N. Gisin, 1996, J. Mod. Opt. {\bf 43}, 2289.
%
\harvarditem{Brun {\em et al}}{1996}{Brun3}
Brun, T.A., I.C. Percival, and R. Schack, 1996, J. Phys. A {\bf 29}, 2077.
%
\harvarditem{Brune {\em et al}}{1996}{Brune1}
Brune, M., F. Schmidt-Kaler, A. Maali, J. Dreyer, E. Hagley, 
J.M. Raimond and S. Haroche, 1996, Phys. Rev. Lett. {\bf 76}, 1800.
%
\harvarditem{Burt and Gea-Banacloche}{1996}{Burt1}
Burt, T.C., and  J. Gea-Banacloche, 1996, Quant. Opt. {\bf 8}, 105.
%
\harvarditem{Calderbank and Shor}{1996}{Calderbank1}
Calderbank, A.R., and P.W. Shor, 1996, Phys. Rev. A {\bf 54}, 1098.
%
\harvarditem{Carmichael}{1993a}{Carmichael2}
Carmichael, H.J., 1993a, {\em An Open Systems 
Approach to Quantum Optics}, Lecture Notes in Physics, (Springer, Berlin).
%
\harvarditem{Carmichael}{1993b}{Carmichael3}
Carmichael, H.J., 1993b, Phys. Rev. Lett. {\bf 70}, 2273.
%
\harvarditem{Carmichael}{1994}{Carmichael4}
Carmichael, H.J., 1994, in {\em Quantum Optics VI},
edited by J. D. Harvey and D. F. Walls (Springer, Berlin), p. 219.
%
\harvarditem{Carmichael {\em et al.}}{1989}{Carmichael1}
Carmichael, H.J., S. Singh, R. Vyas, and P.R. Rice,
1989, Phys. Rev. A {\bf 39}, 1200.
%
\harvarditem{Castin {\em et al.}}{1992}{Castin92}
Castin, Y. , J. Dalibard, and K. M{\o}lmer, (1992), {\em Atomic Physics 
$13$}, eds. H. Walther, T.W. H{\"a}nsch, and B. neizert (A.I.P.,
New York, 1993).
%
\harvarditem{Castin and M{\o}lmer}{1995}{Castin1}
Castin, Y., and K. M{\o}lmer, 1995, Phys. Rev. Lett. {\bf 74}, 3772.
%
%
\harvarditem{Cirac and Zoller}{1995}{Cirac1}
Cirac, J.I., and P. Zoller, 1995, Phys. Rev. Lett. {\bf 74}, 4091.
%
\harvarditem{Cohen-Tannoudji and Dalibard}{1986}{Cohen-Tannoudji1}
Cohen-Tannoudji, C., and J. Dalibard, 1986, Europhys. Lett. {\bf 1}, 441. 
%
\harvarditem{Cohen-Tannoudji {\em et al.}}{1992}{Cohen-Tannoudji2}
Cohen-Tannoudji, C., J. Dupont-Roc, and G. Grynberg, 
1992, {\em Atom-Photon Interactions.} (John Wiley \& Sons Inc.)
%
\harvarditem{Cohen-Tannoudji {\em et al.}}{1992a}{Cohen-Tannoudji4}
Cohen-Tannoudji, C., B. Zambon, and E. Arimondo, 
1992a, Compt. Rend. de L'Academie des Sciences Serie II, {\bf 314}, 1139.
%
\harvarditem{Cohen-Tannoudji {\em et al.}}{1992b}{Cohen-Tannoudji5}
Cohen-Tannoudji, C., B. Zambon, and E. Arimondo, 
1992b, Compt. Rend. de L'Academie des Sciences Serie II, {\bf 314}, 1293.
%
\harvarditem{Cohen-Tannoudji {\em et al.}}{1993}{Cohen-Tannoudji3}
Cohen-Tannoudji, C., B. Zambon, and E. Arimondo,
1993, J. Opt. Soc. Am. {\bf B 10}, 2107. 
%
\harvarditem{Cook}{1981}{Cook1}
Cook, R.J., 1981, Phys. Rev. A {\bf 23}, 1243.
%
\harvarditem{Cook}{1988}{Cook3}
Cook, R.J., 1988, Phys. Scr. {\bf T 21}, 49.
%
\harvarditem{Cook}{1990}{Cook4}
Cook, R.J., 1990, Prog. Opt. {\bf 28}, 361.
%
\harvarditem{Cook and Kimble}{1985}{Cook2}
Cook, R.J., and H.J. Kimble, 1985, Phys. Rev. Lett. {\bf 54}, 1023.
%
\harvarditem{Coppersmith}{1994}{Coppersmith1}
Coppersmith, D., 1994, IBM Research Report No. RC19642 (1994)
%
\harvarditem{Dagenais and Mandel}{1978}{Dagenais1}
Dagenais, M., and L. Mandel, 1978, Phys. Rev. A {\bf 18}, 2217.
%
\harvarditem{Dalibard {\em et al.}}{1992}{Dalibard1}
Dalibard, J., Y. Castin, and K. M{\o}lmer, 1992, Phys. Rev. 
Lett. {\bf 68}, 580.
%
\harvarditem{Davies}{1969}{Davies1}
Davies, E.B., 1969, Comm. Math. Phys. {\bf 15}, 277.
%
\harvarditem{Davies}{1970}{Davies2}
Davies, E.B., 1970, Comm. Math. Phys. {\bf 19}, 83.
%
\harvarditem{Davies}{1971}{Davies3}
Davies, E.B., 1971, Comm. Math. Phys. {\bf 22}, 51.
%
\harvarditem{Davies}{1976}{Davies4}
Davies, E.B., {\em Quantum Theory of Open Systems} (Academic, New York, 1976).
%
\harvarditem{Dehmelt}{1975}{Dehmelt1}
Dehmelt, H.G., 1975, Bull. Am. Phys. Soc. {\bf 20}, 60.
%
\harvarditem{Dehmelt}{1982}{Dehmelt2}
Dehmelt, H.G., 1982, IEEE Trans. Instrum. Meas. {\bf IM-31}, 83.
%
\harvarditem{Dehmelt}{1987}{Dehmelt3}
Dehmelt, H., 1987, Nature {\bf 325}, 581.
%
\harvarditem{Dicke}{1981}{Dicke1}
Dicke, R.H., 1981, Am. J. Phys. {\bf 49}, 925.
%
\harvarditem{Diosi}{1988}{Diosi1}
Di{\`o}si, L., 1988, J. Phys. A {\bf 21}, 2885.
%
%
\harvarditem{Diosi}{1989}{Diosi3}
Di{\`o}si, L., 1989, Phys. Rev. A {\bf 40}, 1165.
%
%
\harvarditem{Diosi {\em et al.}}{1995}{Diosi5}
Di{\`o}si, L., N. Gisin, J.J. Halliwell, I.C. Percival, 
1995, Phys. Rev. Lett. {\bf 74}, 203.
%
%
\harvarditem{Dum {\em et al.}}{1992}{Dum2}
Dum, R., A.S. Parkins, P. Zoller, and C.W. Gardiner, 1992, Phys. Rev.
 A {\bf 46}, 4382.
%
\harvarditem{Dum, Zoller and Ritsch}{1992}{Dum1}
Dum, R., P. Zoller, and H. Ritsch, 1992, Phys. Rev. A {\bf 45}, 4879.
%
\harvarditem{Ekert and Josza}{1996}{Ekert2}
Ekert, A., and R. Josza, 1996, Rev. Mod. Phys. {\bf 68}, 733.
%
\harvarditem{Ekert and Knight}{1995}{Ekert1}
Ekert, A., and P.L. Knight, 1995, Am. Journ. Phys. {\bf 63}, 415.
%
%
\harvarditem{Erber {\em et al.}}{1989}{Erber2}
Erber, T., P. Hammerling, G. Hockney, M. Porrati, and 
S. Putterman, 1989, Ann. Phys. (N. Y.) {\bf 190}, 254.
%
\harvarditem{Erber and Putterman}{1985}{Erber1}
Erber, T., and S. Putterman, 1985, Nature {\bf 318}, 41.
%
%
\harvarditem{Finn {\em et al.}}{1989}{Finn2}
Finn, M. A., G. W. Greenlees, T. W. Hodapp, D. A. Lewis, 1989, 
Phys. Rev. A {\bf 40}, 1704.
%
\harvarditem{Finn {\em et al.}}{1986}{Finn1}
Finn, M.A., G.W. Greenlees, D.A. Lewis, 1986, Opt. Comm. {\bf 60}, 149.
%
\harvarditem{Frerichs and Schenzle}{1991}{Frerichs1}
Frerichs, V., and A. Schenzle, 1991, Phys. Rev. A {\bf 44}, 1962.
%
%
\harvarditem{Gardiner}{1992}{Gardiner4}
Gardiner, C.W., 1992, {\em Quantum Noise} (Springer, Berlin).
%
\harvarditem{Gardiner}{1993}{Gardiner2}
Gardiner, C.W., 1993, Phys. Rev. Lett. {\bf 70}, 2269.
%
%
\harvarditem{Gardiner and Parkins}{1994}{Gardiner3}
Gardiner, C.W., and A.S. Parkins, 1994, Phys. Rev. A {\bf 50}, 1792.
%
\harvarditem{Gardiner {\em et al.}}{1992}{Gardiner1}
Gardiner, C.W., A.S. Parkins, and P. Zoller, 1992, Phys. Rev. A {\bf 46}, 4363.
%
%
\harvarditem{Garraway {\em et al.}}{1995}{Garraway5}
Garraway, B.M., M. S. Kim, and P.L. Knight, 1995, Opt. Comm. {\bf 117}, 560.
%
\harvarditem{Garraway and Knight}{1994a}{Garraway2}
Garraway, B.M., and P.L. Knight, 1994a, Phys. Rev. A {\bf 49}, 1266.
%
\harvarditem{Garraway and Knight}{1994b}{Garraway3}
Garraway, B.M., and P.L. Knight, 1994b, Phys. Rev. A {\bf 50}, 2548.
%
\harvarditem{Garraway, Knight and Steinbach}{1995}{Garraway4}
Garraway, B.M., P.L. Knight, and J. Steinbach, 1995, 
Appl. Phys. {\bf B 60}, 63.
%
\harvarditem{Gell-Mann and Hartle}{1990}{Gell-Mann1}
Gell-Mann, M., and J. B. Hartle, 1990 in 
{\em Complexity, Entropy, and the Physics of Information}, Santa Fe 
Institute Studies in the Science of Complexity, v. VIII, edited by W. Zurek, 
(Addison-Wesley, Reading).
%
\harvarditem{Gell-Mann and Hartle}{1993}{Gell-Mann2}
Gell-Mann, M., and J. B. Hartle, 1993, Phys. Rev. D {\bf 47}, 3345.
%
\harvarditem{Ghirardi {\em et al.}}{1990}{Ghirardi2}
Ghirardi, G.C., Ph. Pearle, and A. Rimini, 1990, 
Phys. Rev. A {\bf 42} 78.
%
\harvarditem{Ghirardi {\em et al.}}{1986}{Ghirardi1}
Ghirardi, G.C., A. Rimini, and T. Weber, 1986, Phys. Rev. D
{\bf 34}, 470.
%
\harvarditem{Gisin}{1984}{Gisin1}
Gisin, N., 1984, Phys. Rev. Lett. {\bf 52}, 1657.
%
\harvarditem{Gisin}{1989}{Gisin2}
Gisin, N., 1989, Helv. Phys. Acta {\bf 62}, 363.
%
\harvarditem{Gisin}{1993}{Gisin8}
Gisin, N., 1993, J. Mod. Optics {\bf 40}, 2313.
%
%
\harvarditem{Gisin {\em et al.}}{1993}{Gisin9}
Gisin, N., P.L. Knight, I.C. Percival, R. C. Thompson, 
D. C. Wilson, 1993, J. Mod. Optics {\bf 40}, 1663.
%
\harvarditem{Gisin and Percival}{1992a}{Gisin3}
Gisin, N., and I.C. Percival, 1992a, J. Phys. A {\bf 25}, 5677.
%
%
\harvarditem{Gisin and Percival}{1993a}{Gisin6}
Gisin, N., and I.C. Percival, 1993a, J. Phys. A {\bf 26}, 2233.
%
\harvarditem{Gisin and Percival}{1993b}{Gisin7}
Gisin, N., and I.C. Percival, 1993b, J. Phys. A {\bf 26}, 2245.
%
%
%
\harvarditem{Goetsch and Graham}{1993}{Goetsch1}
Goetsch, P., and R. Graham, 1993, Ann. Phys. NY {\bf 3}, 706.
%
\harvarditem{Goetsch and Graham}{1994}{Goetsch2}
Goetsch, P., and R. Graham, 1994, Phys. Rev. A {\bf 50}, 5242.
%
\harvarditem{Goetsch {\em et al.}}{1995}{Goetsch3}
Goetsch, P., R. Graham and F. Haake, 1995, Phys. Rev. A {\bf 51},
136.
%
\harvarditem{Granzow}{1996}{Granzow96}
Granzow, C.M., 1996, Diploma thesis (University of Stuttgart)
%
\harvarditem{Griffiths}{1984}{Griffiths1}
Griffiths, R., 1984, J. Stat. Phys. {\bf 36}, 219.
%
%
\harvarditem{Grochmalicki and Lewenstein}{1989a}{Grochmalicki1}
Grochmalicki, J., and M. Lewenstein, 1989a, 
Phys. Rev. A {\bf 40}, 2517.
%
\harvarditem{Grochmalicki and Lewenstein}{1989b}{Grochmalicki2}
Grochmalicki, J., and M. Lewenstein, 1989b, 
Phys. Rev. A {\bf 40}, 2529.
%
\harvarditem{Haake}{1973}{Haake1}
Haake, F., 1973, {\em Statistical Treatment of Open 
Systems by Generalized Master Equations}, 
Springer Tracts of Mod. Physics {\bf 66},  
(Springer, Berlin), p. 98.
%
\harvarditem{Haroche}{1984}{Haroche1}
Haroche, S., J.M. Raimond, 1984, Adv. Atom. Mol. Phys. {\bf 20}, 347.
%
\harvarditem{Hegerfeldt}{1993}{Hegerfeldt3}
Hegerfeldt, G.C., 1993, Phys. Rev. A {\bf 47}, 449.
%
\harvarditem{Hegerfeldt and Plenio}{1992}{Hegerfeldt2}
Hegerfeldt, G.C., and M.B. Plenio, 1992, 
Phys. Rev. A {\bf 46}, 373.
%
\harvarditem{Hegerfeldt and Plenio}{1993}{Hegerfeldt4}
Hegerfeldt, G.C., and M.B. Plenio, 1993, Phys. Rev. A {\bf 47},
2186.
%
\harvarditem{Hegerfeldt and Plenio}{1994}{Hegerfeldt5}
Hegerfeldt, G.C., and M.B. Plenio, 1994, Quantum Opt. {\bf 6}, 15.
%
\harvarditem{Hegerfeldt and Plenio}{1995a}{Hegerfeldt6}
Hegerfeldt, G.C., and M.B. Plenio, 1995a, Z. Phys. B. {\bf 96},
533.
%
\harvarditem{Hegerfeldt and Plenio}{1995b}{Hegerfeldt7}
Hegerfeldt, G.C., and M.B. Plenio, 1995b, Phys. Rev. A {\bf 52},
3333.
%
\harvarditem{Hegerfeldt and Plenio}{1996}{Hegerfeldt8}
Hegerfeldt, G.C., and M.B. Plenio, 1996, Phys. Rev. A {\bf 53},
1164.
%
\harvarditem{Hegerfeldt and Sondermann}{1996}{Hegerfeldt9}
Hegerfeldt, G.C, and D. Sondermann, 1996, Quant. Opt. {\bf 8}, 121.
%
\harvarditem{Hegerfeldt and Wilser}{1991}{Hegerfeldt1}
Hegerfeldt, G.C., and T. S. Wilser, 1991, 
{\em Proceedings of the II. International Wigner Symposium}, edited by 
H. D. Doebner, W. Scherer, and F. Schroeck (World Scientific, Singapore), p. 104.
%
\harvarditem{Hendriks and Nienhuis}{1988}{Hendriks1}
Hendriks, B.H.W., and G. Nienhuis, 1988, 
J. Mod. Opt. {\bf 35}, 1331.
%
\harvarditem{Herkommer {\em et al}}{1996}{Herkommer96}
Herkommer, A.M., H.J. Carmichael, and W.P. Schleich, 1996, 
Quant. Opt. {\bf 8}, 189.
%
\harvarditem{Holland and Cooper}{1996}{Holland1}
Holland, M., and J. Cooper, 1996, private communication
%
\harvarditem{Holland {\em et al}}{1996}{Holland2}
Holland, M., S. Marksteiner, P. Marte, and P. Zoller, 1996, Phys. Rev. Lett.
{\bf 76}, 3683.
%
\harvarditem{Horvath {\em et al}}{1997}{Horvath1}
Horvath, G.Z.K., P.L. Knight, and R.C. Thompson, 1997, Cont. Phys.
{\bf 38}
%
\harvarditem{Imamoglu}{1993}{Imamoglu1}
Imamoglu, A., 1993, Phys. Rev. A {\bf 48}, 770.
%
\harvarditem{Imamoglu}{1994}{Imamoglu2}
Imamoglu, A., 1994, Phys. Rev. A {\bf 50}, 3650.
%
\harvarditem{Itano {\em et al}}{1988}{Itano0}
Itano, W.M., J.C. Bergquist, and D.J. Wineland, 1988, Phys. Rev. A 
{\bf 38}, 559.
%
%
%
\harvarditem{Itano {\em et al}}{1990}{Itano3}
Itano, W.M., D.J. Heinzen, J.J. Bollinger, and
D.J. Wineland, 1990, Phys. Rev. A {\bf 41}, 2295.
%
%
\harvarditem{Javanainen}{1986a}{Javanainen1}
Javanainen, J., 1986a, Phys. Rev. A {\bf 33}, 2121.
%
\harvarditem{Javanainen}{1986b}{Javanainen3}
Javanainen, J., 1986b, Phys. Scr. T{\bf 12}, 67.
%
\harvarditem{Javanainen}{1986c}{Javanainen2}
Javanainen, J., 1986c, Jour. Opt. Soc. B {\bf 3}, 98.
%
\harvarditem{Javanainen}{1992}{Javanainen4}
Javanainen, J., 1992, Europhys. Lett. {\bf 17}, 407.
%
\harvarditem{Jayarao {\em et al.}}{1990}{Jayarao1}
Jayarao, A. S., R. D'Souza, and S.V. Lawande, 1990, 
Phys. Rev. A {\bf 41}, 1533.
%
%
\harvarditem{Kelley and Kleiner}{1964}{Kelley1}
Kelley, P.L., and W. H. Kleiner, 1964, Phys. Rev. {\bf 136}, 316.
%
\harvarditem{Kim}{1987}{Kim3}
Kim, M. S., 1987, PhD Thesis (Imperial College, University of London).
%
\harvarditem{Kim and Knight}{1987}{Kim2}
Kim, M. S., and P.L. Knight, 1987, Phys. Rev. A {\bf 36}, 5265.
%
\harvarditem{Kim {\em et al.}}{1987}{Kim1}
Kim, M. S., P.L. Knight, and K. Wodkiewicz, 1987, Opt. Comm. {\bf 62},
385.
%
\harvarditem{Kim {\em et al.}}{1989}{Kim4}
Kim, M. S., F.A.M. de Oliveira, and P.L. Knight, 1989, Opt. Comm. {\bf 70},
473.
%
\harvarditem{Kimble {\em et al.}}{1986}{Kimble1}
Kimble, H. J., R.J. Cook, and A. L. Wells, 1986, Phys. Rev. A 
{\bf 34}, 3190.
%
\harvarditem{Kimble {\em et al.}}{1977}{Kimble0}
Kimble, H. J., M. Dagenais, and L. Mandel, 1977, Phys. Rev. Lett. 
{\bf 39}, 691.
%
\harvarditem{Knight and Garraway}{1996}{Knight1}
Knight, P.L., and B.M. Garraway, 1996,
{\em Quantum Dynamics of Simple Systems. Proceedings of the 
Forty Fourth Scottish Universities Summer School in Physics Stirling}
edited by G-L. Oppo, S.M. Barnett, E. Riis and M. Wilkinson 
(Institute of Physics Publishing, Bristol), p. 199.
%
\harvarditem{Knight {\em et al.}}{1986}{Knight3}
Knight, P. L., R. Loudon, and D.T. Pegg, 1986, Nature {\bf 323}, 
608.
%
\harvarditem{Knight and Milonni}{1980}{Knight2}
Knight, P. L., and P.W. Milonni, 1980, Phys. Rep. {\bf 66}, 21.
%
\harvarditem{Knight and Pegg}{1982}{Knight4}
Knight, P.L., and D.T. Pegg, 1982, J. Phys. B {\bf 15}, 3211.
%
\harvarditem{Kochan and Carmichael}{1994}{Kochan1}
Kochan, P., and H. J. Carmichael, 1994, Phys. Rev. A {\bf 50}, 1700.
%
\harvarditem{K{\"o}hler}{1996}{Koehler1}
K{\"o}hler, T., Phys. Rev. A {\bf 54}, 4544.
%
%
\harvarditem{Lawande {\em et al.}}{1989a}{Lawande2}
Lawande, S.V., Q. V. Lawande, and B. N. Jagatap, 1989a, 
Phys. Rev. A {\bf 40}, 3434.
%
\harvarditem{Lawande {\em et al.}}{1989b}{Lawande4}
Lawande, S.V., B. N. Jagatap, and Q. V. Lawande, 1989b, Opt. Comm. 
{\bf 73}, 126.
%
\harvarditem{Lawande {\em et al.}}{1990}{Lawande5}
Lawande, Q. V., B. N. Jagatap, and S.V. Lawande, 1990, 
Phys. Rev. A {\bf 42}, 4343.
%
\harvarditem{Lax}{1963}{Lax1}
Lax, M, 1963, Phys. Rev. {\bf 129}, 2342.
%
\harvarditem{Lenstra}{1982}{Lenstra1}
Lenstra, D., 1982, Phys. Rev. A {\bf 26}, 3369.
%
\harvarditem{Lewenstein and Javanainen}{1987}{Lewenstein1}
Lewenstein, M., and J, Javanainen, 1987, 
Phys. Rev. Lett. {\bf 59}, 775.
%
\harvarditem{Lewenstein and Javanainen}{1988}{Lewenstein2}
Lewenstein, M., and J. Javanainen, 1988, 
IEEE J. Quant. Electr. {\bf 24}, 1403.
%
\harvarditem{Ligare}{1988}{Ligare1}
Ligare, M., 1988, Phys. Rev. A {\bf 37}, 3293.
%
\harvarditem{Lindblad}{1976}{Lindblad1}
Lindblad, G., 1976, Comm. Math. Phys. {\bf 48}, 119.
%
\harvarditem{Loudon}{1983}{Loudon1}
Loudon, R., 1983, {\em The Quantum Theory of Light.}
 (Oxford Science Publications).
%
\harvarditem{Loudon and Knight}{1987}{Loudon2}
Loudon, R., and P.L. Knight, 1987, J. Mod. Optics {\bf 34}, 1.
%
\harvarditem{L{\"u}ders}{1951}{Lueders1}
L{\"u}ders, G., 1951, Ann. der Physik {\bf 8}, 323.
%
\harvarditem{Mahler and Weberru{\ss}}{1995}{Mahler1}
Mahler, G., and V. A. Weberru{\ss}, 1995, {\em Quantum networks:
Dynamics of Open Nanostructures} (Springer Berlin).
%
\harvarditem{Mandel}{1979}{Mandel1}
Mandel, L., 1979, Opt. Lett. {\bf 4}, 205.
%
\harvarditem{Manka {\em et al.}}{1993}{Manka1}
Manka, A. S., H. M. Doss, L. M. Narducci, P. Ru, and G.-L. Oppo,
1993, Phys. Rev. A {\bf 47}, 1378.
%
\harvarditem{Marte {\em et al.}}{1993a}{Marte1}
Marte, P., R. Dum, R. Taieb, P.D. Lett, and P. Zoller, 
1993a, Phys. Rev. Lett. {\bf 71}, 1335.
%
\harvarditem{Marte {\em et al.}}{1993b}{Marte2}
Marte, P., R. Dum, R. Taieb, and P. Zoller, 1993b, Phys. Rev. A 
{\bf 47}, 1378.
%
\harvarditem{Meekhof {\em et al.}}{1996}{Meekhof1}
Meekhof, D.M., C. Monroe, B.E. King, W.M. Itano and D.J. Wineland, 1996,
Phys. Rev. Lett. {\bf 76}, 1796.
%
%
%
%
\harvarditem{Milburn}{1991}{Milburn3}
Milburn, G.J., 1991, Phys. Rev. A {\bf 44}, 5401.
%
\harvarditem{Milonni}{1976}{Milonni1}
Milonni, P.W., 1976, Phys. Rep. {\bf 25}, 1.
%
\harvarditem{Misra and Sudarshan}{1977}{Misra1}
Misra, B., and E. C. G. Sudarshan, 1977, 
J. Math. Phys. {\bf 18}, 756.
%
\harvarditem{Mollow}{1969}{Mollow4}
Mollow, B.R., 1969, Phys. Rev. {\bf 188}, 1969.
%
\harvarditem{Mollow}{1972a}{Mollow2}
Mollow, B.R., 1972a, Phys. Rev. A {\bf 5}, 1522.
%
\harvarditem{Mollow}{1972b}{Mollow3}
Mollow, B.R. , 1972b, Phys. Rev. A {\bf 5}, 2217.
%
\harvarditem{Mollow}{1975}{Mollow1}
Mollow, B.R., 1975, Phys. Rev. A {\bf 12}, 1919.
%
\harvarditem{Mollow}{1981}{Mollow5}
Mollow, B.R., 1981, Prog. in Optics {\bf XX}, 1.
%
\harvarditem{M{\o}lmer}{1994}{Molmer2}
M{\o}lmer, K., 1994 {\em Lectures presented at the Trieste 
Winter School on Quantum Optics 1994} (International Centre for 
Theoretical Physics, Trieste).
%
\harvarditem{M{\o}lmer and Castin}{1996}{Molmer3}
M{\o}lmer, K., and Y. Castin, 1996, Quant. Opt. {\bf 8}, 49.
%
\harvarditem{M{\o}lmer {\em et al.}}{1993}{Molmer1}
M{\o}lmer, K., Y. Castin, and J. Dalibard, 1993, J. Opt. Soc. Am. 
{\bf B 10}, 524.
%
\harvarditem{Moya-Cessa {\em et al.}}{1993}{Moya-Cessa1}
Moya-Cessa, H., V. Bu\v{z}ek, M.S. Kim, and P.L. Knight, 
1993, Phys. Rev. A {\bf 48}, 3900.
%
\harvarditem{Mu}{1994}{Mu1}
Mu, Y., 1994, Opt. Comm. {\bf 110}, 334.
%
%
\harvarditem{Nagourney {\em et al.}}{1986a}{Nagourney1}
Nagourney, W., J. Sandberg, and H.G. Dehmelt, 1986a, Phys. Rev.
Lett. {\bf 56}, 2797.
%
\harvarditem{Nagourney {\em et al.}}{1986b}{Nagourney3}
Nagourney, W., J. Sandberg, and H.G. Dehmelt, 1986b,
J. Opt. Soc. B {\bf 3}, 252.
%
\harvarditem{Nakajima}{1958}{Nakajima1}
Nakajima, S., 1958, Prog. Theor. Phys. {\bf 20}, 948.
%
%
\harvarditem{Narducci {\em et al.}}{1990a}{Narducci2}
Narducci, L.M., G.-L. Oppo, M.O. Scully, 1990a, Opt. Comm. 
{\bf 75}, 111.
%
\harvarditem{Narducci {\em et al.}}{1990b}{Narducci1}
Narducci, L.M., M. O. Scully, G.-L. Oppo, P. Ru, and 
J.R. Tredicce, 1990b, Phys. Rev. A {\bf 42}, 1630.
%
\harvarditem{Narozhny {\em et al}}{1981}{Narozhny1}
Narozhny, N.B., J.J. Sanchez-Mondragon and J.H. Eberly, 1981, 
Phys. Rev. A {\bf 23}, 236.
%
%
\harvarditem{Neuhauser {\em et al.}}{1980}{Neuhauser2}
Neuhauser, W., M. Hohenstatt, P. Toschek, and 
H. G. Dehmelt, 1980, Phys. Rev. A {\bf 22}, 1137.
%
\harvarditem{von Neumann}{1955}{Neumann1}
von Neumann, J., 1955, {\em Math. Foundations of Quantum Mechanics}
(Princeton University Press).
%
\harvarditem{Nienhuis}{1987}{Nienhuis1}
Nienhuis, G., 1987, Phys. Rev. A {\bf 35}, 4639.
%
\harvarditem{Omnes}{1988}{Omnes1}
Omn{\'e}s, R., 1988, J. Stat. Phys. {\bf 53}, 893, 933, 957.
%
\harvarditem{Omnes}{1989}{Omnes2}
Omn{\'e}s, R., 1989, J. Stat. Phys. {\bf 57}, 359.
%
\harvarditem{Omnes}{1994}{Omnes3}
Omn{\'e}s, R., 1994, {\em The Interpretation of Quantum Mechanics},
(Princeton University Press, Princeton). 
%
\harvarditem{Paul}{1990}{Paul2}
Paul, W., 1990, Rev. Mod. Phys. {\bf 62}, 531.
%
\harvarditem{Paul {\em et al.}}{1958}{Paul1}
Paul, W., O. Osberghaus, and E. Fischer, 1958, 
{\em Ein Ionenk{\"a}fig.} (Forschungsberichte des Wirtschafts- und 
Verkehrsministeriums Nordrhein-Westfalen), p. 415.
%
\harvarditem{Pearle}{1976}{Pearle1}
Pearle, Ph., 1976, Phys. Rev. D {\bf 13}, 857.
%
\harvarditem{Pegg}{1980}{Pegg4}
Pegg, D.T., 1980, in {\em Laser Physics}, edited by J.D. Harvey and
D.F. Walls (Academic Press, Sydney), p. 33.
%
\harvarditem{Pegg and Knight}{1988a}{Pegg3}
Pegg, D.T., and P.L. Knight, 1988a, Phys. Rev. A {\bf 37}, 4303.
%
\harvarditem{Pegg and Knight}{1988b}{Pegg3b}
Pegg, D.T., and P.L. Knight, 1988b, J. Phys. D {\bf 21}, 128.
%
\harvarditem{Pegg {\em et al.}}{1986a}{Pegg1}
Pegg, D.T., R. Loudon, and P.L. Knight, 1986a, Nature {\bf 323}, 608.
%
\harvarditem{Pegg {\em et al.}}{1986b}{Pegg2}
Pegg, D.T., R. Loudon, and P.L. Knight, 1986b, Phys. Rev. A {\bf 33}, 4085.
%
\harvarditem{Percival}{1994a}{Percival1}
Percival, I.C., 1994a, J. Phys. A {\bf 27}, 1003.
%
\harvarditem{Percival}{1994b}{Percival2}
Percival, I.C., 1994b, Proc. Roy. Soc. London  A {\bf 447}, 189.
%
\harvarditem{Percival}{1995}{Percival4}
Percival, I.C., 1995, Proc. Roy. Soc. London  A {\bf 451}, 503.
%
\harvarditem{Percival and Strunz}{1996}{Percival3}
Percival, I.C., and W.T. Strunz, preprint quant-ph/9607011
submitted to Proc. Roy. Soc. London 1996.
%
\harvarditem{Piraux {\em et al.}}{1990}{Piraux1}
Piraux, B. , R. Bhatt and P.L. Knight, 1990, Phys. Rev. A {\bf 41}, 6296.
%
\harvarditem{Plenio}{1994}{Plenio1}
Plenio, M.B., 1994, Doctoral thesis, (University of 
G{\"o}ttingen).
%
\harvarditem{Plenio}{1996}{Plenio2}
Plenio, M.B., 1996, J. Mod. Optics {\bf 43}, 753.
%
\harvarditem{Plenio {\em et al.}}{1996a}{Plenio4}
Plenio, M.B. and P.L. Knight, 1996a, Phys. Rev. A {\bf 53}, 2986.
%
\harvarditem{Plenio {\em et al.}}{1996b}{Plenio5}
Plenio, M.B. and P.L. Knight, 1996b, submitted to Proc. Roy. Soc. A
%
\harvarditem{Plenio {\em et al.}}{1996}{Plenio3}
Plenio, M.B., P.L. Knight, and R.C. Thompson, 
1996, Opt. Comm. {\bf 123}, 278.
%
\harvarditem{Porrati and Putterman}{1987}{Porrati1}
Porrati, M., and S. Putterman, 1987, Phys. Rev. A {\bf 36}, 929.
%
\harvarditem{Porrati and Putterman}{1989}{Porrati2}
Porrati, M., and S. Putterman, 1989, Phys. Rev. A {\bf 39}, 3010.
%
\harvarditem{Power}{1995a}{Power2}
Power, W.L., 1995a, PhD Thesis (Imperial College, University of London).
%
\harvarditem{Power}{1995b}{Power1}
Power, W.L., 1995b, J. Mod. Optics {\bf 42}, 913.
%
\harvarditem{Power and Knight}{1996}{Power3}
Power, W.L., and P.L. Knight, 1996, Phys. Rev. A {\bf 53}, 1052.
%
\harvarditem{Reibold}{1992}{Reibold1}
Reibold, R.,  1992, Physica A {\bf 190}, 413.
%
\harvarditem{Reibold}{1993}{Reibold2}
Reibold, R., 1993, J. Phys. A {\bf 26}, 179.
%
\harvarditem{Rempe {\em et al}}{1987}{Rempe1}
Rempe, G., and H. Walther, 1987, Phys. Rev. Lett. {\bf 58}, 353.
%
\harvarditem{Reynaud {\em et al.}}{1988}{Reynaud1}
Reynaud, S.,  J. Dalibard, and C. Cohen-Tannoudji, 1988, 
IEEE J. Quant. Electr. {\bf 24}, 1395.
%
\harvarditem{Rigo and Gisin}{1996}{Rigo96}
Rigo, M., and N. Gisin, Quant. Opt. {\bf 8}, 255.
%
\harvarditem{Salama and Gisin}{1993}{Salama1}
Salama, Y., and N. Gisin, 1993, Phys. Lett. A {\bf 181}, 269.
%
\harvarditem{Saleh}{1978}{Saleh1}
Saleh, B., 1978, {\em Photoelectron Statistics}, Springer Series in 
Optical Sciences Vol. 6  (Springer, Berlin).
%
\harvarditem{Sauter, Blatt, Neuhauser, and Toschek}{1986}{Sauter3}
Sauter, T., R. Blatt, W. Neuhauser, and P.E. Toschek, 1986, 
Opt. Comm. {\bf 60}, 287.
%
\harvarditem{Sauter {\em et al.}}{1986b}{Sauter1}
Sauter, T., W. Neuhauser, R. Blatt, and P.E. Toschek, 1986b, Phys.
Rev. Lett. {\bf 57}, 1696.
%
\harvarditem{Sauter {\em et al.}}{1986c}{Sauter2}
Sauter, T., W. Neuhauser, R. Blatt, and P.E. Toschek, 1986c, 
J. Opt. Soc. B {\bf 3}, 252.
%
\harvarditem{Schack {\em et al.}}{1995}{Schack0}
Schack, R., T. A. Brun, and I. C. Percival, 1995, J. Phys. A {\bf 28}, 5401.
%
\harvarditem{Schack {\em et al.}}{1996a}{Schack1}
Schack, R., T. A. Brun, and I. C. Percival, 1996a, 
Phys. Rev. A {\bf 53}, 2694.
%
\harvarditem{Schenzle and Brewer}{1986}{Schenzle2}
Schenzle, A., and R.G. Brewer, 1986, 
Phys. Rev. A {\bf 34}, 3127.
%
%
\harvarditem{Schenzle {\em et al.}}{1986}{Schenzle1}
Schenzle, A., R.G. De Voe, and R.G. Brewer, 
1986, Phys. Rev. A {\bf 33}, 2127.
%
\harvarditem{Schr{\"o}dinger}{1952}{Schroedinger1}
Schr{\"o}dinger, E., British Journal for the Philosophy of 
Science {\bf III}, August 1952.
%
\harvarditem{Schubert {\em et al}}{1992}{Schubert1}
Schubert, M., I. Siemers, R. Blatt, W. Neuhauser, P.E. Toschek, 1992,
Phys. Rev. Lett. {\bf 68}, 3016.
%
\harvarditem{Shor}{1994}{Shor1}
Shor, P.W., 1994, in {\em Proceedings of the 35th Annual Symposium on the
Foundations of Computer Science}, Los Alamitos, CA edited by S. Goldwasser,
(IEEE Computer Society Press, New York), p.124.
%
\harvarditem{Shor}{1996}{Shor2}
Shor, P.W., 1996, Phys. Rev. A {\bf 52}, R2493.
%
\harvarditem{Shore and Knight}{1993}{Shore1} 
Shore, B.W. , and P.L. Knight, 1993, J. Mod. Optics {\bf 40}, 1195.
%
%
\harvarditem{Sondermann}{1995a}{Sondermann1}
Sondermann, D., 1995a, J. Mod. Optics {\bf 42}, 1659.
%
\harvarditem{Sondermann}{1995b}{Sondermann2}
Sondermann, D., 1995b, {\em Nonlinear, Deformed and Irreversible Quantum 
Systems}, edited by H. D. Doebner, V.K. Dobrev, and P. Nattermann (World
Scientific, Singapore), p. 273.
%
\harvarditem{Spiller and Ralph}{1994}{Spiller1}
Spiller, T.P., and J.F. Ralph, 1994, Phys. Lett. A{\bf 194}, 235.
%
\harvarditem{Srinivas and Davies}{1981}{Srinivas1}
Srinivas,  M.D., and E.B. Davies, 1981, Opt. Acta {\bf 28}, 981.
%
\harvarditem{Srinivas and Davies}{1982}{Srinivas2}
Srinivas, M.D. and E.B. Davies, 1982, Opt. Acta {\bf 29}, 235.
%
\harvarditem{Steinbach {\em et al.}}{1995a}{Steinbach1}
Steinbach, J., B.M. Garraway, and P.L. Knight, 1995, Phys. Rev.
A {\bf 51}, 3302.
%
%
%
\harvarditem{Steane}{1996}{Steane1}
Steane, A.M., 1996, Phys. Rev. Lett. {\bf 77}, 793.
%
\harvarditem{Stratonovitch}{1963}{Stratonovitch1}
Stratonovitch, R. L., 1963, {\em Topics in the Theory 
of Random Noise}, Vol.1 (Gordon and Breach, New York).
%
\harvarditem{Swain}{1994}{Swain1}
Swain, S., 1994, Phys. Rev. Lett. {\bf 73}, 1493.
%
\harvarditem{Teich {\em et al}}{1989}{Teich1}
Teich, W. G., G. Anders, and G. Mahler, 1989, Phys. Rev. Lett. 
{\bf 62}, 1.
%
\harvarditem{Teich and Mahler}{1992}{Teich2}
Teich, W. G., and G. Mahler, 1992, Phys. Rev. A {\bf 45}, 3300.
%
\harvarditem{Thompson}{1996}{Thompson}
Thompson, R. C., 1996, private communication
%
\harvarditem{Tian and Carmichael}{1992}{Tian1}
Tian, L., and H. J. Carmichael, 1992, Phys. Rev. A {\bf 46}, 6801.
%
\harvarditem{Vogel and Welsch}{1994}{Vogel1}
Vogel, W., D.-G. Welsch, 1994, {\em Lectures on Quantum Optics}, 
(Akademie Verlag, Berlin)
%
%
\harvarditem{Weisskopf and Wigner}{1930}{Weisskopf1}
Weisskopf, V., and E. Wigner, 1930, Z. Phys. {\bf 63}, 54.
%
\harvarditem{Wilser}{1991}{Wilser1}
Wilser, T. S., 1991, Doctoral thesis (University of G{\"o}ttingen)
%
%
\harvarditem{Wiseman}{1996}{Wiseman4}
Wiseman, H. M., 1996, Quant. Opt. {\bf 8}, 205.
%
\harvarditem{Wiseman and Milburn}{1993a}{Wiseman1}
Wiseman, H. M., and G. J. Milburn, 1993a, 
Phys. Rev. A {\bf 47}, 642.
%
\harvarditem{Wiseman and Milburn}{1993b}{Wiseman2}
Wiseman, H. M., and G. J. Milburn, 1993b, 
Phys. Rev. A {\bf 47}, 1652.
%
\harvarditem{Wiseman and Milburn}{1994}{Wiseman3}
Wiseman, H. M., and G. J. Milburn, 1994, 
Phys. Rev. A {\bf 49}, 1350.
%
%
\harvarditem{Yamada and Berman}{1990}{Yamada1}
Yamada, K., and P. R. Berman, 1990, Phys. Rev. A {\bf 41}, 453.
%
\harvarditem{Yu}{1996}{Yu1}
Yu, T., 1996, Imperial College preprint TP/95-96/44, gr-qc/9605071.
%
\harvarditem{Zoller {\em et al.}}{1987}{Zoller1}
Zoller, P., M. Marte, and D. F. Walls, 1987, Phys. Rev. A {\bf 35},
198.
%
\harvarditem{Zoller and Gardiner}{1995}{Zoller2}
Zoller, P., and C. W. Gardiner, 1995, 
{Lecture notes of the Les Houches Summer School on Quantum Fluctuations.} 
(Elsevier Science Publishers B. V.).
%
\harvarditem{Zwanzig}{1960}{Zwanzig1}
Zwanzig, R., 1960, Lect. Theor. Phys. (Boulder) {\bf 3}, 106.

\end{references}
\end{document}